\documentclass[twocolumn,superscriptaddress,showpacs,prd,
aps,amsmath,amssymb,nofootinbib,eqsecnum]{revtex4}
\usepackage{hyperref}
\usepackage{graphicx}
\usepackage{bm} 
\usepackage{mathrsfs} 

\usepackage{color}
\usepackage{comment}

\newcommand{\be}{\begin{equation}}
\newcommand{\ee}{\end{equation}}
\newcommand{\bea}{\begin{eqnarray}}
\newcommand{\eea}{\end{eqnarray}}
\newcommand{\M}{\mathscr{M}}
\newcommand{\T}{\mathscr{T}}

\newcommand{\w}[1]{\bm{#1}}
\newcommand{\vv}[1]{\bm{#1}}
\newcommand{\vw}[1]{\vec{\w{#1}}}
\newcommand{\uu}[1]{\underline{\bm{#1}}}

\newcommand{\R}{\mathbb{R}}
\newcommand{\Lie}[1]{\pounds_{#1}\,}
\newcommand{\wnab}{\w{\nabla}}

\newcommand{\dd}{\bm{\mathrm{d}}}

\newcommand{\eps}{\epsilon}
\newcommand{\weps}{\w{\eps}}
\newcommand{\vep}{\varepsilon}
\newcommand{\diver}{\wnab\!\cdot}

\usepackage[normalem]{ulem}
\usepackage{amsbsy}
\usepackage{amssymb}
\usepackage{wasysym}
\usepackage{amstext}

\begin{document}

\title{Conservation laws and evolution schemes \\in geodesic, hydrodynamic and magnetohydrodynamic flows}

\author{Charalampos Markakis} 
\email{markakis@illinois.edu}
\affiliation{
NCSA, University of Illinois at
Urbana-Champaign, IL 61801, USA}
\affiliation{
Mathematical Sciences, University of Southampton, Southampton SO17 1BJ, United Kingdom}
\author{K\=oji Ury\=u}
\email{uryu@sci.u-ryukyu.ac.jp}
\affiliation{
Department of Physics, University of the Ryukyus, Senbaru, 
Nishihara, Okinawa 903-0213, Japan}
\author{Eric Gourgoulhon}
\email{eric.gourgoulhon@obspm.fr}
\affiliation{
LUTh, UMR 8102 du CNRS,
Observatoire de Paris, Universit\'e Paris Diderot, F-92190 Meudon, France}

\author{Jean-Philippe Nicolas}
\email{jean-philippe.nicolas@univ-brest.fr}
\affiliation{
D\'epartement de Math\'ematiques,
Universit\'e de Bretagne Occidentale\\
6 avenue Victor Le Gorgeu,
29238 Brest Cedex 3,
France}

\author{Nils Andersson}
\email{N.A.Andersson@soton.ac.uk}
\affiliation{
Mathematical Sciences, University of Southampton, Southampton SO17 1BJ, United
Kingdom}

\author{Athina Pouri}
\email{athpouri@phys.uoa.gr}
\affiliation{
RCAAM, Academy of Athens,
Soranou Efesiou 4, 11527, Athens, Greece}

\author{Vojt\v ech Witzany}
\email{witzany@zarm.uni-bremen.de}
\affiliation{
ZARM, Universit\"at Bremen, Am Fallturm, 28359 Bremen, Germany}

\date{28 December 2016}  

\begin{abstract} 

   Carter and Lichnerowicz have established that  barotropic fluid flows
are conformally geodesic and obey Hamilton's principle. This
variational
approach can accommodate neutral, or charged and poorly conducting,
fluids. We show that, unlike what has been previously thought, this approach
can also accommodate perfectly conducting magnetofluids, via the Bekenstein-Oron
description of ideal magnetohydrodynamics. When  Noether symmetries associated with Killing vectors
or tensors are present in geodesic flows, they lead to constants of motion
polynomial in the momenta. We generalize these concepts to hydrodynamic flows.
Moreover, the   Hamiltonian descriptions of ideal magnetohydrodynamics   allow one to cast the evolution
equations into  a  hyperbolic form useful  for evolving
rotating or binary compact objects with magnetic fields in numerical general relativity. Conserved circulation laws, such as those of Kelvin, Alfv\'en and  Bekenstein-Oron,
 emerge simply   as special cases of the Poincar\'e-Cartan integral invariant of
Hamiltonian systems.
    We use this approach to  obtain an extension of Kelvin's theorem  to baroclinic  (non-isentropic) fluids,  based on   a temperature-dependent
time parameter.
We further extend this result to  perfectly or poorly conducting baroclinic magnetoflows.
Finally, in the barotropic
case, such magnetoflows are shown to also be  geodesic, albeit  in a Finsler
(rather than Riemann) space.



\end{abstract} 


\pacs{04.40.Nr, 47.10.ab, 47.10.Df, 47.10.A-}

\maketitle

\section{Introduction}

A wide variety of compact stellar objects where general relativistic effects are important is currently known. Black holes and neutron stars  are involved in many astrophysical phenomena, including binary mergers and gamma ray bursts, which have observable imprints in the electromagnetic and gravitational wave spectrum.
Many of these phenomena can be explained by means of general relativistic hydrodynamics. In addition, there is a growing number of observed phenomena where electromagnetic effects play a major role. These include observations of accretion disks around black holes \cite{MosciGDSL09}, jets in active galactic nuclei or microquasars \cite{MeliaSTTC10,Komis10}, gamma ray bursts, hypernovae,
pulsars \cite{Beskin2009} and  magnetars \cite{Sotani2007,Sotani2008,Zink2012,Freytsis2016,Watts2016}. Magnetohydrodynamics (MHD) provides a macroscopic continuum approximation to studying such phenomena. General relativistic magnetohydrodynamics (GRMHD)
originates in the works of Lichnerowicz  \cite{Lichnerowicz1967}
and is  a rapidly developing field of 
modern astrophysics \cite{Beski10,Font2008,Anton_al10}. Departures from MHD are discussed in \cite{Glampedakisetal2011,Andersson2016,Andersson2016a} and references therein. Compact objects such as magnetars or the differentially rotating supramassive remnants of binary neutron-star mergers can have magnetic fields of the order of $10^{15}-10^{17}~\text{G}$ which can affect the dynamics and stability \cite{Reisenegger2009} of these objects. A fully relativistic description of magnetized neutron stars is thus desirable.

In this article,  we develop a geometric
treatment of ideal GRMHD. 
To this aim, we  use  Cartan's exterior calculus, relying on 
the  nature of the electromagnetic field as a 2-form and the well known formulation 
of Maxwell's equations by means of the exterior derivative operator. 
We also employ the  formulation of 
hydrodynamics in terms of the fluid vorticity 2-form, following  Synge \cite{Synge37} and Lichnerowicz \cite{Lichn41}. This enables us to formulate
GRMHD entirely in terms of exterior forms.  
Such an approach is not only elegant and fully covariant, but also  simplifies some
calculations which are tedious in the component approach. 
In addition, we obtain particle-like  \textit{Lagrangian and Hamiltonian descriptions of 
ideal MHD}, in Newtonian and relativistic contexts, with several theoretical and  practical advantages. For example, 
schemes for evolution in numerical relativity are straightforward to obtain, and  conserved quantities whose origin  seems \textit{ad hoc}  in the component approach, emerge immediately as Noether-related quantities in this  canonical approach.  
 
In particular, Synge and Lichnerowicz have  shown that barotropic fluid flows may be described via simple variational principles as  geodesic flows in a  manifold conformally related to the spacetime manifold. Carter  \cite{Carte79}  has used this powerful canonical approach to  efficiently derive conservation laws for neutral or charged poorly conducting fluids. 
In this article, we extend the above framework to  perfectly conducting fluids with the aid of the Bekenstein-Oron (hereafter BO) formulation of ideal MHD \cite{Bekenstein:2000sf,Bekenstein2001,Bekenstein2006}.
 In the canonical approach,  conserved circulation integrals, such as those of Alfv\'en, Kelvin and Bekenstein-Oron, emerge simply as special cases of the Poincar\'e-Cartan integral invariant of Hamiltonian systems.  We  further show that the   BO  description  can describe an arbitrary ideal MHD flow without loss of generality and allows one to cast the ideal MHD equations into  a circulation-preserving hyperbolic form, which may be useful in numerical simulations of oscillating stars or radiating binaries with magnetic fields  in numerical relativity. We   generalize the Synge--Lichnerowicz result  to perfectly conducting magnetofluids by showing that ideal MHD flows can be described as geodesic flows in a Finsler space.

Finally,
Kelvin's circulation theorem has been thought to hold only for  barotropic  flows. It has been thought not to hold for baroclinic (non-isentropic)  flows, except in a weak form (i.e. if the circulation is initially computed along rings of constant temperature or specific entropy \cite{Bekenstein1987}). However, using 
a temperature-dependent
time parameter, we obtain a Hamiltonian action principle describing inviscid baroclinic flows within Carter's framework. Moreover, a Poincar\'e-Cartan integral invariant exists iff a system is Hamiltonian. We thus infer that, contrary to common belief, \textit{a  generalization of Kelvin's theorem to baroclinic flows does exist} in the strong form (i.e. the circulation can be initially computed along an \textit{arbitrary} fluid ring). Remarkably, this result can be further extended to  perfectly or poorly conducting baroclinic
magnetoflows.

Symmetries and conservation laws
 are very useful because they  can provide valuable insight of complicated (magneto)hydrodynamic phenomena; the relevant conserved quantities can be extremely useful in constructing initial data in numerical relativity, or significantly simplify solving for the motion. The examples considered below are
applicable, among others, to the mathematical study and numerical simulation of fluid motions in  rotating  or binary relativistic stars 
\cite{Markakis2009,Read2009,Markakis2010,East2012,Farris2012,Giacomazzo2011a,Read2013,Mosta2014,Dionysopoulou2015,Kawamura2016} and their magnetospheres \cite{Contopoulos2016,Nathanail2016},  neutron-star or black-hole accretion rings \cite{Carte79,Zanotti2010,Korobkin2013,Contopoulos2015} and cosmological  dynamics \cite{Kouretsis2010,Kouretsis2013,Barrow2011,Tsagas2011,Dosopoulou2012,Barrow2012,Barrow2012a,Dosopoulou2014,Pen2016}.

\section{Classical Dynamics in Covariant Language }
\label{sec:II}
\subsection{Notation}

We consider a spacetime $(\M,\w{g})$, i.e.~a four-dimensional real manifold $\M$
endowed with a Lorentzian metric $\w{g}$ of
signature $(- + + +)$. We assume that $\M$ is orientable, so that we have at our disposal the
Levi-Civita tensor $\weps$ (also called \emph{volume element}) associated with the metric $\w{g}$. Let $\wnab$ be the covariant derivative associated
with $\w{g}$: $\wnab\w{g} = 0$ and $\wnab\weps=0$. The star operator $\star$ denotes the Hodge dual of a differential form.  For example, the
Hodge dual of the 1-form
$\w \omega$ is a 3-form denoted by $\star  \w \omega$:
\be
\star \! \omega_{\alpha\beta\gamma}:= \eps_{\alpha \beta\gamma\delta}\omega^\delta
\ee
Similarly, the Hodge-dual of the 2-form $\w \Omega$ is a 2-form denoted by $\star \w \Omega$:
\be
  \star\!  \Omega_{\alpha\beta} := \frac{1}{2} \eps_{\alpha\beta\gamma\delta} \Omega^{\gamma\delta}
,
\ee
More details on these
definitions may be found e.g. in Appendix~B of Ref.~\cite{GourgMUE11}.

We shall often use an index-free notation, denoting vectors and tensors on $\M$ by boldface symbols. As in
\cite{Carte79},
given a linear form $\w{\omega}$, we denote by $\vw{\omega}$ the vector
associated to it by the metric tensor: 
\be \label{e:def_vec_form}
  \w{\omega} =: \w{g}(\vw{\omega},  \ \_ \  ). 
\ee
In a given vector basis $(\vv{e}_\alpha)$,  the components of $\w{g}$, $\vw{\omega}$ and $\w{\omega}$
 are  $g_{\alpha\beta}$, $\omega^\alpha= g^{\alpha\beta}
\omega_\beta$ and $\omega_\alpha=g_{\alpha\beta} \omega^\beta$ respectively.

Given a vector $\vw{v}$ and a tensor $\w{T}$ of type $(0,n)$ ($n\geq 1$), i.e. a $n$-linear form (a linear form for $n=1$, a bilinear form for $n=2$, etc.), 
we denote by $\vw{v}\cdot\w{T}$ (resp. $\w{T}\cdot\vw{v}$) the $(n-1)$-linear form
obtained by setting the first (resp. last) argument of $\w{T}$ to $\vw{v}$:
\begin{subequations}
\bea
  \vw{v}\cdot\w{T} & := & \w{T}(\vw{v}, \ \_  \ , \,\ \ldots \,\ ,  \ \_ \  ) \label{e:v_dot_T} \\
  \w{T}\cdot\vw{v} & := & \w{T}(  \ \_ \  , \,\ \ldots \,\ ,  \ \_ \  ,\vw{v}) . 
\eea
\end{subequations}
Thanks to the above conventions, we may write the scalar product of two vectors $\vw{u}$ and $\vw{v}$
as 
\be
  \w{g}(\vw{u},\vw{v}) = \vw{u}\cdot\vv{v} = \vv{u}\cdot\vw{v} . 
\ee
We denote by $\wnab\cdot$ the covariant divergence, with contraction taken on 
the adjacent index.
 For instance, for a tensor field $\vw{T}$ of type $(2,0)$, 
$\wnab\cdot\vw{T}$ is the vector field defined by
\be 
  \wnab\cdot\vw{T} := \nabla_\beta T^{\beta\alpha} \, \vv{e}_\alpha , 
\ee
where $\{\vv{e}_\alpha\}$ is the vector basis with respect to which the components 
$\nabla_\gamma T^{\alpha\beta}$ of $\wnab\vw{T}$ are taken. (Note that the convention 
for the divergence does not follow the rule for the contraction with a vector: 
in (\ref{e:v_dot_T}) the contraction is performed on the \emph{first} index.) 

We use Greek letters $\alpha,\beta,\gamma,\delta,...$ for abstract  and $\mu,\nu,\kappa,\lambda,...$ for concrete spacetime indices. We also use English letters $a,b,c,...$ for abstract and $i,j,k,...$ for concrete spatial indices.
We use geometrized Heaviside-Lorentz units throughout the paper.
We use $\nabla_\alpha$ or $\partial_\alpha$ to denote the (Eulerian) covariant or partial
derivative compatible with a curved or flat metric respectively,
and $\partial /\partial x^\alpha$ to denote the (Lagrangian) partial derivative of a
function $f (x, v)$ with respect to $x$ for fixed $v$.
We   make extensive use of Lie and exterior derivatives:
for a pedagogical introduction to  using these concepts in relativistic hydrodynamics, the reader is referred to  \cite{Gourg06,Gourgoulhon2013}




 

\subsection{Hamiltonian flows}
It is often thought that continuum systems necessarily require an infinite dimensional
manifold for their description, and one often resorts to a classical field-theory approach, based on an action
 integral over  a Lagrangian  density in a spacetime 4-volume. This complicates
  the derivation of conservation laws from symmetries of the action -- one
of the main reasons for using an action functional in the first place. In many cases, however, the very definition of a perfect fluid allows one
 to treat each fluid element as an individual particle interacting with other
fluid elements through pressure terms (in addition to electromagnetic   
or gravitational field terms). If the pressure terms are derivable from a
potential, then a particle-like action principle can be found. This approach
has been utilized by Carter \cite{Carte79}
to derive particle-like conservation laws for neutral perfect fluids and
  for charged poorly conducting fluids.
  Here, we   review Carter's framework and  extend
it to baroclinic fluids and perfectly conducting magnetofluids.

\subsubsection{Lagrangian dynamics}
The results derived in this section will apply to \textit{any} classical motion  
 obeying a Lagrangian variation principle. That is, for any particular (particle, fluid or magneto-fluid) flow, there exists a 
 Lagrangian function 
$L({x},{v})$ of the spacetime coordinates $x^\alpha$
and canonical 4-velocity $v^\alpha$, evaluated at $x^\alpha(\lambda)$
and  $v^\alpha(\lambda)$ where  $\lambda\in\mathbb{R}$ is a   canonical  time parameter (which need  not necessarily coincide with proper time
   $\tau$) in terms of which the 
 (not necessarily unit) vector
\be \label{eq:canonicalvelocity}
v^\alpha=\frac{dx^\alpha}{d\lambda}
\ee
is defined. The
 equations of the (particle or fluid-element) worldlines $x^\alpha(\lambda)$ can be obtained from the action
functional
\be \label{eq:elementaction}
\mathcal{S}=\int_{\lambda_{1}}^{\lambda_{2}} L(x,v)d\lambda.
\ee
Extremizing the action  keeping the endpoints  fixed yields the
Euler-Lagrange equations of motion
\begin{subequations} \label{eq:EulerLagrangeClassNewt}
\bea
\frac{{d{p _\alpha}}}{{d\lambda}}  &=& \frac{{\partial L}}{{\partial {x^\alpha}}}
\\
{p _\alpha}  &=& \frac{{\partial L}}{{\partial {v^\alpha}}}  \label{eq:canonicalmomentum}
\eea
\end{subequations}
where  $p_\alpha $ is the \textit{canonical momentum}
1-form conjugate to  $x^\alpha $. In the context of fluid theory, it is preferable to write the above equations
in the (Eulerian) covariant form  \cite{Carte79}
\begin{equation} \label{eq:eulerlagrangecov1Newt}
\pounds_{\vw{v}}{p}_\alpha = {\nabla _\alpha}L \nonumber
\end{equation}
or, in exterior calculus notation, 
\begin{equation} \label{eq:eulerlagrangecovExtNewt}
 {\pounds_{\vw{v}}}\w{p } = {\bf{d}}L ,
\end{equation}
where $\pounds_{\vw{v}}$ is the Lie derivative along the vector $\vw{v}$    and $\dd$ is the exterior derivative  \cite{Gourg06,Gourgoulhon2013,FriedmanStergioulas2013}. The canonical momentum one-form
$\w p =p_\mu \dd x^\mu$ is 
 also known as the tautological one-form, the Liouville one-form, the Poincar\'e one-form the symplectic potential or simply the canonical one-form \cite{Abraham2008}. 
Using   the definition
of the Lie derivative and the chain rule, the above equation\footnote{In Eq.~\eqref{eq:canonicalmomentum}, 
the Lagrangian $L$ and canonical momentum $\w p$ are regarded   functions
of the time parameter $\lambda$, through $\vw x(\lambda)$
and  $\vw v(\lambda)$, and characterize a single fluid element. In Eq.~\eqref{eq:eulerlagrangecovExtNewt},
the Lagrangian and  canonical momentum are regarded   functions on spacetime
through  $\vw x$ and $\vw v(x)$. They  amount to the Lagrangian and canonical
momentum of the fluid element  located at $\vw x$, and
changing the argument $\vw x$ generally changes the fluid element which $L$
and $\w p $ refer to.
} can be expressed as:
\bea \label{eq:chanruleeulerNewt}
{ {\pounds_{\vw{v}}}{p _\alpha}-\nabla _\alpha}L
 &:=&
   {v^\beta}\frac{{\partial {p_\alpha}}}{{\partial {x^\beta}}} + {p_\beta}\frac{{\partial
{v^\beta}}}{{\partial {x^\alpha}}}
   -\left(
   \frac{{\partial L}}{{\partial {x^\alpha}}} + \frac{{\partial L}}{{\partial
{v^\beta}}}\frac{{\partial {v^\beta}}}{{\partial {x^\alpha}}}
  \right) \nonumber \\
&=& \frac{{d{p
_\alpha}}} {{d\lambda}}-\frac{{\partial L}}{{\partial {x^\alpha}}}
+ \left({p _\beta}-\frac{{\partial L}}{{\partial
{v^\beta}}} \right)\frac{{\partial {v_\beta}}}{{\partial {x^\alpha}}}. 
\eea
This quantity vanishes iff the Euler-Lagrange equations \eqref{eq:EulerLagrangeClassNewt} are satisfied; the latter are thus equivalent  to the covariant equation \eqref{eq:eulerlagrangecov1Newt}.


\subsubsection{Hamiltonian dynamics}
The Legendre transformation
\begin{equation} \label{eq:HamiltonianLagrangianLegendre}
H= v^\alpha p_\alpha - L
\end{equation}
defines the 
 super-Hamiltonian $H({x},{p})$. 
Then, the equations of motion take the form of 
Hamilton's equations
\begin{subequations}
\bea \label{eq:HamiltonsEqsClassNewt}
\frac{{d{p_\alpha}}}{{d\lambda}} &=&  - \frac{{\partial H}}{{\partial {x^\alpha}}} \label{eq:HamiltonsEqsClassNewt1}\\
\frac{{d{x^\alpha}}}{{d\lambda}}  &=& \frac{{\partial H}}{{\partial {p_\alpha}}}
\label{eq:HamiltonsEqsClassNewt2}
\eea
\end{subequations}
The above equations can be written 
covariantly as
\cite{Carte79}\begin{equation} \label{eq:HamiltonsEqsCovNewt}
 {v^\beta}({\nabla _\beta}{p_\alpha} - {\nabla_\alpha}{p_\beta}) =  - {\nabla_\alpha}H
\nonumber
\end{equation}
or, in exterior calculus notation,
\begin{equation} \label{eq:HamiltonsEqsCovExtNewt}
 \vw{v} \cdot {\bf{d}}\w{p} =  - {\bf{d}}H
\end{equation}
One may obtain Eq.~\eqref{eq:HamiltonsEqsCovExtNewt}
using the  Cartan identity
\begin{equation} 
\pounds_{\vw{v}}{p_\alpha}  
= {v^\beta}({\nabla _\beta}{p_\alpha} - {\nabla_\alpha}{p_\beta})+ {\nabla_\alpha}(v^\beta p_\beta)   
\nonumber
\end{equation}
or
\begin{equation} \label{eq:CartanIdentity}
\pounds_{\vw{v}}\w{p
}  
= \vw{v} \cdot {\bf{d}}\w{p}+{\bf{d}}( \vw{v} \cdot \w{p})   
\end{equation}
and the Legendre transformation  \eqref{eq:HamiltonianLagrangianLegendre}
 to write the covariant  Euler-Lagrange equations \eqref{eq:eulerlagrangecovExtNewt} as\begin{equation} 
\pounds_{\vw{v}}\w{p
} - {\bf{d}}L 
= \vw{v} \cdot {\bf{d}}\w{p}+{\bf{d}}H=0. 
\end{equation}
Alternatively, one may prove the equivalence of Eq.~\eqref{eq:HamiltonsEqsCovNewt} to the  Hamilton equations \eqref{eq:HamiltonsEqsClassNewt} by proceeding  analogously to Eq.~\eqref{eq:chanruleeulerNewt}, that is, by using the chain rule to rewrite the (Eulerian) covariant derivative
${\nabla _\alpha}H=\partial H/\partial x^\alpha+ \partial p_\beta / \partial x^\alpha \, \partial H / \partial  p_\beta  $ in terms of (Lagrangian) partial derivatives.  

\subsection{Conservation laws}

\subsubsection{Poincar\'e-Cartan integral invariant}
The 2-form 
\be 
{\Omega}_{\alpha \beta}:= {\nabla _\beta}{p_\alpha} - {\nabla_\alpha}{p_\beta} \nonumber
\ee
or, equivalently,
\be \label{eq:generalizedvorticity}
\w{\Omega}:= {\bf{d}}\w{p}= \dd  p_\mu \wedge \dd x^\mu
\ee
is the \textit{canonical symplectic form}, also known as the  \textit{Poincar\'e two-form} \cite{Abraham2008}. Its physical
content depends on the action \eqref{eq:elementaction}. 
 In Sec.~\ref{sec:ExamplesHamiltonianFlows}
it will be shown that, if
the action describes
a perfect fluid, then $\w{\Omega}$ is    Khalatnikov's
canonical vorticity tensor; if the action describes a purely magnetic field,
then
$\w{\Omega}$ is  the Faraday tensor. Nevertheless, the results
of Sec.~\ref{sec:II}    apply to any generic Hamiltonian flow; no assumptions
on
the
physical content of the action \eqref{eq:elementaction}
will be made prior to Sec.~\ref{sec:ExamplesHamiltonianFlows}.
  
Taking the exterior derivative of 
\eqref{eq:eulerlagrangecovExtNewt}, commuting the exterior derivative  $\bf d$ with the Lie derivative $\pounds_{\vw
v}$ 
and using the identity ${\bf{d}}^2 = 0$,
we immediately deduce that the canonical symplectic form \eqref{eq:generalizedvorticity}
is advected by the flow: 
\begin{equation} \label{eq:ConservCirculationDiffNewt}
 {\pounds_{\vw{v}}}  \w{\Omega} = 0.
\end{equation}
The above equation  also follows directly from the Hamilton equations \eqref{eq:HamiltonsEqsCovExtNewt}
and the Cartan identity
\be \label{eq:CartanIdnentity}
{\pounds_{\vw{v}}}  \w{\Omega}=\vw{v} \cdot {\bf{d}}\w{\Omega} +{\bf{d}}(\vw{v} \cdot \w{\Omega}). 
\ee

The conservation equation~\eqref{eq:ConservCirculationDiffNewt}
 is tied to an important integral invariant: Consider\footnote{This derivation  follows and generalizes Friedman
\& Stergioulas' \cite{FriedmanStergioulas2013} proof of conservation of circulation.} the family   $\Psi_\lambda$ of diffeomorphisms generated by canonical velocity $\vw v$, with $\Psi_{\lambda}^{-1}$
its inverse. Let ${c}$ be a ring in the flow, bounding a 2-surface ${S}$; let 
${c}_\lambda=\Psi_\lambda ({c})$
 be the family of 
 rings dragged along by the flow, bounding the 2-surfaces 
 ${S}_\lambda=\Psi_\lambda ({S})$. That is, each point of 
${S}_\lambda$ is obtained by moving each point of $S$ an affine time $\lambda$  along the flow through that point. The closed line integral of $\w{p}$ around $\mathcal{C}_\lambda$ can then be written as
\begin{equation} \label{eq:CirculationInt}
\mathcal{I}:=\oint_{c_\lambda} \w{p}  =\int_{{S}_{\lambda}} \w{\Omega}=\int_{{S}}
\Psi_{\lambda}^{-1}\w{\Omega}
\end{equation}where we used Green's theorem (relating the circulation integral
$\oint_{{c_\lambda}}p_\alpha dx^\alpha$ with the
  flux integral $\int_{{S_\lambda}}
\Omega_{ \alpha \beta } {dS^{\alpha\beta}}
 $ of $\w{\Omega}$
through   ${S}_\lambda$) and the diffeomorphism
invariance of an integral (i.e. the identity  
$\int_{\Psi_\lambda (S)} \Psi_\lambda \w \Omega = \int_{S} \w \Omega$, 
with    $\w{\Omega}$ replaced by $\Psi_{\lambda}^{-1}\w{\Omega}$,
cf. Eq.~(A.81)\ in 
\cite{FriedmanStergioulas2013}).
Eq.~\eqref{eq:ConservCirculationDiffNewt} implies that the above integral is conserved: 
\begin{equation} \label{eq:ConservCirculationIntNewt}
\frac{d\mathcal{I}}{d\lambda}\mathcal{}=
\int_{{S}}  {\frac{d}{d\lambda}(\Psi^{-1}_\lambda {}} \w{\Omega} )=
\int_{{S}}  {\pounds_{\vw{v}}} \w{\Omega}=0.
\end{equation}The closed line integral
\eqref{eq:CirculationInt} 
 is known in analytical dynamics as the \emph{Poincar\'e-Cartan
integral invariant} associated with Hamiltonian
systems \cite{Poincare1890,Poincare1899,Cartan1922}.
Its existence emerges from the Hamiltonian structure of Eq.~\eqref{eq:HamiltonsEqsCovExtNewt}. In particular,
a dynamical system  possesses a  Poincar\'e-Cartan
 integral invariant \textit{if and only if} it is Hamiltonian 
\cite{Boccaletti2003}. 

Although this result is well-known in analytical dynamics,
  to our knowledge, its applicability to  (magneto)hydrodynamics   was only recognized  by Carter \cite{Carte79} 
Some classical mechanics texts mention that   the integral \eqref{eq:CirculationInt}
corresponds to a conserved circulation in phase space, analogous to  Kelvin's  circulation integral in a barotropic fluid.
In fact, this is more than a mere analogy, albeit in the converse direction: Kelvin's  circulation    \textit{equals}   the integral~\eqref{eq:CirculationInt}
if the Lagrangian is chosen to be that of a perfect barotropic  fluid element, Eq.~\eqref{eq:Lagrangianhguuh} \cite{Carte79,Markakis2014a}. Similarly,  Alfv\'en's magnetic flux theorem,  and the generalizations of Kelvin's theorem to poorly  \cite{Carte79}  or perfectly conducting \cite{Bekenstein:2000sf,Bekenstein2001,Bekenstein2006} magnetofluids,   emerge   also as  \textit{special
cases} of the Poincar\'e-Cartan integral invariant \eqref{eq:CirculationInt}. This will be 
shown in Sec.~\ref{sec:ExamplesHamiltonianFlows} by constructing the appropriate
Lagrangians.

\subsubsection{Irrotational flows}
In general, a flow will be called \emph{irrotational} iff the
canonical vorticity 2-form  vanishes:
\be
\w \Omega = \dd \w p=0 . 
\ee
Then, if the domain $\mathcal{D}$ is simply connected,  the Poincar\'e lemma
implies  the local existence of  
a single-valued scalar field $\mathcal{S}$ such that
\begin{equation} \label{eq:HamiltonsEqsCartanIdentNewt}
\w p={\dd} \mathcal{S} 
\end{equation}
or, equivalently,
\be
p_\alpha=\nabla_\alpha \mathcal{S}.
\nonumber
\ee
The invariance of the Poincar\'e-Cartan  integral 
\eqref{eq:CirculationInt} guarantees that \textit{initially irrotational flows remain
irrotational}\footnote{In the context of barotropic fluids, this is known as Helmholtz's third theorem, which is a corollary of Kelvin's circulation theorem.}. This is very useful when solving the Cauchy problem with irrotational
initial data (cf. \cite{Markakis2014a} for a 3+1 evolution scheme exploiting this property in barotropic fluids). For an  irrotational flow, substituting Eq.~\eqref{eq:HamiltonsEqsCartanIdentNewt} into the  equations of motion \eqref{eq:eulerlagrangecovExtNewt},
\eqref{eq:HamiltonsEqsCovExtNewt}, we find that the latter
have first integrals:
\begin{eqnarray} \label{eq:EulerLagrangeIrrotNewt}
 \pounds_{\vw{v}} \mathcal{S} -L &=& 0 
\\
\label{eq:HamiltonEqsIrrotNewt}
H &=&  0  
\end{eqnarray}
respectively. In general, a system with constant $H$ is called \textit{uniformly
canonical}. This is the case for irrotational flow, and, more generally, for a perfect fluid that is homentropic or barotropic, as will be shown below.

We note that the above first integrals 
hold throughout the flow. Indeed, taking the exterior derivative of the above
equations and 
commuting the operator $\bf d$ with $\pounds_{\w v}$ leads back to the equations
of motion \eqref{eq:eulerlagrangecovExtNewt}, \eqref{eq:HamiltonsEqsCovExtNewt}.
In the above integrals, we have dropped an additive integration
constant  
by absorbing it into  the
definition of the potential  $\mathcal{S}$. 
Note that Eq.~\eqref{eq:EulerLagrangeIrrotNewt} follows directly from 
Eq.~\eqref{eq:HamiltonEqsIrrotNewt} 
with the aid of Eqs.~\eqref{eq:HamiltonianLagrangianLegendre} and
\eqref{eq:HamiltonsEqsCartanIdentNewt} and the definition of the Lie derivative.

Eqs.~\eqref{eq:EulerLagrangeIrrotNewt} and \eqref{eq:HamiltonEqsIrrotNewt}
were derived for an irrotational flow. More generally, the same equations
can be shown to hold for  helicity-free flows  which are representable in
the Clebsch
form $\w p= {\bf d}\mathcal{S}  +\alpha {\bf d} \beta$. This follows by substituting
the latter expression into the equations of motion \eqref{eq:eulerlagrangecovExtNewt},
\eqref{eq:HamiltonsEqsCovExtNewt} and using the fact that the Clebsch potentials
$\alpha,\beta$ are advected by the flow, that is $ \pounds_{\vw{v}}\alpha
 = 0$,
$ \pounds_{\vw{v}}\beta = 0$.

\subsubsection{Poincar\'e
two-form}

Let $u^\alpha=v^\alpha (-v_\beta v^\beta)^{-1/2}$ be the unit vector along $v^\alpha$.
In light of Eq.~\eqref{eq:HamiltonsEqsCovExtNewt}, we can decompose the 2-form \eqref{eq:generalizedvorticity} into `electric' and `magnetic' parts with respect to  $u^\alpha$
as 
\be
\Omega_{\alpha \beta}=u_\alpha
\nabla_\beta H-u_\beta \nabla_\alpha H +u^\delta \eps_{\delta \alpha \beta \gamma} \omega^\gamma \nonumber
\ee
or
\begin{subequations}
\bea
         \w{\Omega} & = & \w{u}\wedge {\bf{d}}H +\star (\w{u}\wedge \w{\omega} ) \label{eq:OmegavdHstarvomega} \\
        \star \w{\Omega} & = & \star(\w{u}\wedge {\bf{d}}H)- \w{u}\wedge \w{\omega}
  ,
\eea
\end{subequations}
where 
\be
\omega_\alpha := \frac{1}{2}u^\delta \eps_{\delta \alpha \beta
\gamma} \Omega^{\beta \gamma} = u^\delta
\star \Omega_{\delta\alpha}   \nonumber
\ee
or
\be \label{e:def_omega}
        \vv{\omega} := \vw{u} \cdot\star\w{\Omega}  
\ee
From the antisymmetry properties of $\w \epsilon$ it follows that 
\be  \label{eq:vdH}
\vw{u} \cdot{\bf{d}}H = 0,
        \qquad
           \vw{u} \cdot \vv{\omega} =0
\ee
and that the scalar invariants of the 2-form $\vv{\Omega}$ are
\bea
\frac{1}{2}\Omega^{\alpha\beta}\Omega_{\alpha\beta}&=&\omega^\alpha \omega_\alpha 
-  \nabla^\alpha H \nabla_\alpha H   \label{eq:OmegaOmega}
\\
\frac{1}{2}(\star \Omega^{\alpha\beta})\Omega_{\alpha\beta}&=&\omega^\alpha \nabla_\alpha H . \label{eq:OmegastarOmega}
\eea
By the definition~\eqref{eq:generalizedvorticity},  $\vv{\Omega}$ is an exact 2-form. Because $\bf{d}^2=0$, any exact 2-form is also closed:
\be \label{eq:div_omega}
{\bf{d}} {\w{\Omega}}=0 \Leftrightarrow   \nabla_\alpha(\star \Omega^{\alpha \beta} )=0
\ee
Given a   scalar field $\phi(x)$ on 
$\M$,
one can construct an exact 1-form
\be
\w l = \bf{d} \phi
\ee  
which is also, by virtue of the identity $\bf{d}^2=0$,  
closed: 
\be \label{eq:dl0}
{\bf{d}} {\w{l}}=0 \Leftrightarrow   \nabla_\alpha l_\beta- \nabla_\beta l_\alpha=0.
\ee
Given a closed 2-form $\w{\Omega}$ and a closed 1-form  $\w l$, one can  construct a  current $j^\alpha:=l_\beta
\star
\Omega^{ \beta\alpha
}  $, or
\be \label{eq:jlOmega}
\w j := \vw l \cdot \star \w \Omega.
\ee
which, by virtue of Eqs.~\eqref{eq:div_omega} and \eqref{eq:dl0}, is conserved:
\be \label{eq:delj}
\nabla_\alpha j^\alpha 
= \nabla_\alpha(\nabla_\beta\phi \star \Omega^{\beta\alpha} )=0.
\ee
This conservation law implies a corresponding global conservation of the integrated flux of $j^\alpha$ across a hypersurface.

An infinite number of (not necessarily independent) conservation laws stem from Eq.~\eqref{eq:delj} since, in general,   $\phi(x)$   can  be \textit{any} differentiable 
function
of the coordinates. For example, in a  chart $\{x^\mu\}=\{t,x^i\}$, if    $\phi$ is chosen to be the spatial coordinate  $x^1$,  the above equation reduces to  the $x^1$-component  of Eq.~\eqref{eq:div_omega}. If  $\phi$  coincides with coordinate time $t$, Eq.~\eqref{eq:delj} yields a spatial constraint equation.
Other combinations of the coordinates give different projections Eq.~\eqref{eq:div_omega}.
Choosing $\phi$ to be the super-Hamiltonian $H$ gives rise to a conserved current
\be \label{eq:GeneralizedErtel}
j^\alpha:=\nabla_\beta H\star \Omega^{\beta\alpha}
\ee
For a baroclinic fluid, the time component of this current is the potential vorticity, as shown in Sec.~\ref{sec:ExamplesHamiltonianFlows}.
The corresponding conservation law, known as Ertel's theorem, is simply a special case of Eq.~\eqref{eq:delj}.

As mentioned earlier, a system with spatially constant super-Hamiltonian $H$ is uniformly canonical. If the uniformity condition $\dd H=0$ holds on an initial hypersurface,
then Eq.~\eqref{eq:vdH} guarantees that the condition is preserved in time.
For such systems, Eqs.~\eqref{eq:OmegavdHstarvomega} and \eqref{eq:div_omega}
yield
the conservation law
\be  \label{eq:VorticityPropagation}
{\bf{d}}\star (\w{u}\wedge \w{\omega})=0 \Leftrightarrow \nabla_\alpha(u^\alpha \omega^\beta-u^\beta
\omega^\alpha)=0
\ee
In 3+1 dimensions, this equation is the curl of Eq.~\eqref{eq:ConservCirculationDiffNewt}.
\subsubsection{Generalized Helicity}

Eqs.~\eqref{eq:generalizedvorticity},  \eqref{eq:div_omega} and \eqref{eq:OmegastarOmega} imply that, for uniformly canonical systems, the generalized helicity
current\be \label{eq:GenHel}
\w h := \vw p \cdot \star \w \Omega  
\ee
is conserved:
\be \label{eq:delh}
\nabla_\alpha h^\alpha =\frac{1}{2}\Omega_{\alpha\beta}
\star \Omega^{\beta\alpha}
 =\omega^\alpha
\nabla_\alpha H=0.
\ee
This conservation law also implies a corresponding global conservation of the
integrated flux of $h^\alpha$ across a hypersurface.
Specific examples are given in Sec.~\ref{sec:ExamplesHamiltonianFlows} \cite{Woltjer1958,MOFFAT1969,Carter1992}.

\subsubsection{Noether's theorem}

 Noether's theorem states that  each continuous symmetry of the action
implies a quantity conserved by the motion. In particular, the \textit{generalized} Noether theorem may be stated as follows \cite{Ioannou2004}. Consider  the $\varepsilon$-family of  infinitesimal coordinate transformations
\be \label{eq:transformers}
\vw x \rightarrow \vw x_{\varepsilon} = \vw x + \varepsilon \, \vw k(x,v)
\ee
generated by the vector field  
$\vw k(x,v)$, which can depend on position \textit{and} velocity, for a small parameter $\varepsilon$. If these  transformations leave the action \eqref{eq:elementaction} unchanged or, equivalently, 
change the Lagrangian $L(x,v)$ by a total  derivative
of some scalar $K(x),$
\be
L\rightarrow L _\varepsilon=L- \varepsilon \, \frac{dK}{d \lambda},
\ee
then the quantity
\be \label{eq:NoetherConstant}
\mathcal{C}(x,v)=\frac{\partial L}{\partial v^\alpha}k^\alpha +K
\ee
is a constant of motion:
\be  \label{eq:NoetherConstantConserva}
\frac{d \mathcal{C}}{d \lambda}={\pounds_{\vw v}}  \mathcal{C}=0
\ee
If $\vw k$ depends on velocity, then the family  \eqref{eq:transformers} of transformations is not generally considered a family of diffeomorphisms. It is, however, a generalized symmetry of the action and Noether-related to an invariant of the form \eqref{eq:NoetherConstant}.

Conversely, the inverse Noether theorem  \cite{Ioannou2004}
 may be stated as follows: if the quantity $\mathcal{C}(x,v)$ is a constant of motion, then the 
$\varepsilon$-family of  infinitesimal  transformations generated by the vector field 
$\vw k(x,v)$, obtained by solving the linear system\be
\frac{\partial^2 L}{\partial v^\alpha \partial v^\beta} k^\beta=
\frac{\partial \mathcal{C}}{\partial v^\alpha } ,
\ee
is a generalized symmetry of the action. 

In the Hamiltonian picture, a scalar quantity ${\cal C}(x,p)$, which does not explicitly depend on
the time parameter $\lambda$, is conserved if it commutes with the super-Hamiltonian, in the sense of a vanishing Poisson bracket:
\be \label{eq:PoissonBra}
\frac{d \mathcal{C}}{d \lambda}={\pounds_{\vw v}}  \mathcal{C}= \{ {\cal C},{ H}\}  \equiv
\frac{{\partial {\cal C}}}{{\partial {x^\gamma}}}\frac{{\partial { H}}}{{\partial
{p _\gamma }}} - \frac{{\partial {\cal C}}}{{\partial {p _\gamma }}}\frac{{\partial
{H}}}{{\partial {x^\gamma }}}=0.
\ee
Conserved quantities polynomial in the momenta are associated with
 Killing vectors or tensors and are Noether-related to symmetries of the
action, as discussed below.
The super-Hamiltonian $H$ does not explicitly depend on the affine 
parameter $\lambda$ and is itself a constant of motion, in agreement with Eq.~\eqref{eq:vdH} (this symmetry is Noether-related to  the metric tensor being a Killing tensor, as discussed in Sec.~\ref{sec:ExamplesHamiltonianFlows}).  

For barotropic fluids,
Eqs.~\eqref{eq:NoetherConstant} and \eqref{eq:NoetherConstantConserva} or \eqref{eq:PoissonBra} give rise to Bernoulli's law, as shown in
the next section.


\section{Examples of Hamiltonian Flows}
\label{sec:ExamplesHamiltonianFlows}
\subsection{Perfect fluids} \label{sec:perfectfluidmodel}


We assume that a part $\mathcal{D}\subset\M$ of spacetime is occupied by
a perfect fluid,
characterized by the  
energy-momentum tensor
\be \label{e:energymomentumfluid}
\w{T}^{\rm fl} = (\eps + p) \, \vv{u}\otimes \vv{u} + p \w{g} ,
\ee
where $\eps$ is the proper energy density, $p$ is the fluid pressure and $u^\alpha=dx^\alpha/d \tau$ is the fluid 4-velocity. 
Moreover, we neglect effects of viscosity or heat conduction and we assume that the fluid is a \emph{simple fluid}, that is, all thermodynamic quantities depend only on the entropy density
 $s$ and  
proper baryon number density $n$. In particular, 
\be \label{e:EOS}
  \eps  = \eps (s,n) . 
\ee
The above relation is called the \emph{equation of state (EOS)}  of the fluid. 
The \emph{temperature} $T$ and the \emph{baryon chemical potential} $\mu$ are then defined by
\be \label{e:def_T_mu}
  T := \frac{\partial\eps }{\partial s} \qquad\mbox{and}\qquad
  \mu := \frac{\partial \eps }{\partial n} .
\ee
Then, the first law of thermodynamics can be written as
\be \label{e:deps1stlaw}
d\eps  =\mu \,dn+Tds
\ee
As a consequence, $p$ is a function of $(s,n)$ 
entirely determined by (\ref{e:EOS}): 
\be \label{e:p_EOS}
  p = -\eps  + T  s + \mu \, n . 
\ee
Let us introduce the \emph{specific enthalpy}, 
\be \label{e:def_h}
  h := \frac{\eps +p}{\rho} =g + T S,
\ee
where $\rho$ is the rest-mass density
\be 
  \rho := m \, n ,
\ee
$g$ is the 
\textit{specific Gibbs free energy}
\be \label{eq:chemicalpotential}
  g  := \frac{\mu}{m} ,
\ee
$m = 1.66\times 10^{-27} {\rm\; kg}$ 
is the baryon rest-mass,
and  $S$ is the \textit{specific entropy}, or entropy per particle:
\be \label{e:def_S}
  S := \frac{s}{\rho} .
\ee
The second equality in (\ref{e:def_h}) is an immediate consequence of (\ref{e:p_EOS}). From Eqs. \eqref{e:deps1stlaw}--\eqref{e:def_S},
we obtain the  thermodynamic relations
\be \label{eq:depsdpthermodynamic}
d\eps =hd\rho+\rho T dS, \quad dp=\rho(dh-TdS)
\ee A simple perfect fluid is  \textit{barotropic} if 
the energy density depends only on the pressure, $\eps = \eps(p)$.
This is the case for a cold or  a homentropic fluid.

With the aid of 
Eqs. \eqref{e:deps1stlaw}--\eqref{eq:depsdpthermodynamic}, the divergence
of the  fluid energy-momentum tensor \eqref{e:energymomentumfluid} can be
decomposed as 
\be \label{eq:fluidenmomediv}
\vw \wnab \cdot {\w T}^{\mathrm{fl}} =h \w{u}[\diver(\rho\, \vw{u})]+ 
  \rho[ \vw{u} \cdot \dd(h \w{u}) - T \dd S ]. 
\ee
Conservation of  rest mass 
\be \label{e:continuity}
\diver(\rho\, \vw{u})=0,
\ee
and the vanishing of \eqref{eq:fluidenmomediv} yield the relativistic Euler equation for \textit{baroclinic fluids}, in the  canonical form:
\bea 
 \pounds_{\vw{u}}(h \w{u})  
+ {\bf{d}}h=  \vw{u} \cdot \dd(h \w{u}) = T \dd S  \label{e:Relativistic-Euler}
\eea
 where the first equality follows from  the Cartan identity \eqref{eq:CartanIdentity} and 
the normalization condition 
\be \label{eq:uuem1}
g_{\alpha \beta}u^\alpha u^\beta=-1.
\ee 
 For \textit{barotropic} \textit{fluids}, the Euler equation  \eqref{e:Relativistic-Euler}
 simplifies to 
\bea 
   \pounds_{\vw{u}}(h \w{u})  
+ {\bf{d}}h=\vw{u} \cdot \dd(h \w{u}) = 0.  \label{e:Relativistic-Euler-Baro}
\eea
Eq.~\eqref{e:Relativistic-Euler-Baro}  was obtained in special relativity
by Synge (1937) \cite{Synge37}
 and in general relativity by Lichnerowicz (1941) \cite{Lichn41}. 
The extension \eqref{e:Relativistic-Euler}
to baroclinic (non-isentropic) fluids was obtained by
Taub (1959) \cite{Taub59} (see also \cite{Carte79,Gourg06,Christodoulou2007}). 
Both of these relativistic hydrodynamic equations are canonical and can be described within the framework
of Sec.~\ref{sec:II},     
 which provides a very efficient approach to the derivation
of conservation laws.

\subsection{Barotropic flows}

\subsubsection{Hamilton's principle for a barotropic-fluid element}

The  Euler equation~\eqref{e:Relativistic-Euler-Baro} for a barotropic fluid is readily in the canonical form \eqref{eq:HamiltonsEqsCovExtNewt}.
Thus, a particle variational principle in the form   described in Sec.~\ref{sec:II} can be found. Indeed the motions of
 fluid elements in a barotropic fluid are
 \textit{conformally geodesic}, that is, they are geodesics of a  manifold 
   with metric $h^2 g_{\alpha \beta}$
 \cite{Synge37,Lichn41,MarkakisRBNSMF2011}. This follows from the fact that Eq.~\eqref{e:Relativistic-Euler-Baro} is the Euler-Lagrange equation of the action functional
\be \label{eq:actionHydrodynamic}
\mathcal{S}=-\int_{\tau_1}^{\tau_2}h  \,d \tau=-\int_{\tau_1}^{\tau_2}  h
\sqrt{-g_{\alpha \beta}\frac{dx^\alpha}{d\tau}
\frac{dx^\beta}{d\tau}} d\tau.
\ee
The Lagrangian 
\be  \label{eq:LagrangianHydrodynamic}
L(x,u)=-h \sqrt{-g_{\alpha \beta}u^\alpha u^\beta}
\ee
is associated with the canonical momentum 1-form 
\be \label{eq:defpihu}
\w p =h \w u ,
\ee
and the canonical vorticity 2-form
\be \label{eq:vorticitydhu}
\w{\Omega} = \dd(h \w{u}).
\ee
On-shell, the condition \eqref{eq:uuem1} is satisfied, 
and the Lagrangian \eqref{eq:LagrangianHydrodynamic} takes the value $L=-h$. Carter
\cite{Carte79}  
introduced a
slightly modified Lagrangian
\be \label{eq:Lagrangianhguuh}
L(x,u)=\frac{1}{2}h g_{\alpha \beta}u^\alpha u^\beta-\frac{1}{2}h ,
\ee
that is associated with the same equations of motion and has the same on-shell value, but its action is \textit{not}  reparametrization invariant. Thus, 
if one wishes, for instance, to use reparametrization invariance to replace proper time $\tau$ by  coordinate time $t$, in order to obtain a  \textit{constrained} Hamiltonian via 3+1 decomposition, as done in \cite{Markakis2014a}, then the action \eqref{eq:actionHydrodynamic}
is the appropriate starting point.
If, on the other hand, one is interested in a \textit{super}-Hamiltonian that describes the dynamics in a 4-dimensional spacetime, then Carter's Lagrangian \eqref{eq:Lagrangianhguuh} is more suitable. Substituting the latter into the Legendre transformation \eqref{eq:HamiltonianLagrangianLegendre}
yields the super-
 Hamiltonian
\be \label{eq:HamiltonianBarotropic}
H(x,p)=\frac{1}{2h}g^{\alpha \beta}p_\alpha p_\beta+\frac{1}{2}h
\ee
which vanishes on-shell (whence Eq.~\eqref{eq:uuem1} holds). Substituting Eqs.~\eqref{eq:defpihu} and \eqref{eq:HamiltonianBarotropic} into the Hamilton equation \eqref{eq:HamiltonsEqsCovExtNewt} yields the barotropic Euler equation~\eqref{e:Relativistic-Euler-Baro}. 

\subsubsection{Conservation of circulation in barotropic flows}

For this system, Eq.~\eqref{eq:ConservCirculationDiffNewt} yields a relativistic generalization of \textit{Helmholtz's vorticity conservation} equation:
\begin{equation} \label{eq:ConservCirculationBarotropic}
 {\pounds_{\vw{u}}}\, {\bf{d}} ( h \w{u}) = 0
\end{equation}
and the Poincar\'e-Cartan integral invariant 
\eqref{eq:CirculationInt}-\eqref{eq:ConservCirculationIntNewt}   gives rise 
a relativistic generalization of  \textit{Kelvin circulation theorem}: 
 the circulation along a  fluid ring 
$c_\tau$ dragged along by the flow
is conserved: 
\begin{equation} \label{eq:ConservCirculationIntKelvin}
\frac{d}{d\tau}\oint_{c_\tau} h \w{u}  
=0.
\end{equation}
Conservation of circulation for the nonrelativistic Euler equations was discovered by Cauchy (1815) \cite{Cauchy1815,Frisch2014a} and independently rediscovered by Kelvin (1869) \cite{Thomson1869}.
The extension of this theorem to relativistic barotropic fluids was obtained by Lichnerowicz and \cite{Lichnerowicz1955} Taub \cite{Taub59}.
The most interesting feature of the above conservation law is that its derivation does not depend on the spacetime metric or spacetime symmetries. Thus, it is
\textit{exact} in time-dependent spacetimes, with gravitational 
waves
carrying energy and angular momentum away from a system. Oscillating stars and radiating binaries, if modeled as
 barotropic fluids with no viscosity or  dissipation other than gravitational waves,
exactly
conserve circulation \cite{FriedmanStergioulas2013}.

\subsubsection{Fluid helicity}

Since the super-Hamiltonian \eqref{eq:HamiltonianBarotropic}
is constant, the system is uniformly canonical, and helicity is conserved. If we we substitute Eq.~\eqref{eq:defpihu} into Eq.~\eqref{eq:GenHel}, then
Eqs.~\eqref{e:def_omega} and \eqref{eq:delh}
imply that the fluid helicity current \cite{Carte79,Prix2004,Prix2005}
\be \label{eq:FluidHel}
{\w{h}_{\rm{fl}}} := h{\vw{u}} \cdot \star {\w {\Omega}} =h {\w \omega}
\ee
is  conserved:
\be \label{eq:FluidHel}
\nabla_\alpha(hu_\beta
\star
\Omega^{\beta\alpha
})=\nabla_\beta(h \omega^\beta)
=0.
\ee
This   implies a corresponding global conservation of
the
integrated flux of $h^\alpha_{\rm{em}}$ across a spatial hypersurface.
In a chart $\{t,x^i\}$, the volume integral of the time component
of ${\vw{h}}_{\rm{fl}}$: 
\be \label{eq:helicityt}
h_{\rm{fl}}^t:= {{\vw{h}}_{\rm{fl}}} \cdot \wnab t
=h \omega^t =hu_i\star \Omega^{it}=-h \omega^i u_i/u_t \,\,\,\ 
\ee
is the relativistic generalization of Moffat's fluid helicity \cite{MOFFAT1969,Bekenstein1987,Carter1992}. The last equality follows from Eq.~\eqref{eq:vdH}. If the vorticity 
$\omega^i$ has sufficient decay, then the total volume integral of the above quantity is conserved by the flow.


\subsubsection{Killing vector fields \& Bernoulli's law}

If there exists 
 a 
vector field $k^\alpha(x)$,  that generates a family of diffeomorphisms~\eqref{eq:transformers}   leaving the Langrangian \eqref{eq:LagrangianHydrodynamic} 
unchanged, then Noether's theorem implies the existence of a streamline  invariant  linear in the momenta, given by Eq.~\eqref{eq:NoetherConstant} (with $K$ set to zero):
\be \label{eq:Cpk}
\mathcal{E}(x,p)= k^\alpha p_\alpha = h u_\alpha k^\alpha .
\ee
As stated by Eq.~\eqref{eq:NoetherConstantConserva}, this quantity is conserved along a streamline (i.e. the trajectory of a fluid element):
\be
\frac{d \mathcal{E}}{d \tau}={\pounds_{\vw u}}  \mathcal{E}= \vw u \cdot \wnab \mathcal{E}=0.
\ee
The above statement is a generalization of Bernoulli's law  to relativistic barotropic fluids. In light of the above, \textit{each Bernoulli-type conservation  law is Noether-related to a continuous symmetry of the flow}.

Given the super-Hamiltonian \eqref{eq:HamiltonianBarotropic}, one may directly verify when   a quantity   of the form \eqref{eq:Cpk}
is conserved by computing the Poisson bracket \eqref{eq:PoissonBra}:
\bea \label{eq:dCdtPoisson}
\frac{d \mathcal{E}}{d \tau}
\! &=& \!
\{ {\cal E},{ H}\}
=\frac{1}{h }{p_\alpha}{p_\beta} {\nabla^\alpha}{k^\beta}-
{k^\gamma}{\nabla_\gamma}h \\ 
&=&
\! \frac{1}{{2h }}{{u^\alpha
}{u^\beta}\pounds_{\vw k}}({h^2}{g_{\alpha \beta}})
\nonumber
\eea
 which   vanishes for all timelike streamlines iff
\be \label{eq:Liekh2g}
\pounds_{\vw{k}}({h^2}{\w g})=0.
\ee
That is, \textit{the necessary and sufficient condition for $\mathcal{E}$ to be a
streamline invariant is that
 $\vw k$ be a Killing vector of a manifold with metric} $h^2{\w g}$. This result is intuitive given the fact that, as mentioned earlier, the fluid streamlines are geodesics of this conformal metric, cf. Eq.~\eqref{eq:actionHydrodynamic}.
We remark that the vanishing of both $\pounds_{\vw{k}}\w g$
and 
$\pounds_{\vw{k}}{h}$, as indicated by the first line of 
Eq.~\eqref{eq:dCdtPoisson}, is a sufficient but not necessary condition for $\mathcal{E}$ to be conserved. 

When the pressure vanishes, i.e. when $h=1$, the condition \eqref{eq:Liekh2g} reduces to the Killing equation
$ \nabla_{( \alpha} k_{\beta )} =0 $,  which is Noether-related to the existence of conserved quantities linear in the momenta for  geodesic motion \cite{MisneTW73,Wald84,Markakis2012}.

As an example, let us consider a helically symmetric, rigidly rotating fluid equilibrium,
such as a  rigidly rotating star (that may be triaxially deformed
\cite{Huang2008}), or a tidally-locked binary on circular orbits. The
flow field may then be written as
\be   \label{eq:ucorrot}
\vw{u}=u^t \, \vw{k},
\ee
where
\be \label{eq:HelicalKillingVector}
\vw{k} = \vw{t} + \Omega\, \vw{\varphi}
\ee
  is a helical Killing vector field,
$\pounds_{\vw{k}}\w g=0$. Here, 
$\Omega$ is the rotation frequency,  
$\vw{t} ={\mathbf{\partial}}_t$ is the generator of time translations:
\be
\vw x \rightarrow \vw x + \delta t \, \vw t
\ee
 and  
$\vw{\varphi} = {\mathbf{\partial}}_\varphi$ is the generator of rotations about the $z$-axis:
\be
\vw x \rightarrow \vw x + \delta \varphi \, \vw \varphi.
\ee
 Let us assume that the fluid configuration is helically symmetric, that is, the Lie derivatives of all fluid variables  (such as $\rho, h, \vv u$) along $\vw{k}$
vanish. Since, by virtue of
Eq.~\eqref{eq:Liekh2g}, the system is stationary in a rotating frame,  Noether's theorem guarantees that the  energy in a rotating frame, given by Eq.~\eqref{eq:Cpk}:\be \label{eq:Cpk2}
\mathcal{E}= k^\alpha p_\alpha = p_t + \Omega p_\varphi
\ee
is conserved along streamlines.

In general, this quantity can differ
from one streamline to the next. However, a stronger result follows from Eq.~\eqref{eq:ucorrot} and the Cartan identity \eqref{eq:CartanIdentity}, which allow one to write the Euler equation \eqref{e:Relativistic-Euler-Baro}  as
\be \label{eq:RigidEuler}
\vw k \cdot \dd \w p=\Lie{\vw k} \w p - \dd (\vw k \cdot \w p)=0.
\ee
Because $\Lie{\vw k} \w p =0$,
the first integral \eqref{eq:Cpk} of the Euler equation  is constant throughout the fluid:
\be \label{eq:vonZeipel}
\wnab \mathcal{E}=0.
\ee
This stronger conservation law is a relativistc generalization of von Zeipel's law \cite{FriedmanStergioulas2013}. 
The energy    \eqref{eq:Cpk} is also a first integral to the Euler equation if a helically symmetric system is irrotational \cite{Bonazzola1997,Teukolsky1998,Marronetti1999,Gourgoulhon1998,Shibata1998a,Bonazzola1999,Uryu2000,Taniguchi2010a,GourgMUE11}. Such first integrals are valuable for solving for obtaining fluid equilibria via the self-consistent field method \cite{PriceMarkakisFriedman2009}. Generalizations of these first integrals have been used to construct equilibria for spinning
\cite{Marronetti2003,Tichy2011,Tichy2012,Tichy2016} or eccentric \cite{Moldenhauer2014,Dietrich2015} compact binaries in numerical relativity.


\subsubsection{Killing tensor fields \&\ the Carter constant}

Geodesic motion of test particles in Kerr (or Kerr-de Sitter) spacetimes is known to admit a fourth constant of motion (in addition to energy, angular momentum, and four-velocity magnitude), known as the Carter constant, which is quadratic in the momenta and is Noether-related to the existence of a Killing tensor field \cite{MisneTW73}. 

To our knowledge, the concept of a Killing tensor for fluid flows has not been defined before, but the framework outlined Sec.~\ref{sec:II}
provides the means to do so. 
Consider a tensor field $K^{\alpha \beta}(x)$ associated with a streamline invariant
 quadratic in the momenta,
\be \label{eq:conservedquantityquadratic}
{\cal E} = K^{\alpha\beta} p_\alpha p_\beta + K,
\ee
where the scalar $K(x)$ is a function of position.
This invariant can be considered a special case of the invariant
\eqref{eq:NoetherConstant} and follows from the  generalized  Noether theorem, with 
$k^\alpha (x,p)= K^{\alpha \beta}(x)p_\beta$
 being the generator of the symmetry transformations \cite{Padmanabhan2010}. For the barotropic fluid super-Hamiltonian~\eqref{eq:HamiltonianBarotropic},
the Poisson bracket \eqref{eq:PoissonBra} is
\be
\{ {\cal E},{ H}\} = ({h ^2}{\nabla_\gamma}{K_{\alpha
\beta}} + 2h {g_{\alpha \beta }}{K_{\gamma \delta}}{\nabla^\delta }h 
- {g^{\alpha \beta }}{\nabla ^\gamma }K){u^\alpha}{u^\beta}{u^\gamma}.
\ee
The above bracket vanishes for all timelike streamlines iff
\be \label{eq:KillingTensorBarotrope}
{h ^2}{\nabla_{(\gamma} }{K_{\alpha \beta) }} + 2h {g_{(\alpha \beta }}{K_{\gamma
)\delta}}{\nabla^\delta }h  - {g_{(\alpha \beta }}{\nabla_{\gamma)} }K=0
\ee
That is, \textit{the quantity \eqref{eq:conservedquantityquadratic}
is conserved along streamlines iff $\w{K}$ is a Killing tensor
of the conformal metric 
 $h^2 \w{g}$.} 

In the case of a reducible Killing tensor of the form 
$K^{\alpha \beta}=k^\alpha k^\beta$, where $k^\alpha$ is a Killing vector
satisfying Eq.~\eqref{eq:Liekh2g}, the condition \eqref{eq:KillingTensorBarotrope}
is automatically satisfied while $K$ again vanishes.

When the pressure vanishes, $h = 1$, the scalar $K$
must vanish and the above condition reduces to the  Killing equation 
$ {\nabla_{(\alpha} }{K_{\beta \gamma)} }=0$, 
which is the necessary and sufficient condition  $K^{\alpha \beta} p_\alpha
 p_\beta$  being conserved along a geodesic of $g_{\alpha\beta}$. This is the condition satisfied by the Killing tensor in the Kerr spacetime, which is Noether-related to the Carter constant \cite{MisneTW73,Wald84}.
In light of this, Eqs.~\eqref{eq:conservedquantityquadratic}-\eqref{eq:KillingTensorBarotrope} generalize the concept of a Carter constant to  test fluids in Kerr spacetime. Note, however, that the fluid configuration must satisfy a generalized symmetry (in particular, the Hamilton-Jacobi equation describing the flow \cite{Markakis2014a} must be separable in Boyer-Lindquist coordinates) in order for this constant to exist.

A geodesic flow can be described by the super-Hamiltonian $H=\frac{1}{2}g^{\alpha
\beta}p_\alpha p_\beta$, with $p_\alpha=u_\alpha$, which is conserved
by virtue of the normalization condition \eqref{eq:uuem1}. This conserved quantity arises from  $g^{\alpha\beta}$ being 
covariantly constant and thus a Killing
tensor, and is Noether-related to  the super-Hamiltonian being  independent
of the
affine parameter $\tau$.
For barotropic
flow, however,  $g^{\alpha \beta}$ is not  a Killing tensor, as
it does not satisfy  the condition \eqref{eq:KillingTensorBarotrope} except in the
geodesic limit. 
(If $g^{\alpha \beta}$ were a Killing tensor, then  $g^{\alpha \beta} p_\alpha
p_\beta=-h^2$ would be a streamline constant, but this is not true unless $h =1$). However,  $K^{\alpha \beta}= g^{\alpha
\beta}/h$
is a Killing tensor, since it satisfies the condition  \eqref{eq:KillingTensorBarotrope}
provided that $K=h$. The quadratic streamline constant  \eqref{eq:conservedquantityquadratic}
associated with this Killing tensor is simply the super-Hamiltonian \eqref{eq:HamiltonianBarotropic}.


\subsection{Baroclinic flows}

\subsubsection{Hamilton's principle for a baroclinic-fluid element}

The possibility of expressing the equations of baroclinic (non-isentropic) fluid flows in canonical form has been demonstrated by Carter \cite{Carte79}. An intuitively simple action principle (different from but equivalent
to Carter''s) may be obtained as follows. 

A  free test particle of rest mass $m$,  moving along a geodesic of spacetime, extremizes the action $\mathcal{S}=-m\int_{\tau_1}^{\tau_2} d\tau$ \cite{Landau1975}. For barotropic flows, as indicated by Eq.~\eqref{eq:actionHydrodynamic},
the pressure force on a fluid element can be accounted for by replacing rest mass by the specific enthalpy $h \,m $. For baroclinic flows, in light of Eqs.~\eqref{e:def_h}, \eqref{eq:chemicalpotential} and \eqref{eq:depsdpthermodynamic}, the  natural generalization is to replace rest mass in the above action by the chemical potential $\mu = g \,m$ or, equivalently, the specific Gibbs free energy $g$ (the rest mass can be dropped without affecting the equations of motion).  Upon inspection, it becomes immediately clear that Eq.~\eqref{e:Relativistic-Euler}
is indeed the Euler-Lagrange equation of the  action functional
\be \label{eq:actionHydrodynamicBaroclinic}
\mathcal{S}=-\int_{\tau_1}^{\tau_2}g  \,d \tau 
=-\int_{\lambda_1}^{\lambda_2} { \left( {  h\sqrt { - {g_{\alpha \beta }}\frac{{d{x^\alpha }}}{{d\lambda }}\frac{{d{x^\beta }}}{{d\lambda }}}  - S} \right)}d \lambda
\ee
provided that the (non-affine) canonical time parameter 
\be \label{eq:lamda}
\lambda(\tau):=\int^\tau T(\tau') d\tau'
\ee
is used to parametrize the action.
Note that entropy breaks time-parametrization invariance: unlike the barotropic fluid action~\eqref{eq:actionHydrodynamic}, the baroclinic fluid action 
\eqref{eq:actionHydrodynamicBaroclinic} is \textit{not} parametrization invariant. Consequently, parameter choices other than \eqref{eq:lamda}, such as proper time $\tau$ or coordinate time $t$,  lead to incorrect equations of motion. 
The Lagrangian 
\be \label{eq:LagrangianBaroclinic}
L(x,v)=-h\sqrt { - {g_{\alpha \beta }} v^\alpha v^\beta} + S
\ee
is associated, by virtue of Eqs.~\eqref{eq:canonicalvelocity} and \eqref{eq:canonicalmomentum},
with the canonical velocity and canonical momentum
\begin{subequations} \label{eq:canonicalvelocitymomentum}
\bea
{v^\alpha} &=& \frac{{d{x^\alpha}}}{{d\lambda }} =\frac{1}{T}\frac{{d{x^\alpha}}}{{d\tau }}= \frac{1}{T}{u^\alpha}  \label{eq:canonicalvelocityT} \\
{p_\alpha} &=& \frac{{\partial L}}{{\partial {{v}^\alpha }}} =T \,h v_\alpha= h{u_\alpha}  \label{eq:canonicalmomentumT}
.\eea
\end{subequations}
On-shell, by virtue of Eq.~\eqref{eq:uuem1}, one has $v^\alpha v_\alpha=-T^{-2}$  and 
 the Lagrangian takes the value $L  =  - g /T =  - h/T + S$ and, by virtue of Eq.~\eqref{eq:HamiltonianLagrangianLegendre}, the super-Hamiltonian takes the value $H=-S$. Then, the Euler-Lagrange equation \eqref{eq:eulerlagrangecovExtNewt}
becomes
\begin{equation} \label{eq:eulerlagrangeBaroclinic}
  {\pounds_{\vw{u}/T}}(h\w{u}) = {\bf{d}}(S-h/T) 
\end{equation}
and the Hamilton equation \eqref{eq:HamiltonsEqsCovExtNewt} becomes
\begin{equation} \label{eq:HamiltonsEqsBaroclinic}
  \frac{\vw{u}}{T}\cdot {\bf{d}}(h\w{u}) =   {\bf{d}}S.
\end{equation}
Both of these equations  are  equivalent expressions of the relativistic Euler equation \eqref{e:Relativistic-Euler} for baroclinic fluids.

Carter
\cite{Carte79}  
introduced a
different Lagrangian
analogous to Eq.~\eqref{eq:Lagrangianhguuh}
\be \label{eq:LagrangianBaroclinicCarter}
L(x,v)=\frac{1}{2}Th g_{\alpha \beta}v^\alpha v^\beta-\frac{1}{2}\left(\frac{g}{T}-S\right) ,
\ee that is associated with
the same canonical velocity and momentum \eqref{eq:canonicalvelocitymomentum},  has the same on-shell
value as our Lagrangian \eqref{eq:LagrangianBaroclinic}, and leads to the same equation of motion \eqref{eq:eulerlagrangeBaroclinic}.  The Legendre transformation~\eqref{eq:HamiltonianLagrangianLegendre} yields the super-Hamiltonian
\be \label{eq:HamiltonianBaroclinic}
H(x,p)=\frac{1}{2Th}g^{\alpha \beta}p_\alpha p_\beta+\frac{h}{2T}-S,
\ee
which 
 has the same on-shell value and leads
to the canonical equation of motion
\eqref{eq:HamiltonsEqsBaroclinic}.


\subsubsection{Conservation of circulation in baroclinic flows}

The canonical momentum and canonical vorticity are given  by the same expressions \eqref{eq:defpihu} and \eqref{eq:vorticitydhu}
as for barotropic flows. However, the vorticity is no longer Lie-dragged by the fluid four-velocity $\vw u$: the exterior derivative of Eq.~\eqref{e:Relativistic-Euler} reads
\begin{equation} \label{eq:ConservCirculationBarotropic}
 {\pounds_{\vw{u}}}\, {\bf{d}} ( h \w{u}) =  {\bf{d}} T \wedge  {\bf{d}} S.
\end{equation}
Thus, the circulation around a fluid ring $c_\tau=\Psi_\tau (c)$ dragged along by the flow (where $\Psi_\tau $ is the family of diffeomorphisms generated by fluid four-velocity $\vw u$) is not generally conserved:
\bea \label{eq:ConservCirculationIntKelvin}
\frac{d}{d\tau}\oint_{c_\tau} h \w{u}  
&=&\frac{d}{d\tau}\int_{{S}_{\tau}}
\, {\bf{d}} ( h \w{u})\nonumber\\
&=&\int_{{S}}
{\pounds_{\vw{u}}}\, {\bf{d}} ( h \w{u})
=\int_{{S}} {\bf{d}} T \wedge  {\bf{d}} S.
\eea
Hence, Kelvin's theorem has been commonly thought to not hold for baroclinic flows, except in a weaker form: the circulation  computed initially
along a fluid ring of constant temperature or specific entropy is conserved \cite{Friedman1978a,Bekenstein1987}.

In lieu of  a conserved circulation law, one may introduce the \textit{potential vorticity},  defined in general relativity by selecting the scalar field in Eq.~\eqref{eq:jlOmega}, or  the negative Hamiltonian in Eq.~\eqref{eq:GeneralizedErtel}, to coincide with specific entropy $S$ (i.e. setting $\w l = {\bf{d}} S$), to obtain a flux conservation law of the form \eqref{eq:delj}. 
This law can also be written in terms of a Lie derivative along  fluid velocity, and is the relativistic generalization of \textit{Ertel's theorem} obtained by   Friedman \cite{Friedman1978a} (see also Katz
\cite{Katz1984}).
 
Here, we take a different route,  and show that 
Carter's framework \cite{Carter1989}
implies the existence of a strong  circulation law.
We have just shown above that an inviscid baroclinic fluid is a Hamiltonian system and, as such, must  possess a Poincar\'e-Cartan
integral invariant. Indeed, the exterior derivative of Eq.~\eqref{eq:eulerlagrangeBaroclinic}   implies that the canonical vorticity \eqref{eq:vorticitydhu}
is  Lie-dragged by the \textit{canonical fluid velocity} \eqref{eq:canonicalvelocity}:
\begin{equation} \label{eq:ConservVorticityBaroclinic}
 {\pounds_{\vw{u}/T}}\, {\bf{d}} ( h \w{u}) = 0
\end{equation}
as dictated by Eq.~\eqref{eq:ConservCirculationDiffNewt}. Hence,
the circulation around a fluid ring $c_\lambda=\Psi_\lambda (c)$,
obtained by moving each point of $c$ a thermal time $\lambda$ (cf. Eq.~\eqref{eq:lamda}) 
 along the flow through that point, is indeed conserved:
\be \label{eq:ConservCirculationIntKelvinBaroclinic}
\frac{d}{d\lambda}\oint_{c_\lambda} h \w{u}  
=\frac{d}{d\lambda}\int_{{S}_{\lambda}}
\, {\bf{d}} ( h \w{u})
=\int_{{S}}
{\pounds_{\vw{u}/T}}\, {\bf{d}} ( h \w{u})
=0  
\ee
as dictated by Eqs.~\eqref{eq:CirculationInt}-\eqref{eq:ConservCirculationIntNewt}. Here, the  circulation
can be initially computed along an \textit{arbitrary} fluid ring $c$.
Thus,
unlike the previous weak form, this  circulation theorem  is a \textit{strong} form of Kelvin's  theorem, applicable to  baroclinic fluids.
 
It will be shown below that
this circulation theorem can be further extended to  barotropic
or baroclinic,
perfectly or poorly conducting, magnetofluids.
These (new and old) circulation theorems are again special cases of the Poincar\'e-Cartan integral invariant 
\eqref{eq:CirculationInt}.
The fluid helicity, on the other hand, is not conserved for baroclinic fluids, as these systems are not uniformly canonical.
\subsection{Ideal magnetoflows} \label{s:perfect_cond}

\subsubsection{Maxwell equations}

Consider an electromagnetic field in $\M$, described by the electromagnetic 2-form $\w{F}$, known as the \emph{Faraday tensor},  satisfying   the Maxwell equations which, in natural Heaviside-Lorentz units, read
$\nabla_\alpha(\star F^{\alpha \beta})=0$, $\nabla_\alpha F^{\alpha \beta} =J^\beta$ or
\begin{subequations} \label{e:max12}
\bea  
   \dd \w{F} &=& 0 \label{e:max1} \\
   \dd  \star\!\w{F} &=&  \star\!\vv{J} , \label{e:max2}
\eea 
\end{subequations}
where
$\star\w{F}$ is the 2-form Hodge-dual of $\w{F}$, namely
$
  \star\!F_{\alpha\beta} := \frac{1}{2} \eps_{\alpha\beta\gamma\delta} F^{\gamma\delta} $,
and $\star\vv{J}$ is the 3-form Hodge-dual of the 1-form $\vv{J}$ associated with the electric 4-current $\vw{J}$, namely
$\star J_{\alpha\beta\gamma}:= \eps_{\alpha \beta\gamma\delta}J^\delta$. 

The electric 4-current may be decomposed as $\vw{J}=e \vw{u}+\vw{j}$
where $e=-\vv{u}\cdot\vw{J}$ is the proper charge density, $e \vw{u}$ is the convection current and $\vw{j}$ is the conduction current, satisfying $\vv{u}\cdot\vw{j}=0$.
For an isotropically conducting medium, Ohm's law can be written as 
\be \label{eq:Ohmmm}
\w{j}=\sigma {\w{E}}
\ee
 where $\sigma$ is the conductivity of the medium and ${\w{E}}$ is the electric field measured by an observer comoving with the fluid, given by
Eq.~\eqref{e:def_e_b} below.
In the perfect conductivity limit, $\sigma \rightarrow \infty$, the electric field vanishes, ${\w{E}} \rightarrow 0$. In the poor conductivity limit, 
$\sigma \rightarrow 0$,  the conduction current vanishes, 
$\w{j} \rightarrow 0$.


\subsubsection{Magnetohydrodynamic Euler equation}

The relativistic MHD-Euler equation can be obtained from the conservation law of energy-momentum, 
\be \label{e:cons_enermom}
  \diver (\w{T}^{\rm fl} + \w{T}^{\rm em}) = 0 , 
\ee
where
 $\w{T}^{\rm em}$ is the energy-momentum tensor of the electromagnetic field:
\be \label{e:Tem}
        T_{\alpha\beta}^{\rm em} = 
         F_{\gamma\alpha} F^\gamma_{\ \, \beta}
        - \frac{1}{4} F_{\gamma\delta} F^{\gamma\delta} \; g_{\alpha\beta}  . 
\ee
This tensor is trace-free: $g^{\alpha\beta}{T}_{\alpha\beta}^{\rm em}=0$.
Taking the divergence of Eq.~(\ref{e:Tem}) and using the Maxwell equations  \eqref{e:max12}, one obtains the  well known relation 
\be \label{e:divTem}
  \diver \w{T}^{\rm em} = - \w{F}\cdot\vw{J} . 
\ee
Substituting Eqs.~\eqref{eq:fluidenmomediv} and \eqref{e:divTem} into the conservation 
law (\ref{e:cons_enermom}) yields the  \emph{MHD-Euler} equation for \textit{baroclinic magnetofluids}:
\bea
   \vw{u} \cdot \dd(h \w{u}) = T \dd S + \frac{1}{\rho} \w{F}\cdot\vw{J} . \label{e:MHD-Euler}
\eea
As shown in Ref.~\cite{GourgMUE11}, the specific form (\ref{e:MHD-Euler})
is well adapted to the cases where the spacetime exhibits some symmetries. 
Projecting the MHD-Euler equation along $\vw{u}$ yields
$T \Lie{\vw{u}} S = \frac{1}{\rho} \vv{E} \cdot\vw{J}$. The right hand side of this equation, which represents Joule 
heating,  vanishes in the limit of  perfect conductivity, whence the flow is adiabatic:
\bea
  && \Lie{\vw{u}} S =\vw u \cdot
\wnab S= 0 . \label{e:S_conserved} 
\eea
 For \textit{barotropic
magnetofluids}, the above equation simplifies to
\bea
   \vw{u} \cdot \dd(h \w{u}) = \frac{1}{\rho} \w{F}\cdot\vw{J}
. \label{e:MHD-Euler-Barotropic}
\eea 
In the abscence of pressure and currents ($h\rightarrow1 $ and $\vv J\rightarrow0$), this equation reduces to  the geodesic equation, $\vw{u} \cdot \dd \w{u} =0$, as expected.
 
\subsubsection{Perfectly conducting magnetoflows } \label{s:perfect_cond}


The electric field 1-form
$\w{E}$ and the magnetic field vector $\vw{B}$ measured in the fluid rest-frame, by an observer
of 4-velocity $\vw{u}$,
are given in terms of $\w{F}$ by 
\be \label{e:def_e_b}
        \vv{E} = - \vw{u} \cdot \w{F},
        \qquad
        \vv{B} = \vw{u} \cdot\star\!\w{F}  
\ee
and satisfy
\be  \label{eq:EdotuBdotu}
\vv{E} \cdot\vw{u} = 0,
        \qquad
        \vv{B} \cdot  \vw{u}  =0.
\ee
Equivalently, we can decompose $\w{F}$ into electric and magnetic parts with respect to  the rest frame defined by the vector $\vw{u}$,
as
\begin{subequations}
\bea
         \w{F} & = & \w{u}\wedge \w{E} +\star (\w{u}\wedge \w{B})  \\
        \star \!\w{F} & = & \star(\w{u}\wedge \w{E})- \w{u}\wedge \w{B}.
\eea
\end{subequations}
The scalar invariants of the field are given by
\bea
\frac{1}{2}F^{\alpha\beta}F_{\alpha\beta}&=&\vw{B} \cdot \w{B}
-  \vw{E} \cdot \w{E}  \label{eq:FdotwedgeF} \\
\frac{1}{2}(\star F^{\alpha\beta})F_{\alpha\beta}&=&\vw{B} \cdot \w{E} .
\eea

In ideal MHD, one assumes that the fluid occupying the part $\mathcal{D}\subset\M$ of
spacetime is  a 
\emph{perfect conductor}. By this, we mean that the observers comoving with
the fluid measure a vanishing electric field. By virtue of Ohm's law \eqref{eq:Ohmmm}, this
expresses the \emph{infinite conductivity} condition.  From (\ref{e:def_e_b}),
this condition amounts to 
\be  \label{e:perfect_cond}
       \vv{E} = \w{F}\cdot \vw{u} = 0 . 
\ee
The electromagnetic field then reduces to
\begin{subequations}
\bea
         \w{F} & = &  \star ( \w{u}\wedge \w{B}) 
         \label{e:sF_u_a}
           \\
        \star \!\w{F} & = & - \w{u}\wedge \w{B}   \label{e:sF_u_b}
\eea
\end{subequations}
and the Maxwell equation \eqref{e:max1} simplifies to 
\bea \label{eq:MaxwellBfluxfreeze}
 \dd\star ( \w{u}\wedge \w{B})=0\Leftrightarrow \nabla_\alpha(u^\alpha
B^\beta-u^\beta B^\alpha)=0.
\eea
This equation is a special case of Eq.~\eqref{eq:VorticityPropagation}, for reasons that will become clear below. In ideal MHD, one only has to evolve
the magnetic field equation 
\eqref{eq:MaxwellBfluxfreeze}. The current has no dynamical degrees of freedom and is merely
\textit{defined} in terms of the magnetic field via Eq.~\eqref{e:sF_u_a} and the Maxwell equation
 \eqref{e:max2}.  
    One then  evolves the MHD-Euler equation \eqref{e:MHD-Euler} after evaluating
the Lorentz force term in its right-hand side.

Alternatively, by writing $\w F$ in terms of the electromagnetic potential 1-form $\w{A}$,
\be \label{eq:FeqsdA}
\w F = \dd \w{A},
\ee
one automatically satisfies the Maxwell
equation (\ref{e:max1}). The perfect conductivity condition \eqref{e:perfect_cond}
is then used to  evolve the electromagnetic potential \cite{Etienneetal2010,Etienneetal2011}: 
\be  \label{eq:MHDevol_a}
\vv{u} \cdot \dd \vv{A} = 0\Leftrightarrow u^\alpha
(\nabla_\alpha A_\beta-\nabla_\beta A_\alpha)=0
\ee
In 3+1 dimensions, Eq.~\eqref{eq:MaxwellBfluxfreeze} is the curl of Eq.~\eqref{eq:MHDevol_a}, as shown in Sec.~\ref{subseq:CanonicalEvolutionScheme}.

\subsubsection{Action of a magnetic field frozen into the flow}

A  magnetic field frozen into the fluid, as defined by the perfect conductivity condition~\eqref{e:perfect_cond},
 is characterized by the action functional
 \be \label{eq:actionPurelyMagneticField}
\mathcal{S}=\int_{\tau_1}^{\tau_2} A_\alpha  \,\frac{dx^\alpha}{d\tau}d \tau
\ee
where the electromagnetic potential $\w A$ is considered a function of $x$
only. From the Lagrangian \cite{Carte79}
\be \label{eq:LuA}
L(x,u)=u^\alpha A_\alpha
\ee
we finds that the canonical momentum 1-form \eqref{eq:canonicalmomentum} is the electromagnetic potential
\be   \label{eq:MomentumPurelyMagnetic}
p_\alpha =\frac{\partial L}{\partial u^\alpha}= A_\alpha
\ee
and the canonical vorticity 2-form \eqref{eq:generalizedvorticity} is simply the Faraday tensor
\be \label{eq:GeneralizedVorticityFaraday}
 \w F = \dd \w{A} . 
\ee  
Because the super-Hamiltonian \eqref{eq:HamiltonianLagrangianLegendre} vanishes,
\be \label{eq:superHamPurelyMagnetic}
H=0,
\ee
the canonical equation of motion
\eqref{eq:HamiltonsEqsCovExtNewt} takes the form of the perfect conductivity
condition \eqref{eq:MHDevol_a}.

\subsubsection{Alfv\'en's theorem: conservation of magnetic flux}
If we express the Lie derivative of $\w{F}$ along $\vw{u}$ via the Cartan
identity, 
\be
\pounds_{\vw{u}} \w{F} = \vw{u}\cdot \dd \w{F} + \dd(\vw{u}\cdot\w{F}),
\ee
and take into account the Maxwell equation (\ref{e:max1})
and the perfect conductivity  condition (\ref{e:perfect_cond}), we get
\be  \label{eq:Alfventheoremdif}
        \pounds_{\vw{u}} \w{F} = 0 . 
\ee
This result, which  also follows from Eq.~\eqref{eq:ConservCirculationDiffNewt}, is the
geometrical expression of
\emph{Alfv\'en's magnetic flux theorem}: 
 the magnetic flux through a fluid ring 
$c_\tau$ dragged along by the flow
is conserved
\begin{equation} \label{eq:AlfvenTheoremNewt}
\frac{d}{d\tau}\oint_{c_\tau} \w{A} 
=\frac{d}{d\tau}\int_{{S}_{\tau}}  \w{F} =\int_{{S}}  \pounds_{\vw{u}} \w{F}=0.
\end{equation}
This follows directly from Eq.~\eqref{eq:ConservCirculationIntNewt} for the Lagrangian \eqref{eq:LuA} and is therefore simply a  special case of the Poincar\'e-Cartan
integral invariant \eqref{eq:CirculationInt}. Intuitively, Alfv\'en's theorem is a consequence of perfect conductivity. If one attempts to change the magnetic field and thus the magnetic flux 
 through the ring $c_\tau$  of fluid, then, in accordance with
Lenz's law,   induced currents will
generate a compensatory magnetic field in an attempt to cancel the change
of flux. In the limit of perfect conductivity, this  cancellation is perfect
and the flux is exactly conserved. 

\subsubsection{Magnetic helicity}
 
 Since the super-Hamiltonian \eqref{eq:superHamPurelyMagnetic}
is constant, the system is uniformly canonical, and the magnetic helicity, \be \label{eq:FluidHel}
{\w{h}_{\rm{em}}} := {\vw{A}} \cdot \star {\w {F}} ,
\ee
obtained by substituting Eqs.~\eqref{eq:MomentumPurelyMagnetic} and \eqref{eq:GeneralizedVorticityFaraday}
 into \eqref{eq:GenHel},  
is  conserved
\be \label{eq:FluidHel}
\nabla_\alpha {h}^\alpha_{\rm{em}}=0
\ee
by virtue of
Eq.~\eqref{eq:delh}. This   implies a corresponding global conservation of
the
integrated flux of $h^\alpha_{\rm{em}}$ across a spatial hypersurface, which amounts to  the relativistic generalization of Woltjer's magnetic helicity \cite{Woltjer1958,Bekenstein1987,Carter1992}.



\subsubsection{Einstein-Maxwell-Euler spacetimes}
 The classical action describing 
an Einstein-Maxwell-Euler spacetime $(\M,\w{g})$, coupled with a perfect fluid 
carrying an electric current, is given by \cite{FriedmanStergioulas2013}
\begin{equation} \label{eq:AReFFAj}
\mathcal{A} \,=\,
\int d^4 x \sqrt{-g}\Big( - \eps +\frac1{16\pi} R
\,-\,\frac1{4} F_{\alpha \beta} F^{\alpha \beta} + A_\alpha J^\alpha
\Big),
\end{equation}
where $R$ is the Ricci scalar. By writing $\w F$ in terms of a 1-form potential $\w{A}$, Eq.~\eqref{eq:FeqsdA}, one satisfies the Maxwell
equation (\ref{e:max1}). Varying the action with respect to the 
metric $\w g$ yields the Einstein equations; varying with respect to the electromagnetic 4-potential $\w A$ yields the Maxwell equations
\eqref{e:max2};
and varying with respect to the fluid variables 
yields the MHD Euler equation \eqref{e:MHD-Euler}.

Instead of imposing the perfect MHD condition after varying the action, one may incorporate it into the action. 
This can be done by  replacing the action \eqref{eq:AReFFAj} with 
\bea  \label{eq:AReBBadelBuuB}
\mathcal{A}\ \,=\,
\int d^4 x \sqrt{-g}\Big[-\eps +\frac1{16\pi} R
\,-\,\frac1{2} B_\alpha B^\alpha  \,
\nonumber\\
+\, a_\alpha \nabla_\beta(B^\alpha u^\beta-B^\beta u^\alpha)
\,\Big]
\eea
where the 1-form $\w{a}$ is a Lagrange multiplier used to enforce the flux freezing condition \eqref{eq:MaxwellBfluxfreeze}. In writing the action functional above, we have taken into
account Eq.~\eqref{eq:FdotwedgeF} in order to evaluate the magnetic energy
term. 
 This action functional differs by a
surface term from that  of Bekenstein-Oron \cite{Bekenstein:2000sf} which, in our notation, reads
\bea  \label{eq:AReBB}
\mathcal{A}\ \,=\,
\int d^4 x \sqrt{-g}\Big(- \eps +\frac1{16\pi} R 
-\frac1{2} B_\alpha B^\alpha  
+ b^\alpha F_{\alpha \beta} u^\beta
\Big).\nonumber
\eea
Here, the Lagrange multiplier $\w b$ is used to enforce the perfect conductivity condition \eqref{e:perfect_cond} and is shown to be the curl of $\w a$ as indicated by Eq.~\eqref{eq:beqsstarda} below. Our action \eqref{eq:AReBBadelBuuB} closely resembles  the non-relativistic action of  Bekenstein-Oron \cite{Bekenstein:2000sf}, which is a more natural starting point and  simplifies the  discussion below. 
Variation of the action \eqref{eq:AReBBadelBuuB} with respect to the multiplier $\w{a}$ yields the Maxwell equation \eqref{eq:MaxwellBfluxfreeze}, while variation with respect to the magnetic field $\w B$ and integration by parts yields the equation
\be \label{eq:udotdaeqB}
\vw{u} \cdot \dd \vv{a}=-\w{B} .
\ee
The multiplier $\w{a}$ may thus be thought of as an auxiliary field, with $\w{B} $  the \textit{electric} part of the 2-form 
\be  \label{eq:fda}
\vv{f}=\dd \vv{a}
\ee
[compare Eq.~(\ref{eq:udotdaeqB}) with (\ref{e:def_e_b})]. 
Note that the  above equation automatically satisfies the orthogonality condition
\eqref{eq:EdotuBdotu}. Comparing Eqs.~\eqref{eq:udotdaeqB}
and \eqref{e:def_e_b}, we infer that the Faraday tensor $\w F$ must be related to the 2-form $\vv{f}$ via a relation  $\star \w F =- \vv{f}+ {\w w}$ where $\w w$ is some 2-form satisfying $  \vw{u} \cdot \vv{w} =0$. Since 
$\w w$ has no electric part, it can be written in terms of its magnetic part, $\w{b} = \vw{u} \cdot\star\w{w} $, as $\w w = \star ( \w{u}\wedge \w{b}) $. Taking Eq.~\eqref{e:sF_u_b} into account, we  infer that
\be \label{eq:dadecomposition}
\vv{f}= \w{u}\wedge \w{B} + \star(\w{u}\wedge \w{b} )
\ee
That is, the 2-form \eqref{eq:fda} has an electric part given by Eq. \eqref{eq:udotdaeqB} and a magnetic part given by the 1-form 
\be  \label{eq:beqsstarda}
\w{b} = \vv{u} \cdot\star\dd \vv{a} .
\ee
As pointed out by Bekenstein and Oron  \cite{Bekenstein:2000sf}, the theory has a U(1)$\times$U(1)
symmetry, since the observable field $\w B$ remains invariant under gauge transformations $\w A \rightarrow \w A + \dd \Lambda$ and
 $\w a \rightarrow \w a + \dd \lambda$.

Taking the exterior derivative of Eq. \eqref{eq:dadecomposition} yields the Maxwell equation \eqref{e:max2}, with the Faraday tensor given by Eq. \eqref{e:sF_u_a}
and the current `defined' by
\be
J^\alpha=\nabla_\beta(u^\alpha b^\beta-u^\beta b^\alpha)
\ee
or
\be \label{eq:BOcur}
\vv{J} = \star \dd\!\star(\w{u}\wedge \w{b}) .
\ee
This expression has been obtained in \cite{Bekenstein:2000sf} via a lengthy  route and will be referred to as the \textit{Bekenstein-Oron current}. 
Note that the  above expression automatically satisfies the  continuity equation
\be
\diver\vw{J} =-\star\dd\!\star\vv{J} =0
\ee
regardless of any assumption about $\w u$ and $\w b$. Physically, the above equation expresses the  conservation of electric charge. The operator $\star\dd\star$ is the \emph{codifferential} and has  been expressed as the 
\emph{divergence} taken with the $\wnab$ connection. For  convenience, let us introduce an auxiliary vector $\vw q$ and an auxiliary 1-form $\w \eta$ defined by 
\be
q^\alpha:= b^\alpha/\rho , \quad
{\eta_\alpha}:= {F_{\alpha \beta}} {q^\beta} \nonumber
\ee
or
\be
\vw q:=\vw b/\rho , \quad
\w{\eta}:= \w{F} \cdot \vw{q} .\label{eq:defeta}
\ee
One may then use the continuity equation \eqref{e:continuity} to write the Bekenstein-Oron current \eqref{eq:BOcur} as 
\be
\vw J
 \,=\, \pounds_{\vw q} (\rho \vw u) 
 \,+\,\rho \vw u (\diver\vw{q}) .
\label{eq:BOcur2}
\ee
This expression can be used to write the Lorentz force term in \eqref{eq:MHDevol_b} as 
\be
\frac{1}{\rho} \w{F}\cdot\vw{J}=\frac{1}{\rho} \w{F}\cdot \Lie{\vw q} (\rho \vw u) =-\vw{u} \cdot \dd \vv{\eta} \label{eq:BO_LFterm} .
\ee
The last equality follows from projecting the Cartan identity, 
$\Lie{\vw q}\w F=\vw q\cdot \dd \w F + \dd(\vw q \cdot \w F)$, along the vector $\rho \,\vw u$ and using Eq.  
\eqref{e:perfect_cond}. By virtue of the above equality, the   MHD-Euler equation \eqref{e:MHD-Euler} takes the canonical form
\be
\pounds_{\vw{u}}(h \w{u}+\,\w\eta)  
+ {\bf{d}}h=\vw{u} \cdot \dd (h\w{u} \,+\,\w\eta) =  T \dd S 
\label{eq:MHD-Euler_form1}
\ee
%
which is valid for \textit{baroclinic magnetofluids}. For \textit{barotropic magnetofluids}, the above equation simplifies  to 
\be
\pounds_{\vw{u}}(h \w{u}+\,\w\eta)  
+ {\bf{d}}h=\vw{u} \cdot \dd (h\w{u} \,+\,\w\eta) =  0.
\label{eq:MHD-Euler_form1Barotropic}
\ee 
The last equality was obtained by Bekenstein et al. \cite{Bekenstein:2000sf,Bekenstein2006}. 

The tensor or vector calculus-based derivations in Ref.~\cite{Bekenstein:2000sf,Bekenstein2006} did not clarify the generality of this approach. In particular, one may  question whether the Bekenstein-Oron ansatz \eqref{eq:BOcur} for the current is
generic enough to accommodate any given ideal MHD flow. This question boils
down to whether  Eq. \eqref{eq:udotdaeqB} can be solved for any given magnetofluid
configuration with  magnetic field $\vv B$ and 4-velocity $\vw u$. The answer
 may be obtained   by   using the Cartan identity to write Eq.~\eqref{eq:udotdaeqB}
as $\pounds_{\vw{u}} \vv{a}-\dd(\vw{u} \cdot  \vv{a})=-\w{B}$ and  using
the gauge freedom in $\w a$ to set $\vw u \cdot \w a=0$ (this gauge condition can be shown to be preserved by the flow if satisfied initially). The resulting differential
equation, $\pounds_{\vw{u}} \vv{a}=-\w{B}$, is always solvable along the
integral curves of $\vw u$. We have thus shown that \textit{no loss of generality
is entailed in the  Bekenstein-Oron description of ideal MHD flows}.
For perfectly conducting magnetofluids, the Einstein-Maxwell-Euler action~\eqref{eq:AReFFAj}  may always be replaced by the  action~\eqref{eq:AReBBadelBuuB}, and the MHD-Euler equation \eqref{e:MHD-Euler} may always be replaced by Eq.~\eqref{eq:MHD-Euler_form1}.

\subsubsection{Hamilton's principle for a barotropic  magnetofluid element}

Carter \cite{Carte79}
has allowed the possibility that the perfect fluid be charged. His approach
is valid for poorly conducting fluids, but has been  considered inapplicable
to  conducting magnetofluids \cite{Bekenstein2001}.
 Nevertheless, it is shown below that Carter's framework can in  fact accommodate
perfectly conducting fluids in the context of  Bekenstein-Oron magnetohydrodynamics.
For a  barotropic,  perfectly conducting magnetofluid, we  generalize the action \eqref{eq:actionHydrodynamic} as follows:
\bea \label{eq:ActionMHDranders}
\mathcal{S}=
\int_{\tau_1}^{\tau_2}\left(-h 
\sqrt{-g_{\alpha \beta}\frac{dx^\alpha}{d\tau}
\frac{dx^\beta}{d\tau}}  +\eta_\alpha \frac{dx^\alpha}{d\tau}
\right)d\tau \qquad
\eea
 with Lagrangian
 \be \label{eq:Lagrangianhguuhetau0}
L(x,u)=-h 
\sqrt{-g_{\alpha \beta}{u^\alpha}{u^\beta}}
 +\eta_\alpha u^{\alpha}
\ee
and with $\w \eta$   given by Eq.~\eqref{eq:defeta}. The canonical velocity and momentum
of a magnetofluid element are given by
\begin{subequations} \label{eq:canonicalvelocitymomentumPerfectMagnetofluid}
\bea
{u^\alpha} &=& \frac{{d{x^\alpha}}}{{d\tau }}   \label{eq:canonicalvelocityPerfectMagnetofluid} \\
{p_\alpha} &=& \frac{{\partial L}}{{\partial {{u}^\alpha }}} =hu_\alpha+\eta_\alpha.
\label{eq:canonicalmomentumPerfectMagnetofluid}
\eea
\end{subequations}

Alternatively, one may introduce a Lagrangian which generalizes that of Carter, Eq.~\eqref{eq:Lagrangianhguuh}:
\be \label{eq:Lagrangianhguuhetau}
L(x,u)=\frac{1}{2}h g_{\alpha \beta}u^\alpha u^\beta-\frac{1}{2}h+\eta_\alpha
u^\alpha.
\ee
The associated Hamiltonian,
\be \label{eq:superHamMHD}
H(x,\pi)=\frac{1}{2h}g^{\alpha \beta}(p_\alpha-\eta_\alpha)(p_\beta-\eta_\beta)+\frac{1}{2}h,
\ee
 vanishes on-shell, so the Hamilton equation 
\eqref{eq:HamiltonsEqsCovExtNewt} 
yields the MHD--Euler equation in the Bekenstein-Oron form, 
Eq.~\eqref{eq:MHD-Euler_form1Barotropic}.

\subsubsection{Conservation of circulation in barotropic magnetoflows}

The canonical momentum 1-form of a barotropic ideal magnetofluid-element is given by Eq.~\eqref{eq:canonicalmomentumPerfectMagnetofluid}. 
Then, the Poincar\'e 2-form \eqref{eq:generalizedvorticity} amounts to  the 
\emph{canonical vorticity 2-form}:
\be
\w \Omega \,=\, \dd (h\w{u} \,+\,\w\eta) .
\label{eq:def_vorticity} 
\ee
Then, the Cartan identity, combined with Eq.~(\ref{eq:MHD-Euler_form1}) and
the identity
$\dd^2=0$, yields 
\be
\Lie{\vw u} \w \Omega = 0. 
\label{eq:Lieuomega}
\ee
This equation implies that the canonical vorticity of a  barotropic, perfectly
conducting magnetofluid is preserved by the flow. This leads to a generalization
of Kelvin's theorem to magnetized fluids.

Indeed, for the system \eqref{eq:ActionMHDranders}, the    Poincar\'e-Cartan  theorem 
\eqref{eq:ConservCirculationIntNewt}   implies that the circulation through a ring ${c}_\tau $ dragged along by the flow is conserved:
\begin{equation} \label{eq:ConservCirculationBOIntNewt}
\frac{d}{d\tau}\oint_{c_\tau}  (h \w u \,+\,\w\eta)
=0.
\end{equation}
This law    follows directly from Eq.~\eqref{eq:Lieuomega} and was first obtained  by Bekenstein and Oron \cite{Bekenstein:2000sf,Bekenstein2001}.      It
 is a generalization  of the relativistic Kelvin circulation theorem \eqref{eq:ConservCirculationIntKelvin}
(which is recovered in the non-magnetic limit $\w \eta=0$)
 to ideal MHD.
The most interesting feature of this conservation law is that it is
\textit{exact} in time-dependent spacetimes, with gravitational and electromagnetic
waves
carrying energy and angular momentum away from a system. In
particular, oscillating stars and radiating binaries, if modeled as
 barotropic magnetofluids with no viscosity, resistivity or other dissipation,
exactly
conserve circulation. 

\subsubsection{Ideal magnetofluid helicity}

Since the super-Hamiltonian \eqref{eq:superHamMHD}
is constant (zero), the system is uniformly canonical, and helicity is conserved:\ Substituting Eq.~\eqref{eq:canonicalmomentumPerfectMagnetofluid} into Eq.~\eqref{eq:GenHel} yields the magnetolfuid helicity \be \label{eq:FluidHel}
{\w{h}_{\rm{mfl}}} :=( h{\vw{u}} +\w \eta)\cdot \star {\w {\Omega}} 
\ee
which, by virtue of 
Eq.~\eqref{eq:delh}, 
is  conserved:
\be \label{eq:FluidHel}
\nabla_\alpha[(hu_\beta+\eta_\alpha)
\star
\Omega^{\beta\alpha
}]=0.
\ee
This   implies a corresponding global conservation of
the
integrated flux of $h^\alpha_{\rm{mfl}}$ across a spatial hypersurface. One may proceed analogously to Eq.~\eqref{eq:helicityt} to obtain a conserved volume integral, which 
amounts to  the  generalization of Moffat's fluid helicity \cite{MOFFAT1969,Bekenstein1987,Carter1992} to ideal GRMHD. 

\subsubsection{A canonical evolution scheme for ideal MHD}   \label{subseq:CanonicalEvolutionScheme}

In binary neutron-star inspiral,  the temperature is much lower than
the Fermi temperature, and heat conduction, viscosity and resistivity can
be neglected 
\cite{FriedmanStergioulas2013}. The fluid may then be approximated as \textit{barotropic},
\textit{adiabatic, inviscid} and \textit{perfectly conducting}. In general
relativity, such fluids are  described by the ideal MHD equations   \eqref{eq:MHDevol_a}
and \eqref{e:MHD-Euler-Barotropic}:
\begin{subequations} \label{eq:MHDevol}
\bea 
\label{eq:MHDevol_a2}
u^\alpha
(\nabla_\alpha A_\beta-\nabla_\beta A_\alpha) &=& 0
\\   \label{eq:MHDevol_b}
u^\alpha
[\nabla_\alpha (hu_\beta)-\nabla_\beta (hu_\alpha)]&=& \frac{1}{\rho} F_{\beta
\alpha} J^\alpha.
\eea
\end{subequations}
coupled to the continuity equation \eqref{e:continuity}. One can evolve 
Eq.~\eqref{eq:MHDevol_a} for the
electromagnetic potential and compute the Faraday tensor via Eq.~\eqref{eq:FeqsdA}.
 In ideal MHD, as mentioned earlier, the current lacks dynamical degrees
of freedom and is merely
`defined' in terms of the electromagnetic potential via the Maxwell equation
 \eqref{e:max2}.  
    One then  evolves the MHD-Euler equation \eqref{eq:MHDevol_b} after evaluating
the Lorentz force term in its right-hand side.

As shown earlier, the Bekenstein-Oron description  of ideal MHD allows one
to
replace the  MHD-Euler equation \eqref{e:MHD-Euler-Barotropic} by the system
of equations
  \eqref{eq:udotdaeqB} and \eqref{eq:MHD-Euler_form1Barotropic},
namely
\begin{subequations} \label{eq:circulationpreservingsystem}
\bea 
u^\alpha
(\nabla_\alpha A_\beta-\nabla_\beta A_\alpha) &=& 0 \label{eq:circulationpreservingsystem_a}
\\
u^\alpha
(\nabla_\alpha a_\beta-\nabla_\beta a_\alpha) &=& -B_{\beta} \label{eq:circulationpreservingsystem_b}
\\
u^\alpha
(\nabla_\alpha p_\beta-\nabla_\beta p_\alpha) &=& 0 \label{eq:circulationpreservingsystem_c}
\eea
\end{subequations}
where 
\be
\w p=h\w{u} \,+\,\w\eta
\label{eq:defw}
\ee
is the canonical momentum 1-form of a magnetofluid element, as shown in the
next section. 

In a chart $\{t,x^i\}$, the above system can be written in 3+1 canonical
hyperbolic form as
\begin{subequations} \label{eq:circulationpreservingsystem2}
\bea 
\partial_t A_i-\partial_i A_t+ \upsilon^j
(\partial_j A_i-\partial_i A_j) &=& 0 \label{eq:circulationpreservingsystem_a2}
\\
\partial_t a_i-\partial_i a_t+ \upsilon^j
(\partial_j a_i-\partial_i a_j) &=& -B_{i} \label{eq:circulationpreservingsystem_b2}
\\
\partial_t p_i-\partial_i p_t+ \upsilon^j
(\partial_j p_i-\partial_i p_j)  &=& 0 \label{eq:circulationpreservingsystem_c2}
\eea
\end{subequations}
where  $\upsilon^i=u^i/u^t=dx^i/dt$ is the fluid velocity measured in local
coordinates.
This system may be evolved analogously to the  system \eqref{eq:MHDevol}.
One evolves the first equation for $\w A$ and computes the magnetic field
$\w{B} = \vw{u} \!\cdot\!\star\dd\w{A}$. With this source, one evolves the
second equation for $\w a$ and  computes the auxiliary field  $\w{b} = \vw{u}
\cdot\star\dd \vv{a}$. Finally, one solves the last equation of the above
system, taking Eqs.~\eqref{eq:defeta} and \eqref{eq:defw} into account, to
evolve the hydromagnetic flow.

 The curl of  the evolution equation~\eqref{eq:circulationpreservingsystem_a2}
 is an evolution equation  for
the magnetic field. Explicitly, the spatial exterior derivatives
of the system \eqref{eq:circulationpreservingsystem2} yield an evolution  system for the spatial parts of the 2-forms \eqref{eq:GeneralizedVorticityFaraday}, \eqref{eq:fda} and \eqref{eq:def_vorticity}.
In flux-conservative form, this system reads:
\begin{subequations} \label{eq:circulationpreservingsystem3}
\bea 
\partial_t F_{jk}+ {\partial _i}(\delta_{jk}^{\;\;\,\, il}{\upsilon^m F_{ml}}) &=& 0 \qquad \label{eq:circulationpreservingsystem_a3}
\\
\partial_t f_{jk}+ {\partial _i}[\delta_{jk}^{\;\;\,\, il}({\upsilon^m f_{ml}}+B_l)]  &=& 0  \qquad \label{eq:circulationpreservingsystem_b3}
\\
\partial_t \Omega_{jk}+ {\partial _i}(\delta_{jk}^{\;\;\,\, il}{\upsilon^m \Omega_{ml}}) &=& 0  \qquad \label{eq:circulationpreservingsystem_c3}
\eea
\end{subequations}
where $\delta_{jk}^{\;\;\,\, il}=\epsilon_{jkn}\epsilon^{iln}=\delta _j^i\delta _k^l - \delta
_k^i\delta _j^l$ is the generalized Kronecker delta.
 Eq.~\eqref{eq:circulationpreservingsystem_a3} is an evolution equation, equivalent\footnote{Unlike Eq.~\eqref{eq:MaxwellBfluxfreeze} which contains
the metric and its connection, Eq.~\eqref{eq:circulationpreservingsystem_a3}
contains no such dependence, yet both equations are equivalent and exact in curved spacetime.} 
to  Eq.~\eqref{eq:MaxwellBfluxfreeze}, for the magnetic field. Numerical evolution of the latter typically  requires techniques such as
hyperbolic divergence cleaning or constrained transport to avoid error accumulation  from 
a finite magnetic divergence \cite{Gammie2003}.
 Such numerical schemes can also
be
applied  to evolving the system \eqref{eq:circulationpreservingsystem3}.  Etienne et al. \cite{Etienneetal2010,Etienneetal2011}
have performed GRMHD simulations that directly evolve the electromagnetic
potential $\w A$ by means of Eq.~\eqref{eq:circulationpreservingsystem_a2}.
The magnetic field  is then computed from the curl of the vector potential
and has zero divergence by construction. This numerical scheme can also be
applied  to evolving the system  \eqref{eq:circulationpreservingsystem2}. 

Note that the  systems  \eqref{eq:circulationpreservingsystem2} and  \eqref{eq:circulationpreservingsystem3}
 were obtained from equations involving only exterior spatial derivatives,
and thus do not involve the spacetime metric or its connection.
Thus, these systems are \textit{independent of gravity theory} and can be shown to be \textit{valid as written even in the Newtonian limit}. For nonmagnetic
fluids, Eq.~\eqref{eq:circulationpreservingsystem_c2} was obtained  from
a  3+1 constrained Hamiltonian formulation of the Euler equation in Ref.~\cite{Markakis2014a},
where it was shown to be strongly hyperbolic. Other strongly hyperbolic  formulations of the relativistic Euler equation
include the Valencia formulation \cite{Font2008} and the symmetric hyperbolic
Fraudendiner-Walton formulation \cite{Frauendiener2003,Walton2005,Oliynyk2012,Oliynyk2014}. The hyperbolicity of the evolution
system   \eqref{eq:circulationpreservingsystem2} is the subject of future
work.
 A notable feature of the  canonical evolution system \eqref{eq:circulationpreservingsystem}
is that it  manifestly preserves magnetic flux and circulation, owing to
its symplecic
structure. 
  Eq.~\eqref{eq:circulationpreservingsystem_a2} can also be obtained from a constrained Hamiltonian.
Symplectic evolution schemes based on the Hamiltonians of Eqs.~\eqref{eq:circulationpreservingsystem_a2}
and \eqref{eq:circulationpreservingsystem_c2}
are expected to numerically preserve such properties.
Moreover, if the
system admits a Noether symmetry, this canonical form quickly gives rise
to first integrals as  discussed below.

\subsubsection{Magnetars with helical symmetry}
As an example, let us consider a helically symmetric rigidly rotating system,
such as a  rigidly rotating magnetar triaxially deformed by its off-axis
frozen magnetic field. The
flow field may then be written in the form of Eq.~\eqref{eq:ucorrot}. 
Let us assume that all observable fields (such as $h, \vv u, \vv B, \w F,\w
g$) are helically symmetric, that is, their Lie derivatives along the helical Killing vector  $\vw{k}$, given by Eq.~\eqref{eq:HelicalKillingVector}, vanish. 

Using gauge freedom, one can always find a gauge class for which the electromagnetic
potential $\w A$ inherits the Killing symmetries of $\w F=\dd \w A$ 
\cite{Uryu2010,Gourgoulhon2011,Uryu2014}. Then,
using Eq. \eqref{eq:ucorrot} and the Cartan identity, $\Lie{\vw k} \w A =
\vw k \cdot \dd \w A + \dd (\vw k \cdot \w A)=0$, we find that
Eq.~\eqref{eq:circulationpreservingsystem_a} has the first integral
\be  \label{eq:Adotkeq0}
\w A \cdot \vw k=A_t+\Omega A_\varphi= \mathrm{constant}. 
\ee
Similarly, using $\Lie{\vw k} \w a = \vw k \cdot \dd \w a+ \dd (\vw k \cdot
\w a)$ and imposing the gauge condition $\vw k \cdot \w a=0$ allows one to
write  
Eq.~\eqref{eq:circulationpreservingsystem_b} as 
$\Lie{\vw k} \w a = - \w B/u^t$. 
This equation has the simple solution
\be \label{eq:asolBtut}
\w a = - \w B \,t/u^t , 
\ee
 where the scalar field $t$ satisfies  $t^\alpha \nabla_\alpha t=1$, so that
 $\Lie{\vw k} t=(\partial_t + \Omega \partial_\varphi)t=1$.  
  Note that the auxiliary  fields $\w a$ and
 $\w{b} = \vw{u} \cdot\star\dd \vv{a}$ are not observable and need not satisfy
helical symmetry (cf. Appendix \ref{sec:AppendixB}). Finally, Eq. \eqref{eq:ucorrot}
and the Cartan identity allow one to write Eq.~\eqref{eq:circulationpreservingsystem_c}
  in the form of Eq.~\eqref{eq:RigidEuler},
 which has the first integral
\be \label{eq:pidotkplusfconst}
\w p \cdot \vw k + f =-h/u^t +f = \mathrm{constant}. 
\ee
The first integrals \eqref{eq:Adotkeq0} and \eqref{eq:pidotkplusfconst} are
consequences of  stationarity in an inertial  ($\Omega = 0$) or rotating
($\Omega>0$) frame and, like Eq.~\eqref{eq:vonZeipel}, can be considered generalizations of \textit{von Zeipel's law} to relativistic magnetoflows.
The scalar $f$ is such that $\dd f = - \Lie{\vw k} \w p$ or, by virtue
of Eq. \eqref{eq:defw},
\be \label{eq:dfLieketa}
\dd f = - \Lie{\vw k} \w \eta .
\ee 
The right-hand side of this equation is proportional to the Lorentz force.
One way to  see this is to  act with $ \Lie{\vv k}$ on Eq.~\eqref{eq:defeta}, 
\be
\w
\eta = \dd \w A \cdot \vw b / \rho = ( B^2/\rho) \w u- \dd \w a \cdot \vw
B / \rho,
\ee
and use
 Eq. \eqref{eq:asolBtut},  yielding
\be \label{eq:LieketaLorentzForce}
\pounds_{\vw k} \w \eta = 
 \dd(\w B/u^t)\cdot \vw B / \rho.
\ee 
Eq. \eqref{eq:dfLieketa} then implies that the Lorentz force must be the
gradient of a scalar potential $f$ in order for helically symmetric corotating
configurations  solutions to exist. This equation is subject to the integrability
condition
\be  \label{eq:intcontLorentzForce}
\dd \Lie{\vw k} \w \eta = - \,\dd^2 f= 0. 
\ee
which constitutes  a restriction on the magnetic field $\w B$ on which $\w
\eta$ depends.  By virtue of Eq. \eqref{eq:LieketaLorentzForce},  the above
condition becomes
\be  \label{eq:intcontLorentzForce2}
\dd(\w B/u^t) \wedge \dd(\w B / \rho)=0.
\ee
For corotating helically symmetric magnetoflows,  the system of nonlinear
partial differential equations 
\eqref{eq:circulationpreservingsystem}  has been reduced to the system of
 algebraic equations \eqref{eq:Adotkeq0}--\eqref{eq:pidotkplusfconst} 
 and  the partial differential equation \eqref{eq:dfLieketa}.
The Newtonian analogue of Eq.~\eqref{eq:intcontLorentzForce}
 has been considered in Ref.~\cite{Haskelletal2008}.

\subsubsection{Hamilton's principle for a baroclinic magnetofluid element}

For a  baroclinic,  perfectly conducting magnetofluid, we consider the
action
\bea \label{eq:ActionMHDrandersBaroclinic}
\mathcal{S}=
\! \int_{\lambda_1}^{\lambda_2} \! \left(-h 
\sqrt{-g_{\alpha \beta}\frac{dx^\alpha}{d\lambda}
\frac{dx^\beta}{d\lambda}}  +\eta_\alpha \frac{dx^\alpha}{d\lambda}+S
\right) \! d\lambda \qquad \quad
\eea
with $\w \eta$   given by Eq.~\eqref{eq:defeta}. Like its non-magnetic limit \eqref{eq:actionHydrodynamicBaroclinic}, the above functional is parametrized in terms of  thermal time
 $\lambda$, cf. Eq.~\eqref{eq:lamda}.
 The Lagrangian of a magnetofluid element
\be \label{eq:Lagrangianhguuhetau}
L(x,v)=-h 
\sqrt{-g_{\alpha \beta} v^\alpha v^\beta}  +\eta_\alpha v^\alpha+S
\ee
is associated with a canonical velocity
and canonical momentum
\begin{subequations} \label{eq:canonicalvelocitymomentumPerfectMagnetofluidS}
\bea
{v^\alpha} &=& \frac{{d{x^\alpha}}}{{d\lambda }}=\frac{1}{T}\frac{{d{x^\alpha}}}{{d\tau
}}= \frac{1}{T}{u^\alpha}   \label{eq:canonicalvelocityPerfectMagnetofluidS}
\\
{p_\alpha} &=& \frac{{\partial L}}{{\partial {{v}^\alpha }}} =T \,h v_\alpha+\eta_\alpha=hu_\alpha+\eta_\alpha.
\label{eq:canonicalmomentumPerfectMagnetofluidS}
\eea
\end{subequations}
On-shell, by virtue of Eqs.~\eqref{eq:uuem1},  
 \eqref{e:perfect_cond}
and \eqref{eq:defeta}, the Lagrangian takes the value $L  =  - g /T =  - h/T + S$ and, by virtue
of Eq.~\eqref{eq:HamiltonianLagrangianLegendre},  the super-Hamiltonian takes
the value $H=-S$. The Euler-Lagrange equation \eqref{eq:eulerlagrangecovExtNewt}
thus becomes
\begin{equation} \label{eq:eulerlagrangeBaroclinicMHD}
  {\pounds_{\vw{u}/T}}(h\w{u}+\w\eta) = {\bf{d}}(S-h/T) 
\end{equation}
and the Hamilton equation \eqref{eq:HamiltonsEqsCovExtNewt} becomes
\begin{equation} \label{eq:HamiltonsEqsBaroclinicMHD}
  \frac{\vw{u}}{T}\cdot {\bf{d}}(h\w{u}+\w\eta) =   {\bf{d}}S.
\end{equation}
These equations are related via the Cartan identity and are  equivalent expressions of the  MHD-Euler equation \eqref{eq:MHD-Euler_form1}.

Alternatively, one may generalize Carter's Lagrangian  
\eqref{eq:LagrangianBaroclinicCarter}
to perfectly conducting baroclinic magnetofluids: the resulting Lagrangian
\be \label{eq:LagrangianBaroclinicCarterMHD}
L(x,v)=\frac{1}{2}Th g_{\alpha \beta}v^\alpha v^\beta+\eta_\alpha v^\alpha-\frac{1}{2}\left(\frac{g}{T}-S\right)
\ee  is associated with
the same canonical velocity and momentum \eqref{eq:canonicalvelocitymomentumPerfectMagnetofluid}
  and leads also to the
 equation of motion \eqref{eq:eulerlagrangeBaroclinicMHD}.  The Legendre
transformation~\eqref{eq:HamiltonianLagrangianLegendre} yields the super-Hamiltonian
\be \label{eq:HamiltonianBaroclinic}
H(x,p)=\frac{1}{2Th}g^{\alpha \beta}(p_\alpha-\eta_\alpha) (p_\beta-\eta_\beta)+\frac{h}{2T}-S,
\ee
which  leads
to the canonical equation of motion
\eqref{eq:HamiltonsEqsBaroclinicMHD}.

\subsubsection{Conservation of circulation in baroclinic magnetoflows}
Like their nonmagnetic counterparts, baroclinic magnetoflows do not Lie-drag the vorticity \eqref{eq:def_vorticity}: the exterior derivative of Eq.~\eqref{eq:MHD-Euler_form1}
reads
\begin{equation} \label{eq:ConservCirculationBarotropic}
 {\pounds_{\vw{u}}}\, {\bf{d}} ( h \w{u}+\w{\eta} ) =  {\bf{d}} T \wedge
 {\bf{d}}
S.
\end{equation} 
Thus, as in Eq.~\eqref{eq:ConservCirculationIntKelvin}, the circulation around a magnetofluid ring dragged
along by the flow is not generally conserved, except in a weak sense, i.e. for rings of constant specific entropy or temperature.

Nevertheless, like their nonmagnetic counterparts, ideal baroclinic magnetoflows are 
 Lie-dragged by the canonical fluid velocity \eqref{eq:canonicalvelocityPerfectMagnetofluid}:
\begin{equation} \label{eq:ConservVorticityBaroclinic}
 {\pounds_{\vw{u}/T}}\, {\bf{d}} ( h \w{u}+\w{\eta}) = 0
\end{equation}
as dictated by Eq.~\eqref{eq:eulerlagrangeBaroclinicMHD}, and this leades to a strong conservation law. In particular,
the circulation around a magnetofluid ring $c_\lambda=\Psi_\lambda (c)$,
obtained by moving each point of $c$ a thermal time $\lambda$ (cf. Eq.~\eqref{eq:lamda})
along the flow through that point, is indeed conserved:
\bea \label{eq:ConservCirculationIntKelvinBaroclinic}
\frac{d}{d\lambda}\oint_{c_\lambda} h \w{u}  +\w{\eta}
&=&\frac{d}{d\lambda}\int_{{S}_{\lambda}}
\, {\bf{d}} ( h \w{u}+\w{\eta}) \nonumber \\
&=&\int_{{S}}
{\pounds_{\vw{u}/T}}\, {\bf{d}} ( h \w{u}+\w{\eta})
=0.  \qquad
\eea
Here, the  circulation
can be initially computed along an \textit{arbitrary} fluid ring $c$.
This conservation of circulation law generalizes the Bekenstein-Oron law~\eqref{eq:ConservCirculationBOIntNewt} to baroclinic magnetofluids. The conserved circulation is 
the Poincar\'e-Cartan integral invariant of the Hamiltonian system 
described by the action \eqref{eq:ActionMHDrandersBaroclinic}. Although it has not appeared in the literature before, it is a special case of Eqs.~\eqref{eq:CirculationInt} and \eqref{eq:ConservCirculationIntNewt}, like all 
circulation integrals presented earlier. 

A very similar conservation of circulation law can be obtained    for \textit{a poorly conducting
fluid}, simply by replacing $\eta_\alpha$ with $e A_\alpha$, where $e$ is the net charge per fluid element, in the action \eqref{eq:ActionMHDrandersBaroclinic} and all equations that follow from it (cf. \cite{Carte79} for poorly conducting barotropic fluids).
 Although conservation of circulation holds in the limits of infinite or zero conductivity, we have not been able to obtain such a law for finite conductivity. This may be attributed to the fact that, for finite conductivity, 
the MHD-Euler equation \eqref{e:MHD-Euler} does  not follow from a Hamiltonian and, equivalently,  does not possess  a Poincar\'e-Cartan  integral invariant.



 
\subsection{The geometry  of barotropic  flows  }


\subsubsection{Hydrodynamic flows as geodesics in a Riemann space}

In Riemann geometry,  the line element  
 is given
by the quadratic expression  
\be \label{eq:lineelementRiemann}
d \mathcal{S}^2={-\gamma_{\alpha \beta}(x)dx^\alpha
dx^\beta}.
\ee
 where
$\gamma_{\alpha \beta}(x)$ is a Lorentzian metric on a Riemannian manifold $\mathcal{M}$. The distance between two points (or events) 1 and 2 is then given by the
integral
 \be \label{eq:lengthRiemann}
\mathcal{S} \!= \! - \! \int_1^2 \sqrt{-\gamma_{\alpha \beta}(x)dx^\alpha dx^\beta}
=- \! \int_{\tau_1}^{\tau_2} \sqrt{-\gamma_{\alpha \beta}(x) \dot x^\alpha \dot x^\beta}
d \tau 
\ee
where $\dot x^\alpha=dx^\alpha/d\tau$ is the velocity.
This functional  is independent of the parameter $\tau$. 

It was demonstrated above that if a perfect fluid is barotropic, then the
motion of a fluid element is conformally geodesic. In particular, Synge \cite{Synge37} and Lichnerowicz \cite{Lichn41} have shown that the motions of
 fluid elements in a barotropic fluid are
 geodesics of a  manifold 
$\mathcal{M}$ with metric
\bea \label{eq:metricRiemannHydrodynamic}
{\gamma _{\alpha \beta }}(x)
 &=&
    {h(x)^2}{g_{\alpha \beta }(x)} 
\eea
conformally related to the spacetime metric $g_{\alpha \beta}(x)$.
As shown earlier, such fluid motions  can indeed be obtained from the action \eqref{eq:actionHydrodynamic}, which represents the arc length \eqref{eq:lengthRiemann} between two events, and is  independent of the parameter $\tau$.

\subsubsection{Magnetohydrodynamic flows as geodesics in a Finsler space}
 One may think that the above result of Synge and  Lichnerowicz ceases to apply in MHD, due to the highly complicated nature of the MHD-Euler equation \eqref{e:MHD-Euler}. 
Surprisingly, however, the above results can be extended to   magnetofluids that are barotropic and perfectly conducting. 
Such
flows are described by the action \eqref{eq:ActionMHDranders}, which is  independent of the parameter 
$\tau$, and are  geodesic in a Finsler (rather than Riemann) space \cite{ShiingShenChern1996,Asanov1989,Lovelock1989}. In particular, in the context
of Finsler spaces, Eq.~\eqref{eq:ActionMHDranders} has similarities with  the Randers metric \cite{Randers1941,Basilakos2013}.

As pointed out by Chern
\cite{ShiingShenChern1996}, Finsler geometry is simply  Riemann geometry without
the quadratic restriction \eqref{eq:lineelementRiemann}. In Finsler geometry,  the line-element is replaced by the  general expression
\be  \label{eq:lineelementFinsler}
d\mathcal{S}=L(x,dx).
\ee
where $L: \mathbb{R}^2 \rightarrow \mathbb{R}$ is an arbitrary function that can be identified with the Lagrangian. 
Then, the distance between two points  is given by
\be \label{eq:lengthFinsler}
\mathcal{S}=\int_1^2 L(x,dx)=\int_{\tau_1}^{\tau_2} L(x,\dot x)d\tau
\ee
where the last equality holds iff  the function $L(x,\dot x)$ is  homogeneous of degree 1 in the velocity $\dot x^\alpha=dx^\alpha/d\tau$:
\be  \label{eq:Lagrangianhom1}
L(x,\kappa \, \dot x)=\kappa L(x,\dot x) \quad \forall \kappa > 0.
\ee
Lagrangians with this homogeneity property give rise to a parametrization-independent action functional, and lay at the foundation of  Finsler geometry.

The Lagrangian in the perfect magnetofluid  action functional  \eqref{eq:ActionMHDranders} satisfies the above homogeneity property and 
can thus be expressed in the form of arc length in a Finsler space. To show this explicitly, we proceed as follows. Following Chern \cite{ShiingShenChern1996}, we consider the projectivized tangent bundle $\mathcal{PTM}$ (i.e. the bundle of line elements) of the manifold $\mathcal{M}$.
All geometric quantities constructed from the Lagrangian $L$ are homogeneous of degree zero in $\dot{{x}}^\alpha$ and thus naturally live on $\mathcal{PTM}$, although $L$ itself does not. Let $\{x^\mu\}$  be local coordinates on   $\mathcal{M}$. Express tangent vectors as $\dot x^\mu \partial_\mu$  so that $\{x^\mu,\dot x^\mu \}$ can be used as local coordinates of $\mathcal{TM}$ and, with $\dot x^\mu$ homogeneous, as local coordinates on $\mathcal{PTM}$. 
Euler's theorem of homogeneous functions (c.f. Appendix~\ref{sec:EulerFinsler})
can be used  
to show that 
\be \label{eq:LagrangianEulerEqualities}
L(x,\dot x)=\underbrace{\frac{{\partial L}}{{\partial {\dot x ^\alpha }}}}_{p_\alpha} \dot x^\alpha=- \Big( \underbrace{\frac{1}{2}\frac{{{\partial ^2}{L^2}}}{{\partial {\dot x^\alpha }\partial {\dot x^\beta }}}}_{-\gamma_{\alpha\beta}} \dot x^\alpha \dot x^\beta \Big)^{1/2} . 
\ee
The Hessian
\be \label{eq:Finslermetricdef}
{\gamma_{\alpha \beta }(x,\dot x)} :=- \frac{1}{2}\frac{{{\partial ^2}{L^2}}}{{\partial {\dot x^\alpha }\partial {\dot x^\beta }}}
\ee
plays the role of a metric on $\mathcal{PTM}$. This is a metric in a \emph{Finsler} (rather than \textit{Riemann}) space, as it depends on velocity in addition to position.  
A Finslerian metric is  homogeneous of degree zero in the velocity:
\be  \label{eq:Finsmetrichom0}
\gamma_{\alpha \beta }(x,\kappa\dot x)=\gamma_{\alpha \beta }(x,\dot x) \quad \forall \kappa > 0,
\ee 
as implied by Eqs.~\eqref{eq:Lagrangianhom1} and \eqref{eq:Finslermetricdef}.
That is, the Finslerian metric $\gamma_{\alpha \beta }(x,\dot x)$ depends on the direction, but not   magnitude, of the velocity  $\dot x^\alpha$. The line element \eqref{eq:lineelementFinsler} can then be written as
\be \label{eq:lineelementFinsler2}
d\mathcal{S}^2={-\gamma_{\alpha \beta }(x,\dot x)dx^\alpha
dx^\beta}.
\ee
and the functional \eqref{eq:lengthFinsler}
becomes
\bea \label{eq:ActionMHDFinsler0}
\mathcal{S}&=&-\int_1^2 \sqrt{-\gamma_{\alpha \beta }(x,\dot x)dx^\alpha dx^\beta}\nonumber\\
&=&-\int_{\tau_1}^{\tau_2} \sqrt{-\gamma_{\alpha \beta }(x,\dot x) \dot x^\alpha \dot x^\beta}
d \tau 
\eea

For our particular application, substituting the ideal MHD Lagrangian \eqref{eq:Lagrangianhguuhetau0} into the definition \eqref{eq:Finslermetricdef} yields
\bea \label{eq:metricFinslerHydromagnetic}
{\gamma _{\alpha \beta }}(x,\dot x)
 &=&
    {h^2}{g_{\alpha \beta }} - {\eta _\alpha }{\eta _\beta } - h({\eta _\alpha }{u_\beta } + {\eta _\beta }{u_\alpha }) 
\nonumber\\& &
- h q_{\alpha \beta}\,{\eta _\gamma }{u^\gamma} ,
\eea
where $u^\alpha =\dot x^\alpha 
{( - {g_{\beta \gamma}}{\dot x^\beta}{\dot x^\gamma })}^{-1/2}$
 is the unit vector along $\dot x^\alpha$,
 $q_{\alpha \beta}={g_{\alpha \beta }}
+ {u_\alpha }{u_\beta } $ is the projection tensor orthogonal to that vector, and $g_{\alpha \beta }$ is the Riemannian metric in the spacetime $\M$.
As required by the homogeneity condition \eqref{eq:Finsmetrichom0}, the expression \eqref{eq:metricFinslerHydromagnetic} gives a metric  that 
 depends  on the direction, 
 but not the magnitude, of the velocity. Eq. \eqref{eq:metricFinslerHydromagnetic}
may be compactly written as ${\gamma _{\alpha \beta }} =- {p_\alpha }{p_\beta } - h 
q_{\alpha \beta}\,{p_\gamma}{u^\gamma}$
 where $p_\alpha=h{u_\alpha}+\eta_\alpha$.  On shell, we have 
${u^\alpha }={\dot x^\alpha}$ and, by virtue of Eqs. \eqref{e:perfect_cond} and \eqref{eq:defeta},  ${\eta _\alpha}{u^\alpha}=0$, i.e. the last term in Eq.~\eqref{eq:metricFinslerHydromagnetic} vanishes.
The Finsler metric $\gamma _{\alpha \beta}$  plays the role of an effective metric felt by a magnetofluid element.

With the aid of Eqs.~\eqref{eq:LagrangianEulerEqualities} and \eqref{eq:metricFinslerHydromagnetic}, the action functional \eqref{eq:ActionMHDranders}
takes the form of the length \eqref{eq:ActionMHDFinsler0}.
 This functional is independent of $\tau$ and represents the arc length between events 1 and 2. 
That is,    \textit{the motions of  fluid elements
 in a barotropic, perfectly conducting flow are geodesics 
in  a Finsler space} with  metric given by Eq.  \eqref{eq:metricFinslerHydromagnetic}. 

 The  geodesic equation is obtained by minimizing the  functional  \eqref{eq:ActionMHDFinsler0}
and using 
Eqs.~\eqref{eq:udotpartialgamma}--\eqref{eq:CabcdeqspCabcdpud}. This yields 
\be
\frac{{{d^2}{x^\lambda }}}{{d{\tau ^2}}} + \Gamma _{\mu \nu }^\lambda \frac{{d{x^\mu }}}{{d\tau }}\frac{{d{x^\nu }}}{{d\tau }} = 0 ,
\ee
where
\be
\Gamma _{\mu \nu }^\lambda:  = \frac{1}{2}{\gamma^{\lambda \kappa }}\left( {\frac{{\partial {\gamma_{\kappa \mu }}}}{{\partial {x^\nu }}} + \frac{{\partial {\gamma_{\kappa \nu }}}}{{\partial {x^\mu }}} - \frac{{\partial {\gamma_{\mu \nu }}}}{{\partial {x^\kappa }}}} \right)
\ee
denote the  \emph{Finslerian Christoffel symbols} \cite{Asanov1989}. Although the above equations are identical to those of Riemannian geometry, the transformation 
law of the symbols $\Gamma^\lambda_{\mu \nu}$ is more complicated since it involves  the \emph{Cartan torsion tensor}:
\be
C_{\alpha \beta \gamma}:=\frac{1}{2}\frac{\partial \gamma_{\alpha \beta}}{\partial {\dot x}^\gamma}
=\frac{3h}{{( - {g_{\epsilon \zeta}}{{\dot x}^\epsilon }{{\dot x}^\zeta })}^{1/2}} q_{(\alpha \beta} q_{\gamma) \delta} 
\eta^\delta .
\ee

By extending the notion of a metric in $\mathcal{M}$ to allow for Finsler geometry, the problem of
 ideal MHD becomes one of pure geometry. We note that the  geometry of the spacetime $\mathscr{M}$ remains Riemannian:
no deviation from  general relativity has been assumed. In the limit $\eta_\alpha \rightarrow 0$, the Cartan torsion tensor vanishes, the geometry of $\mathcal{M}$  also becomes Riemannian, and we recover the Synge-Lichnerowicz result on barotropic fluids.

We note that a   similar approach may be used   for \textit{poorly conducting fluids}, by replacing $\eta_\alpha$ with $e A_\alpha$ in the equations above, where $e$ is the net charge per fluid element \cite{Carte79}. Furthermore, with the replacements  $h \rightarrow 1$, $\eta_\alpha \rightarrow e A_\alpha$, we recover the motion of a charged particle under the influence of an electromagnetic field in curved spacetime \cite{Randers1941,Asanov1977,Gibbons2011a,Gibbons2009,Gibbons2010}.
We note, however, that for \textit{baroclinic fluids}, the action is not parametrization invariant, and thus cannot be described within Riemann or Finsler geometry. 

\section{Discussion } \label{s:concl}
We have illustrated that barotropic flows and  magnetoflows without viscosity, resistivity or other dissipation  can be described via simple variational principles.
These action principles can be written in terms of a Lagrangian density integrated over spacetime, as done traditionally for fluids, or in terms of a particle-like Lagrangian integrated over a proper-time or affine parameter. The latter approach paves the way for deriving  simple 
\textit{Lagrangian and Hamiltonian descriptions of 
ideal MHD}, in Newtonian and relativistic contexts.  These descriptions   are as valuable for fluids as they have been for classical mechanics and carry  the  same advantages over approaches focused on the equation-of-motion level. 

For instance, 
 certain  conserved quantities -- whose origin  seems \textit{ad hoc}  when  obtained by tedious algebraic manipulation of the equations of motion -- emerge directly from the action   in this geometric canonical approach. In particular, when the ideal MHD  Lagrangians \eqref{eq:LuA} and \eqref{eq:Lagrangianhguuhetau}
admit  continuous symmetries, Noether's theorem immediately yields the associated quantity conserved along streamlines  \cite{MarkakisRBNSMF2011}. 
As shown by Carter and Lichnerowicz, the relativistic hydrodynamics and magnetohydrodynamics are most naturally expressed in the language of differential forms. Cartan's identity can then be used to simplify calculations tremendously compared to the usual tensor or vector calculus, as demonstrated above. This  approaches to MHD is not yet very  widely
known, but this has been changing in recent years, and it is being used to obtain new results  \cite{Brennan2013,Gralla2014,Brennan2014,Gralla2015,Gralla2015a,Freytsis2016,Gralla2016,Gralla2016a,Gralla2016b,Gralla2016c,Compere2016}. For stationary  and irrotational or corotating magnetoflows,   Cartan's identity implies that these quantities, given by Eqs.~\eqref{eq:Adotkeq0} and \eqref{eq:pidotkplusfconst}, are constant throughout the fluid. These equations  represent   relativistic, magnetized generalizations of Bernoulli's principle and provide a way to construct equilibrium solutions via iterative methods \cite{Gourg10,PriceMarkakisFriedman2009}.
Such results can be extended to the case of generalized Noether symmetries generated by Killing tensors (cf. \cite{MarkakisRBNSMF2011} for details) and applied to the theory of black hole accretion rings \cite{Carte79,Contopoulos2015}.

Several theoretical insights arise from this formulation. The symplectic geometry of phase space gives rise to various circulation theorems  that stem from the Poincar\'e-Cartan integral invariant. 
The symplectic  structure of the perfect MHD equations can be exploited in
 numerical simulations that use smoothed-particle hydrodynamic (SPH) methods
\cite{Rosswog2015}. For instance, symplectic or time-symetric methods can
be used to conserve phase-space volume, circulation, and energy. 

Geometric considerations have led to deeper understanding of magnetic
phenomena in fluids in curved soacetime. Exploring the similarities
of geodesic motion to hydrodynamic and magnetohydrodynamic motion, Lasota
et al. \cite{Lasota2014}   generalized
the Penrose process \cite{PENROSE1971} from point particles to fluid particles
and jets. Moreover, the Finsler geometry described by the metric 
\eqref{eq:metricFinslerHydromagnetic}  allows one to represent ideal MHD flows as purely geodesic flows with no loss of generality. A notable feature of both pictures is that they are exact in time-dependent spacetimes, with gravitational and electromagnetic waves carrying energy and angular momentum away from the system. Although such geometrical insights 
have been sometimes used to construct
first integrals for non-magnetized initial data  \cite{Gourg10}, 
they have  not so far been  used   for magnetized initial data or for evolving hydrodynamic and magnetohydrodynamic flows in numerical general relativity. The integrals  \eqref{eq:Adotkeq0}, \eqref{eq:pidotkplusfconst} and the evolution system  \eqref{eq:circulationpreservingsystem}
provide avenues for exploiting such geometric properties in the future.

\acknowledgments

We thank Brandon Carter for pointing out the second line of 
Eq.~\eqref{eq:dCdtPoisson}. We thank Theocharis Apostolatos, Jacob Bekenstein, Brandon Carter, Greg Comer, John Friedman,  Roland Haas, David Hilditch and David Kaplan
   for  very fruitful discussions. This work was supported by 
JSPS Grant-in-Aid for Scientific Research(C) 20540275, 
MEXT Grant-in-Aid for Scientific Research
on Innovative Area 20105004, 
the  Greek State Scholarships Foundation (IKY), NSF Grant PHY1001515, DFG grant SFB/Transregio 7
``Gravitational Wave Astronomy', STFC grant PP/E001025/1
and ANR grant 06-2-134423 \emph{M\'ethodes
math\'ematiques pour la relativit\'e g\'en\'erale}.  
KU and EG acknowledge  support from a JSPS Invitation Fellowship 
for Research in Japan (Short-term) and the invitation program 
of foreign researchers at the Paris observatory.
CM and JPN thank the Paris Observatory for hospitality during the course of this work.%


\appendix
\section{Finsler geometry and Euler's theorem}  \label{sec:EulerFinsler}
The homogeneity property \eqref{eq:Lagrangianhom1}
plays a fundamental role in Finsler geometry. This property  gives rise to many important relations by means of  \textit{Euler's homogeneous function theorem}: Consider a function $Z(x,v)$ that is positively homogeneous of degree $r$ with respect to $v^\alpha$, that is,
\be
Z(x,\kappa \,v) = {\kappa ^r}Z(x,v) \quad \forall \kappa > 0.
\ee
Differentiating with respect to $\kappa$ and setting $\kappa=1$ 
yields
\be \label{eq:EulersTheoremHomFun}
{v^\alpha }\frac{{\partial Z(x,v)}}{{\partial {v^\alpha}}} = rZ(x,v) .
\ee
This is the mathematical statement of Euler's theorem. 
Applying the above theorem to the case of the Lagrangian  \eqref{eq:Lagrangianhom1}
yields
\be  \label{eq:EulerthmLagrang}
\dot x^\alpha\frac{{\partial L(x,\dot x)}}{{\partial {{\dot x}^\alpha }}} = L(x,\dot x) .
\ee
Differentiating this expression with respect to $\dot x^\alpha$ yields
\be \label{eq:EulerthmLagrangOrthogonal}
{\dot x^\alpha}\frac{{{\partial ^2}L(x,\dot x)}}{{\partial {{\dot x}^\alpha }\partial {{\dot x}^\beta }}} = 0 .
\ee  
Then, differentiating the relation
\be
\frac{1}{2}\frac{{\partial {L^2}(x,\dot x)}}{{\partial {{\dot x}^\alpha }}} =L(x, \dot x)\frac{{\partial L(x, \dot x)}}{{\partial {{ \dot x}^\alpha }}}
\ee
with respect to $\dot x^\beta$, contracting with $\dot x^\alpha \dot x^\beta$ and using 
Eqs. \eqref{eq:EulerthmLagrang} and \eqref{eq:EulerthmLagrangOrthogonal} yields
\be \label{eq:EulerthmLagrangSquared}
{L^2}(x,\dot x) = \underbrace{\frac{1}{2}\frac{{{\partial ^2}{L^2}(x,\dot x)}}{{\partial {\dot x^\alpha }\partial {\dot x^\beta }}}}_{-\gamma_{\alpha\beta}} {\dot x^\alpha}{\dot x^\beta} .
\ee
Equations~\eqref{eq:EulerthmLagrang} and \eqref{eq:EulerthmLagrangSquared} 
reproduce \eqref{eq:LagrangianEulerEqualities}. 
From Eqs.  \eqref{eq:Lagrangianhom1} and \eqref{eq:EulerthmLagrangSquared} we infer that the  metric $\gamma_{\alpha \beta}(x,\dot x)$ is homogeneous of degree zero in the velocity, Eq. \eqref{eq:Finsmetrichom0}. Then, applying Euler's theorem \eqref{eq:EulersTheoremHomFun} for $\gamma_{\alpha \beta}$ with $r=0$ yields
\be \label{eq:udotpartialgamma}
\dot x^\gamma C_{\alpha \beta \gamma}=0 ,
\ee
where
\be \label{eq:CartanTorsionTensorDef}
C_{\alpha \beta \gamma}:=\frac{1}{2}\frac{\partial \gamma_{\alpha \beta}}{\partial\dot x^\gamma}=\frac{1}{4}\frac{{{\partial ^3}{L^2}}}{{\partial {\dot x^\alpha }\partial {\dot x^\beta\partial\dot x^\gamma }}}
\ee
is the \emph{Cartan torsion tensor}. The last equality, which follows from  Eq. \eqref{eq:Finslermetricdef}, implies that the above tensor is fully symmetric.
From the above definition we infer that $C_{\alpha \beta \gamma}$
is homogeneous of degree $r=-1$ in the velocity. Then,  Euler's theorem 
\eqref{eq:EulersTheoremHomFun} yields
\be
{\dot x^\delta }C_{\alpha \beta \gamma \delta} =  -C_{\alpha \beta \gamma} ,
\ee
where
\be \label{eq:CabcdeqspCabcdpud}
C_{\alpha \beta \gamma \delta} (x,\dot x)=\frac{{\partial C_{\alpha \beta \gamma}(x,\dot x)}}{{\partial {{\dot x}^\delta }}} .
\ee
The geodesic equation in Finsler space can be obtained with the same variational methods as in a Riemann space, with additional use of Eqs.~\eqref{eq:udotpartialgamma}-\eqref{eq:CabcdeqspCabcdpud}.
Finsler geometry reduces to Riemann geometry iff the Cartan torsion tensor and its derivatives vanish, whence the  metric $\gamma_{\alpha \beta}$ is independent of velocity \cite{Asanov1989}.

\section{Beckenstein-Oron current with one symmetry} \label{sec:AppendixB}

Assuming that the system obeys a Killing symmetry, i.e. that there exists a vector field $\vw k$ such that
\begin{gather}
\Lie{\vw{k}} {\w{g}} = 0, \, ~ \Lie{\vw{k}} {\w{u}} =0, \, ~ \Lie{\vw{k}} {\vv j} =0, \,  \label{eq:Symm1} \\
\Lie{\vw k} {\w{F}} =0, \,  ~ \Lie{\vw k} {h} =0, \, ~ \Lie{\vw k} {\rho} =0, \,  \label{eq:Symm2}
\end{gather}
a natural question is whether or not one can impose the same symmetry on the auxiliary quantities $\w{a}$ and $\w{b}$. 
First, note that
\begin{eqnarray}
\Lie{\vw k}  {\w{b}} &=&   \Lie{\vw k} ({\vw{u}} \cdot \star\dd {\w{a}} ) \\
&=& {\vw{u}} \cdot \Lie{\vw k} (\star\dd {\w{a}} ) \\
&=&  {\vw{u}} \cdot \star\Lie{\vw k} (\dd {\w{a}} )
 \mbox{ (since $\vw k$ is Killing)}\\
&=&  {\vw{u}} \cdot \star \dd (\Lie{\vw k}  {\w{a}} )
\mbox{ (since $\dd$ and $\Lie{\vw k}$ commute) \,\,  }
\end{eqnarray}
%
%
%
In addition, using equation \eqref{eq:dadecomposition}, as well as the symmetries \eqref{eq:Symm1}, \eqref{eq:Symm2},
\begin{equation}
\Lie{\vw k} \dd \vv{a}= \Lie{\vw k} [\w{u}\wedge \w{B} + \star(\w{u}\wedge \w{b} )] = \star(\w{u}\wedge \Lie{\vw k} \w{b} ) \, .
\end{equation}
We have therefore that
\begin{equation}
\Lie{\vw k}  {\w{b}} = 0 \Leftrightarrow \Lie{\vw k} \dd \vv{a}=0 \Leftrightarrow \Lie{\vw k} \vv{a} \mbox{ closed}
\end{equation}
and of course $\Lie{\vw k}  {\w{a}} =0$ implies $\Lie{\vw k}  {\w{b}} =0$. So in effect, assuming that the auxiliary quantities $\w{a}$ and $\w{b}$ satisfy the same symmetry as the physical quantities is equivalent to assuming merely $\Lie{\vw k}  {\w{a}} =0$. If on the other hand we are ready to sacrifice $\Lie{\vw k}  {\w{a}} =0$ and to assume only that $\Lie{\vw k}  {\w{b}} =0$, we must still impose that $\Lie{\vw k}  {\w{a}}$ is closed.

We first notice that $\Lie{\vw k}  {\w{a}} =0$ is not systematically compatible with the gauge condition $\vw{u} \cdot \w{a} =0$. Indeed, let us consider the case where $\vw{u}$ and $\vw{k}$ are parallel, i.e.
\begin{equation}
\vw{u} = f \vw{k} \, .
\end{equation}
The question is whether we can impose consistently the three equations
\begin{align}
\vw{u} \cdot \w{a} &=0 \, , \label{eq:Syst1} \\
\Lie{\vw u}  {\w{a}} &=-\w{B} \, ,\\
\Lie{\vw k}  {\w{a}} &=0 \label{eq:Syst2} \, .
\end{align}
Using the Cartan identity, we have
\begin{equation}
-\w{B} = \Lie{\vw u}  {\w{a}} = f \Lie{\vw k}  {\w{a}} + (\vw{k}\cdot\w{a} ) \dd f =  (\vw{k}\cdot\w{a} ) \dd f = 0
\end{equation}
since $\vw{u} \cdot \w{a} =0$ implies $\vw{k} \cdot \w{a} =0$. This is in general inconsistent.

Giving up the gauge condition $\vw{u} \cdot \w{a} =0$ does not improve things. Let us put $\phi = \vw{u} \cdot \w{a}$ and still assume that $\vw{u}$ and $\vw{k}$ are colinear. Now we have $\Lie{\vw u}  {\w{a}} = \vw{u} \cdot \dd \w{a} + \dd (\vw{u}\cdot \w{a} )$ and instead of \eqref{eq:Syst1}-\eqref{eq:Syst2} we must consider
\begin{align}
\vw{u} \cdot \w{a} &= \phi \, , \\
\Lie{\vw u}  {\w{a}} &=-\w{B} + \dd \phi \, ,\\
\Lie{\vw k}  {\w{a}} &=0  \, .
\end{align}
Then
\begin{equation}
\Lie{\vw u}  {\w{a}} = \phi \, \dd (\mathrm{log} f) = -\w{B} + \dd \phi \, ,
\end{equation}
i.e.
\begin{equation}
\w{B} = \phi  \,  \dd \left( \mathrm{log} \left\vert \frac{\phi}{f} \right\vert \right) \, .
\end{equation}
This forces the magnetic field $\w{B}$ to be exact modulo multiplication by a scalar function, which is not a generic property. Indeed consider the 1-form on $\R^4$
\begin{equation}
\alpha = -y \dd x + x\dd y
\end{equation}
whose divergence vanishes. Can we find a globally defined smooth function $\psi$ such that $\psi \alpha$ be closed? This amounts to
\begin{equation}
2 \psi + x \partial_x \psi + y \partial_y \psi =0 \, ,
\end{equation}
which imposes that $\psi$ be homogeneous of degree $-2$ and contradicts the fact that $\psi$ be globally defined and smooth.

We conclude that we cannot in all generality assume that the auxiliary fields $\w{a}$ and $\w{b}$ satisfy the same symmetry as the physical quantities.

\section{Fluid super-Hamiltonians} \label{sec:AppendixC}
The canonical form of the Euler equation
\eqref{e:Relativistic-Euler} 
involves only the thermodynamic variables $T, S, h$. We thus assert that the super-Hamiltonian for this equation has the general
 form
\begin{equation}
H= H(h,S,T,N)\,,
\end{equation} 
where $N := g^{\alpha \beta} p_\alpha p_\beta$ is the norm of the (generally non-normalized) canonical momenta $p_\alpha$, whose nature
is to be determined. 
Furthermore, we assume that the Hamiltonian  generates a reparametrization
with respect to the proper time of the fluid which we  denote by a parameter
$\mathrm{d}\lambda = \mathrm{d} \tau/\mathcal{A}$, where $\mathcal{A}$ is
some function of the variables involved. 

Computing Hamilton's equations and comparing them with the Euler equation
we deduce that we are `on-shell' only if $p_\alpha = h  u_\alpha$
and thus $N=-h^2$. Additionally, the following equalities 
must be satisfied by the Hamiltonian on-shell in order to reproduce the Euler equation:
\bea
\frac{\partial H}{\partial T}  & =& 0 \label{eq:HT}\\ 
\frac{\partial H}{\partial h} &=& 2h\frac{\partial H}{\partial N}  \label{eq:Hh}\\
\frac{\partial H}{\partial S} &=& -T \frac{\partial H}{\partial h} -2 Th \frac{\partial H}{\partial N}    \label{eq:Hs}\\
 \mathcal{A} &=& 2h\frac{\partial H}{\partial N}  \, \label{eq:Acond}
\eea 
One way  to satisfy this set of constraints on the form of the Hamiltonian
is via the expression
\begin{equation}
H= \frac{\mathcal{C}'(S)}{2T h} ( g^{\alpha \beta} p_\alpha p_\beta
+ h^2) - \mathcal{C}(S), \label{eq:hamclinic}
\end{equation}
where $\mathcal{C}(S)$ is an arbitrary function of the specific entropy with $\mathcal{C}'(S)\neq
0$ for $S \ge 0$. The on-shell value of the conserved super-Hamiltonian is then $-\mathcal{C}(S)$ and the  canonical time parameter
$\lambda$ satisfies $\mathrm{d} \lambda = T  \mathrm{d} \tau/\mathcal{C}'$. Carter's
baroclinic Hamiltonian \eqref{eq:HamiltonianBaroclinic}
 is  obtained simply by setting 
$\mathcal{C}(S)=S$.

For barotropic fluids one can use a similar approach to obtain a set of Hamiltonians of the form 
\begin{equation}
H = -\frac{\mathcal{D}(h)}{2h}(g^{\alpha \beta} p_\alpha p_\beta
+ h^2 ) \label{eq:hamtropic}
\end{equation}
where  $\mathcal{D}(h)$ is an arbitrary function
of $h$, and the parametrization corresponding to this Hamiltonian is $\mathrm{d}\lambda
= \mathrm{d}\tau /\mathcal{D}$. The transition between the Hamiltonians
(\ref{eq:hamclinic}) and (\ref{eq:hamtropic}) for baroclinic and barotropic
fluids depends on the form of temperature expressed as a function of entropy
and enthalpy $T=T(h,S)$. 

For baroclinic magnetofluids, we see from Eq.~\eqref{eq:MHD-Euler_form1Barotropic}
 that the streamlines of a perfectly
conducting fluid behave as if under the influence of a vector potential
$\boldsymbol{\eta}$. We thus assume that there is a canonical momentum $p_\alpha$
such that the Hamiltonian depends only on the normalization ${N}= g^{\alpha
\beta}({p}_\alpha - \kappa_\alpha)( {p}_\beta - \kappa_\beta)$
with $\kappa_\alpha$ some vector. In that case, we obtain the on-shell values
${p}_\alpha = h u_\alpha + \eta_\alpha,\, \kappa_\alpha= \eta_\alpha
$ ${N} = -h^2$ and the same set of constraints as in \eqref{eq:HT}-\eqref{eq:Acond}.
This means that one class of super-Hamiltonians
which reproduce the ideal MHD-Euler equation 
\eqref{eq:MHD-Euler_form1Barotropic}
is 
\begin{equation}
H= \frac{\mathcal{C}'(S)}{2T h} \left[ g^{\alpha \beta}({p}_\alpha
- \eta_\alpha)({p}_\beta - \eta_\beta) + h^2 \right] - \mathcal{C}(S),
\label{eq:hamclinicEM}
\end{equation}
where the on-shell value of the super-Hamiltonian is again $-\mathcal{C}(S)$  and the  canonical time parameter $\lambda$ satisfies
$\mathrm{d} \lambda = T  \mathrm{d} \tau/\mathcal{C}'$.

\bibliographystyle{apsrev4-1}

\bibliography{library}

\begin{thebibliography}{129}%
\makeatletter
\providecommand \@ifxundefined [1]{%
 \@ifx{#1\undefined}
}%
\providecommand \@ifnum [1]{%
 \ifnum #1\expandafter \@firstoftwo
 \else \expandafter \@secondoftwo
 \fi
}%
\providecommand \@ifx [1]{%
 \ifx #1\expandafter \@firstoftwo
 \else \expandafter \@secondoftwo
 \fi
}%
\providecommand \natexlab [1]{#1}%
\providecommand \enquote  [1]{``#1''}%
\providecommand \bibnamefont  [1]{#1}%
\providecommand \bibfnamefont [1]{#1}%
\providecommand \citenamefont [1]{#1}%
\providecommand \href@noop [0]{\@secondoftwo}%
\providecommand \href [0]{\begingroup \@sanitize@url \@href}%
\providecommand \@href[1]{\@@startlink{#1}\@@href}%
\providecommand \@@href[1]{\endgroup#1\@@endlink}%
\providecommand \@sanitize@url [0]{\catcode `\\12\catcode `\$12\catcode
  `\&12\catcode `\#12\catcode `\^12\catcode `\_12\catcode `\%12\relax}%
\providecommand \@@startlink[1]{}%
\providecommand \@@endlink[0]{}%
\providecommand \url  [0]{\begingroup\@sanitize@url \@url }%
\providecommand \@url [1]{\endgroup\@href {#1}{\urlprefix }}%
\providecommand \urlprefix  [0]{URL }%
\providecommand \Eprint [0]{\href }%
\providecommand \doibase [0]{http://dx.doi.org/}%
\providecommand \selectlanguage [0]{\@gobble}%
\providecommand \bibinfo  [0]{\@secondoftwo}%
\providecommand \bibfield  [0]{\@secondoftwo}%
\providecommand \translation [1]{[#1]}%
\providecommand \BibitemOpen [0]{}%
\providecommand \bibitemStop [0]{}%
\providecommand \bibitemNoStop [0]{.\EOS\space}%
\providecommand \EOS [0]{\spacefactor3000\relax}%
\providecommand \BibitemShut  [1]{\csname bibitem#1\endcsname}%
\let\auto@bib@innerbib\@empty
\bibitem [{\citenamefont {Mo{\'{s}}cibrodzka}\ \emph
  {et~al.}(2009)\citenamefont {Mo{\'{s}}cibrodzka}, \citenamefont {Gammie},
  \citenamefont {Dolence}, \citenamefont {Shiokawa},\ and\ \citenamefont
  {Leung}}]{MosciGDSL09}%
  \BibitemOpen
  \bibfield  {author} {\bibinfo {author} {\bibfnamefont {M.}~\bibnamefont
  {Mo{\'{s}}cibrodzka}}, \bibinfo {author} {\bibfnamefont {C.~F.}\ \bibnamefont
  {Gammie}}, \bibinfo {author} {\bibfnamefont {J.~C.}\ \bibnamefont {Dolence}},
  \bibinfo {author} {\bibfnamefont {H.}~\bibnamefont {Shiokawa}}, \ and\
  \bibinfo {author} {\bibfnamefont {P.~K.}\ \bibnamefont {Leung}},\ }\href
  {\doibase 10.1088/0004-637X/706/1/497} {\bibfield  {journal} {\bibinfo
  {journal} {Astrophys. J.}\ }\textbf {\bibinfo {volume} {706}},\ \bibinfo
  {pages} {497} (\bibinfo {year} {2009})}\BibitemShut {NoStop}%
\bibitem [{\citenamefont {Meliani}\ \emph {et~al.}(2010)\citenamefont
  {Meliani}, \citenamefont {Sauty}, \citenamefont {Tsinganos}, \citenamefont
  {Trussoni},\ and\ \citenamefont {Cayatte}}]{MeliaSTTC10}%
  \BibitemOpen
  \bibfield  {author} {\bibinfo {author} {\bibfnamefont {Z.}~\bibnamefont
  {Meliani}}, \bibinfo {author} {\bibfnamefont {C.}~\bibnamefont {Sauty}},
  \bibinfo {author} {\bibfnamefont {K.}~\bibnamefont {Tsinganos}}, \bibinfo
  {author} {\bibfnamefont {E.}~\bibnamefont {Trussoni}}, \ and\ \bibinfo
  {author} {\bibfnamefont {V.}~\bibnamefont {Cayatte}},\ }\href {\doibase
  10.1051/0004-6361/200912920} {\bibfield  {journal} {\bibinfo  {journal}
  {Astron. Astrophys.}\ }\textbf {\bibinfo {volume} {521}},\ \bibinfo {pages}
  {A67} (\bibinfo {year} {2010})}\BibitemShut {NoStop}%
\bibitem [{\citenamefont {Komissarov}(2011)}]{Komis10}%
  \BibitemOpen
  \bibfield  {author} {\bibinfo {author} {\bibfnamefont {S.~S.}\ \bibnamefont
  {Komissarov}},\ }in\ \href {\doibase 10.1111/j.1365-2966.2007.12050.x} {\emph
  {\bibinfo {booktitle} {Mem SAIt}}},\ Vol.~\bibinfo {volume} {82}\ (\bibinfo
  {year} {2011})\ pp.\ \bibinfo {pages} {95--103},\ \Eprint
  {http://arxiv.org/abs/1006.2242} {arXiv:1006.2242} \BibitemShut {NoStop}%
\bibitem [{\citenamefont {Beskin}(2009)}]{Beskin2009}%
  \BibitemOpen
  \bibfield  {author} {\bibinfo {author} {\bibfnamefont {V.}~\bibnamefont
  {Beskin}},\ }\href {\doibase 10.1007/978-3-642-01290-7} {\emph {\bibinfo
  {title} {{MHD Flows in Compact Astrophysical Objects: Accretion, Winds and
  Jets}}}}\ (\bibinfo  {publisher} {Springer},\ \bibinfo {year}
  {2009})\BibitemShut {NoStop}%
\bibitem [{\citenamefont {Sotani}\ \emph
  {et~al.}(2008{\natexlab{a}})\citenamefont {Sotani}, \citenamefont
  {Colaiuda},\ and\ \citenamefont {Kokkotas}}]{Sotani2007}%
  \BibitemOpen
  \bibfield  {author} {\bibinfo {author} {\bibfnamefont {H.}~\bibnamefont
  {Sotani}}, \bibinfo {author} {\bibfnamefont {A.}~\bibnamefont {Colaiuda}}, \
  and\ \bibinfo {author} {\bibfnamefont {K.~D.}\ \bibnamefont {Kokkotas}},\
  }\href {\doibase 10.1111/j.1365-2966.2008.12977.x} {\bibfield  {journal}
  {\bibinfo  {journal} {Mon. Not. R. Astron. Soc.}\ }\textbf {\bibinfo {volume}
  {385}},\ \bibinfo {pages} {2161} (\bibinfo {year} {2008}{\natexlab{a}})},\
  \Eprint {http://arxiv.org/abs/0711.1518} {arXiv:0711.1518} \BibitemShut
  {NoStop}%
\bibitem [{\citenamefont {Sotani}\ \emph
  {et~al.}(2008{\natexlab{b}})\citenamefont {Sotani}, \citenamefont
  {Kokkotas},\ and\ \citenamefont {Stergioulas}}]{Sotani2008}%
  \BibitemOpen
  \bibfield  {author} {\bibinfo {author} {\bibfnamefont {H.}~\bibnamefont
  {Sotani}}, \bibinfo {author} {\bibfnamefont {K.~D.}\ \bibnamefont
  {Kokkotas}}, \ and\ \bibinfo {author} {\bibfnamefont {N.}~\bibnamefont
  {Stergioulas}},\ }\href {\doibase 10.1111/j.1745-3933.2007.00420.x}
  {\bibfield  {journal} {\bibinfo  {journal} {Mon. Not. R. Astron. Soc. Lett.}\
  }\textbf {\bibinfo {volume} {385}},\ \bibinfo {pages} {L5} (\bibinfo {year}
  {2008}{\natexlab{b}})}\BibitemShut {NoStop}%
\bibitem [{\citenamefont {Zink}\ \emph {et~al.}(2012)\citenamefont {Zink},
  \citenamefont {Lasky},\ and\ \citenamefont {Kokkotas}}]{Zink2012}%
  \BibitemOpen
  \bibfield  {author} {\bibinfo {author} {\bibfnamefont {B.}~\bibnamefont
  {Zink}}, \bibinfo {author} {\bibfnamefont {P.~D.}\ \bibnamefont {Lasky}}, \
  and\ \bibinfo {author} {\bibfnamefont {K.~D.}\ \bibnamefont {Kokkotas}},\
  }\href {\doibase 10.1103/PhysRevD.85.024030} {\bibfield  {journal} {\bibinfo
  {journal} {Phys. Rev. D}\ }\textbf {\bibinfo {volume} {85}},\ \bibinfo
  {pages} {024030} (\bibinfo {year} {2012})},\ \Eprint
  {http://arxiv.org/abs/1107.1689} {arXiv:1107.1689} \BibitemShut {NoStop}%
\bibitem [{\citenamefont {Freytsis}\ and\ \citenamefont
  {Gralla}(2016)}]{Freytsis2016}%
  \BibitemOpen
  \bibfield  {author} {\bibinfo {author} {\bibfnamefont {M.}~\bibnamefont
  {Freytsis}}\ and\ \bibinfo {author} {\bibfnamefont {S.~E.}\ \bibnamefont
  {Gralla}},\ }\href {\doibase 10.1088/1475-7516/2016/05/042} {\bibfield
  {journal} {\bibinfo  {journal} {J. Cosmol. Astropart. Phys.}\ }\textbf
  {\bibinfo {volume} {2016}},\ \bibinfo {pages} {042} (\bibinfo {year}
  {2016})}\BibitemShut {NoStop}%
\bibitem [{\citenamefont {Watts}\ \emph {et~al.}(2016)\citenamefont {Watts},
  \citenamefont {Andersson}, \citenamefont {Chakrabarty}, \citenamefont
  {Feroci}, \citenamefont {Hebeler}, \citenamefont {Israel}, \citenamefont
  {Lamb}, \citenamefont {Miller}, \citenamefont {Morsink}, \citenamefont
  {{\"{O}}zel}, \citenamefont {Patruno}, \citenamefont {Poutanen},
  \citenamefont {Psaltis}, \citenamefont {Schwenk}, \citenamefont {Steiner},
  \citenamefont {Stella}, \citenamefont {Tolos},\ and\ \citenamefont {van~der
  Klis}}]{Watts2016}%
  \BibitemOpen
  \bibfield  {author} {\bibinfo {author} {\bibfnamefont {A.~L.}\ \bibnamefont
  {Watts}}, \bibinfo {author} {\bibfnamefont {N.}~\bibnamefont {Andersson}},
  \bibinfo {author} {\bibfnamefont {D.}~\bibnamefont {Chakrabarty}}, \bibinfo
  {author} {\bibfnamefont {M.}~\bibnamefont {Feroci}}, \bibinfo {author}
  {\bibfnamefont {K.}~\bibnamefont {Hebeler}}, \bibinfo {author} {\bibfnamefont
  {G.}~\bibnamefont {Israel}}, \bibinfo {author} {\bibfnamefont {F.~K.}\
  \bibnamefont {Lamb}}, \bibinfo {author} {\bibfnamefont {M.~C.}\ \bibnamefont
  {Miller}}, \bibinfo {author} {\bibfnamefont {S.}~\bibnamefont {Morsink}},
  \bibinfo {author} {\bibfnamefont {F.}~\bibnamefont {{\"{O}}zel}}, \bibinfo
  {author} {\bibfnamefont {A.}~\bibnamefont {Patruno}}, \bibinfo {author}
  {\bibfnamefont {J.}~\bibnamefont {Poutanen}}, \bibinfo {author}
  {\bibfnamefont {D.}~\bibnamefont {Psaltis}}, \bibinfo {author} {\bibfnamefont
  {A.}~\bibnamefont {Schwenk}}, \bibinfo {author} {\bibfnamefont {A.~W.}\
  \bibnamefont {Steiner}}, \bibinfo {author} {\bibfnamefont {L.}~\bibnamefont
  {Stella}}, \bibinfo {author} {\bibfnamefont {L.}~\bibnamefont {Tolos}}, \
  and\ \bibinfo {author} {\bibfnamefont {M.}~\bibnamefont {van~der Klis}},\
  }\href {\doibase 10.1103/RevModPhys.88.021001} {\bibfield  {journal}
  {\bibinfo  {journal} {Rev. Mod. Phys.}\ }\textbf {\bibinfo {volume} {88}},\
  \bibinfo {pages} {021001} (\bibinfo {year} {2016})}\BibitemShut {NoStop}%
\bibitem [{\citenamefont {Lichnerowicz}(1967)}]{Lichnerowicz1967}%
  \BibitemOpen
  \bibfield  {author} {\bibinfo {author} {\bibfnamefont {A.}~\bibnamefont
  {Lichnerowicz}},\ }\href@noop {} {\emph {\bibinfo {title} {{Relativistic
  hydrodynamics and magnetohydrodynamics}}}}\ (\bibinfo  {publisher} {W. A.
  Benjamin},\ \bibinfo {address} {New York, NY},\ \bibinfo {year}
  {1967})\BibitemShut {NoStop}%
\bibitem [{Bes()}]{Beski10}%
  \BibitemOpen
  \href@noop {} {\enquote {\bibinfo {title} {{No Title}},}\ }\BibitemShut
  {NoStop}%
\bibitem [{\citenamefont {Font}(2008)}]{Font2008}%
  \BibitemOpen
  \bibfield  {author} {\bibinfo {author} {\bibfnamefont {J.~A.}\ \bibnamefont
  {Font}},\ }\href {http://relativity.livingreviews.org/Articles/lrr-2008-7/}
  {\bibfield  {journal} {\bibinfo  {journal} {Living Rev. Relativ.}\ }\textbf
  {\bibinfo {volume} {11}} (\bibinfo {year} {2008})}\BibitemShut {NoStop}%
\bibitem [{\citenamefont {Ant{\'{o}}n}\ \emph {et~al.}(2010)\citenamefont
  {Ant{\'{o}}n}, \citenamefont {Miralles}, \citenamefont {Mart{\'{i}}},
  \citenamefont {Ibanez}, \citenamefont {Aloy},\ and\ \citenamefont
  {Mimica}}]{Anton_al10}%
  \BibitemOpen
  \bibfield  {author} {\bibinfo {author} {\bibfnamefont {L.}~\bibnamefont
  {Ant{\'{o}}n}}, \bibinfo {author} {\bibfnamefont {J.}~\bibnamefont
  {Miralles}}, \bibinfo {author} {\bibfnamefont {J.}~\bibnamefont
  {Mart{\'{i}}}}, \bibinfo {author} {\bibfnamefont {J.}~\bibnamefont {Ibanez}},
  \bibinfo {author} {\bibfnamefont {M.}~\bibnamefont {Aloy}}, \ and\ \bibinfo
  {author} {\bibfnamefont {P.}~\bibnamefont {Mimica}},\ }\href {\doibase
  10.1088/0067-0049/188/1/1} {\bibfield  {journal} {\bibinfo  {journal}
  {Astrophys. J. Suppl.}\ }\textbf {\bibinfo {volume} {188}},\ \bibinfo {pages}
  {1} (\bibinfo {year} {2010})},\ \Eprint {http://arxiv.org/abs/0912.4692}
  {arXiv:0912.4692 [astro-ph.IM]} \BibitemShut {NoStop}%
\bibitem [{\citenamefont {Glampedakis}\ \emph {et~al.}(2012)\citenamefont
  {Glampedakis}, \citenamefont {Andersson},\ and\ \citenamefont
  {Lander}}]{Glampedakisetal2011}%
  \BibitemOpen
  \bibfield  {author} {\bibinfo {author} {\bibfnamefont {K.}~\bibnamefont
  {Glampedakis}}, \bibinfo {author} {\bibfnamefont {N.}~\bibnamefont
  {Andersson}}, \ and\ \bibinfo {author} {\bibfnamefont {S.~K.}\ \bibnamefont
  {Lander}},\ }\href {\doibase 10.1111/j.1365-2966.2011.20112.x} {\bibfield
  {journal} {\bibinfo  {journal} {Mon. Not. R. Astron. Soc.}\ }\textbf
  {\bibinfo {volume} {420}},\ \bibinfo {pages} {1263} (\bibinfo {year}
  {2012})}\BibitemShut {NoStop}%
\bibitem [{\citenamefont {Andersson}\ \emph
  {et~al.}(2016{\natexlab{a}})\citenamefont {Andersson}, \citenamefont {Hawke},
  \citenamefont {Dionysopoulou},\ and\ \citenamefont {Comer}}]{Andersson2016}%
  \BibitemOpen
  \bibfield  {author} {\bibinfo {author} {\bibfnamefont {N.}~\bibnamefont
  {Andersson}}, \bibinfo {author} {\bibfnamefont {I.}~\bibnamefont {Hawke}},
  \bibinfo {author} {\bibfnamefont {K.}~\bibnamefont {Dionysopoulou}}, \ and\
  \bibinfo {author} {\bibfnamefont {G.~L.}\ \bibnamefont {Comer}},\ }\href
  {http://arxiv.org/abs/1610.00448} {\  (\bibinfo {year}
  {2016}{\natexlab{a}})},\ \Eprint {http://arxiv.org/abs/1610.00448}
  {arXiv:1610.00448} \BibitemShut {NoStop}%
\bibitem [{\citenamefont {Andersson}\ \emph
  {et~al.}(2016{\natexlab{b}})\citenamefont {Andersson}, \citenamefont
  {Dionysopoulou}, \citenamefont {Hawke},\ and\ \citenamefont
  {Comer}}]{Andersson2016a}%
  \BibitemOpen
  \bibfield  {author} {\bibinfo {author} {\bibfnamefont {N.}~\bibnamefont
  {Andersson}}, \bibinfo {author} {\bibfnamefont {K.}~\bibnamefont
  {Dionysopoulou}}, \bibinfo {author} {\bibfnamefont {I.}~\bibnamefont
  {Hawke}}, \ and\ \bibinfo {author} {\bibfnamefont {G.~L.}\ \bibnamefont
  {Comer}},\ }\href {http://arxiv.org/abs/1610.00449} {\  (\bibinfo {year}
  {2016}{\natexlab{b}})},\ \Eprint {http://arxiv.org/abs/1610.00449}
  {arXiv:1610.00449} \BibitemShut {NoStop}%
\bibitem [{\citenamefont {Reisenegger}(2009)}]{Reisenegger2009}%
  \BibitemOpen
  \bibfield  {author} {\bibinfo {author} {\bibfnamefont {A.}~\bibnamefont
  {Reisenegger}},\ }\href@noop {} {\bibfield  {journal} {\bibinfo  {journal}
  {Volume}\ }\textbf {\bibinfo {volume} {499}} (\bibinfo {year}
  {2009})}\BibitemShut {NoStop}%
\bibitem [{\citenamefont {Synge}(2002)}]{Synge37}%
  \BibitemOpen
  \bibfield  {author} {\bibinfo {author} {\bibfnamefont {J.~L.}\ \bibnamefont
  {Synge}},\ }\href@noop {} {\bibfield  {journal} {\bibinfo  {journal} {London
  Math}\ }\textbf {\bibinfo {volume} {43}} (\bibinfo {year}
  {2002})}\BibitemShut {NoStop}%
\bibitem [{\citenamefont {Lichnerowicz}(1941)}]{Lichn41}%
  \BibitemOpen
  \bibfield  {author} {\bibinfo {author} {\bibfnamefont {A.}~\bibnamefont
  {Lichnerowicz}},\ }\href {www.numdam.org} {\bibfield  {journal} {\bibinfo
  {journal} {Ann. Sci. {\'{E}}cole Norm. Sup}\ }\textbf {\bibinfo {volume}
  {58}},\ \bibinfo {pages} {285} (\bibinfo {year} {1941})}\BibitemShut
  {NoStop}%
\bibitem [{\citenamefont {Carter}(1979)}]{Carte79}%
  \BibitemOpen
  \bibfield  {author} {\bibinfo {author} {\bibfnamefont {B.}~\bibnamefont
  {Carter}},\ }\href {http://adsabs.harvard.edu/abs/1979agn..book..273C}
  {\bibfield  {journal} {\bibinfo  {journal} {Act. Galact. Nucl.}\ }\textbf
  {\bibinfo {volume} {1}},\ \bibinfo {pages} {273} (\bibinfo {year}
  {1979})}\BibitemShut {NoStop}%
\bibitem [{\citenamefont {Bekenstein}\ and\ \citenamefont
  {Oron}(2000)}]{Bekenstein:2000sf}%
  \BibitemOpen
  \bibfield  {author} {\bibinfo {author} {\bibfnamefont {J.~D.}\ \bibnamefont
  {Bekenstein}}\ and\ \bibinfo {author} {\bibfnamefont {A.}~\bibnamefont
  {Oron}},\ }\href {\doibase 10.1103/PhysRevE.62.5594} {\bibfield  {journal}
  {\bibinfo  {journal} {Phys. Rev. E}\ }\textbf {\bibinfo {volume} {62}},\
  \bibinfo {pages} {5594} (\bibinfo {year} {2000})}\BibitemShut {NoStop}%
\bibitem [{\citenamefont {Bekenstein}\ and\ \citenamefont
  {Oron}(2001)}]{Bekenstein2001}%
  \BibitemOpen
  \bibfield  {author} {\bibinfo {author} {\bibfnamefont {J.}~\bibnamefont
  {Bekenstein}}\ and\ \bibinfo {author} {\bibfnamefont {A.}~\bibnamefont
  {Oron}},\ }\href {\doibase 10.1023/A:1017507917267} {\bibfield  {journal}
  {\bibinfo  {journal} {Found. Phys.}\ }\textbf {\bibinfo {volume} {31}},\
  \bibinfo {pages} {895} (\bibinfo {year} {2001})}\BibitemShut {NoStop}%
\bibitem [{\citenamefont {Bekenstein}\ and\ \citenamefont
  {Betschart}(2006)}]{Bekenstein2006}%
  \BibitemOpen
  \bibfield  {author} {\bibinfo {author} {\bibfnamefont {J.}~\bibnamefont
  {Bekenstein}}\ and\ \bibinfo {author} {\bibfnamefont {G.}~\bibnamefont
  {Betschart}},\ }\href {\doibase 10.1103/PhysRevD.74.083009} {\bibfield
  {journal} {\bibinfo  {journal} {Phys. Rev. D}\ }\textbf {\bibinfo {volume}
  {74}} (\bibinfo {year} {2006}),\ 10.1103/PhysRevD.74.083009}\BibitemShut
  {NoStop}%
\bibitem [{\citenamefont {Bekenstein}(1987)}]{Bekenstein1987}%
  \BibitemOpen
  \bibfield  {author} {\bibinfo {author} {\bibfnamefont {J.~D.}\ \bibnamefont
  {Bekenstein}},\ }\href {\doibase 10.1086/165447} {\bibfield  {journal}
  {\bibinfo  {journal} {Astrophys. J.}\ }\textbf {\bibinfo {volume} {319}},\
  \bibinfo {pages} {207} (\bibinfo {year} {1987})}\BibitemShut {NoStop}%
\bibitem [{\citenamefont {Markakis}\ \emph {et~al.}(2009)\citenamefont
  {Markakis}, \citenamefont {Read}, \citenamefont {Shibata}, \citenamefont
  {Uryu}, \citenamefont {Creighton}, \citenamefont {Friedman},\ and\
  \citenamefont {Lackey}}]{Markakis2009}%
  \BibitemOpen
  \bibfield  {author} {\bibinfo {author} {\bibfnamefont {C.}~\bibnamefont
  {Markakis}}, \bibinfo {author} {\bibfnamefont {J.~S.}\ \bibnamefont {Read}},
  \bibinfo {author} {\bibfnamefont {M.}~\bibnamefont {Shibata}}, \bibinfo
  {author} {\bibfnamefont {K.}~\bibnamefont {Uryu}}, \bibinfo {author}
  {\bibfnamefont {J.~D.~E.}\ \bibnamefont {Creighton}}, \bibinfo {author}
  {\bibfnamefont {J.~L.}\ \bibnamefont {Friedman}}, \ and\ \bibinfo {author}
  {\bibfnamefont {B.~D.}\ \bibnamefont {Lackey}},\ }in\ \href {\doibase
  10.1088/1742-6596/189/1/012024} {\emph {\bibinfo {booktitle} {J. Phys. Conf.
  Ser.}}},\ Vol.\ \bibinfo {volume} {189},\ \bibinfo {editor} {edited by\
  \bibinfo {editor} {\bibfnamefont {N.}~\bibnamefont {Stergioulas}}}\ (\bibinfo
   {publisher} {IOP Science},\ \bibinfo {address} {Philadelphia},\ \bibinfo
  {year} {2009})\ p.\ \bibinfo {pages} {012024}\BibitemShut {NoStop}%
\bibitem [{\citenamefont {Read}\ \emph {et~al.}(2009)\citenamefont {Read},
  \citenamefont {Markakis}, \citenamefont {Shibata}, \citenamefont {Uryu},
  \citenamefont {Creighton},\ and\ \citenamefont {Friedman}}]{Read2009}%
  \BibitemOpen
  \bibfield  {author} {\bibinfo {author} {\bibfnamefont {J.~S.}\ \bibnamefont
  {Read}}, \bibinfo {author} {\bibfnamefont {C.}~\bibnamefont {Markakis}},
  \bibinfo {author} {\bibfnamefont {M.}~\bibnamefont {Shibata}}, \bibinfo
  {author} {\bibfnamefont {K.}~\bibnamefont {Uryu}}, \bibinfo {author}
  {\bibfnamefont {J.~D.~E.}\ \bibnamefont {Creighton}}, \ and\ \bibinfo
  {author} {\bibfnamefont {J.~L.}\ \bibnamefont {Friedman}},\ }\href {\doibase
  10.1103/PhysRevD.79.124033} {\bibfield  {journal} {\bibinfo  {journal} {Phys.
  Rev. D}\ }\textbf {\bibinfo {volume} {79}},\ \bibinfo {pages} {12} (\bibinfo
  {year} {2009})},\ \Eprint {http://arxiv.org/abs/0901.3258} {arXiv:0901.3258}
  \BibitemShut {NoStop}%
\bibitem [{\citenamefont {Markakis}\ \emph {et~al.}(2011)\citenamefont
  {Markakis}, \citenamefont {Read}, \citenamefont {Shibata}, \citenamefont
  {Uryu}, \citenamefont {Creighton},\ and\ \citenamefont
  {Friedman}}]{Markakis2010}%
  \BibitemOpen
  \bibfield  {author} {\bibinfo {author} {\bibfnamefont {C.}~\bibnamefont
  {Markakis}}, \bibinfo {author} {\bibfnamefont {J.~S.}\ \bibnamefont {Read}},
  \bibinfo {author} {\bibfnamefont {M.}~\bibnamefont {Shibata}}, \bibinfo
  {author} {\bibfnamefont {K.}~\bibnamefont {Uryu}}, \bibinfo {author}
  {\bibfnamefont {J.~D.~E.}\ \bibnamefont {Creighton}}, \ and\ \bibinfo
  {author} {\bibfnamefont {J.~L.}\ \bibnamefont {Friedman}},\ }in\ \href
  {http://arxiv.org/abs/1008.1822} {\emph {\bibinfo {booktitle} {12th Marcel
  Grossman Meet.}}},\ \bibinfo {editor} {edited by\ \bibinfo {editor}
  {\bibfnamefont {T.}~\bibnamefont {Damour}}, \bibinfo {editor} {\bibfnamefont
  {R.}~\bibnamefont {Jantzen}}, \ and\ \bibinfo {editor} {\bibfnamefont
  {R.}~\bibnamefont {Ruffini}}}\ (\bibinfo  {publisher} {World Scientific},\
  \bibinfo {address} {Singapore},\ \bibinfo {year} {2011})\ pp.\ \bibinfo
  {pages} {5--7}\BibitemShut {NoStop}%
\bibitem [{\citenamefont {East}\ \emph {et~al.}(2012)\citenamefont {East},
  \citenamefont {Pretorius},\ and\ \citenamefont {Stephens}}]{East2012}%
  \BibitemOpen
  \bibfield  {author} {\bibinfo {author} {\bibfnamefont {W.~E.}\ \bibnamefont
  {East}}, \bibinfo {author} {\bibfnamefont {F.}~\bibnamefont {Pretorius}}, \
  and\ \bibinfo {author} {\bibfnamefont {B.~C.}\ \bibnamefont {Stephens}},\
  }\href {\doibase 10.1103/PhysRevD.85.124010} {\bibfield  {journal} {\bibinfo
  {journal} {Phys. Rev. D}\ }\textbf {\bibinfo {volume} {85}},\ \bibinfo
  {pages} {124010} (\bibinfo {year} {2012})}\BibitemShut {NoStop}%
\bibitem [{\citenamefont {Farris}\ \emph {et~al.}(2012)\citenamefont {Farris},
  \citenamefont {Gold}, \citenamefont {Paschalidis}, \citenamefont {Etienne},\
  and\ \citenamefont {Shapiro}}]{Farris2012}%
  \BibitemOpen
  \bibfield  {author} {\bibinfo {author} {\bibfnamefont {B.~D.}\ \bibnamefont
  {Farris}}, \bibinfo {author} {\bibfnamefont {R.}~\bibnamefont {Gold}},
  \bibinfo {author} {\bibfnamefont {V.}~\bibnamefont {Paschalidis}}, \bibinfo
  {author} {\bibfnamefont {Z.~B.}\ \bibnamefont {Etienne}}, \ and\ \bibinfo
  {author} {\bibfnamefont {S.~L.}\ \bibnamefont {Shapiro}},\ }\href {\doibase
  10.1103/PhysRevLett.109.221102} {\bibfield  {journal} {\bibinfo  {journal}
  {Phys. Rev. Lett.}\ }\textbf {\bibinfo {volume} {109}},\ \bibinfo {pages}
  {221102} (\bibinfo {year} {2012})}\BibitemShut {NoStop}%
\bibitem [{\citenamefont {Giacomazzo}\ \emph {et~al.}(2011)\citenamefont
  {Giacomazzo}, \citenamefont {Rezzolla},\ and\ \citenamefont
  {Baiotti}}]{Giacomazzo2011a}%
  \BibitemOpen
  \bibfield  {author} {\bibinfo {author} {\bibfnamefont {B.}~\bibnamefont
  {Giacomazzo}}, \bibinfo {author} {\bibfnamefont {L.}~\bibnamefont
  {Rezzolla}}, \ and\ \bibinfo {author} {\bibfnamefont {L.}~\bibnamefont
  {Baiotti}},\ }\href {\doibase 10.1103/PhysRevD.83.044014} {\bibfield
  {journal} {\bibinfo  {journal} {Phys. Rev. D}\ }\textbf {\bibinfo {volume}
  {83}},\ \bibinfo {pages} {044014} (\bibinfo {year} {2011})}\BibitemShut
  {NoStop}%
\bibitem [{\citenamefont {Read}\ \emph {et~al.}(2013)\citenamefont {Read},
  \citenamefont {Baiotti}, \citenamefont {Creighton}, \citenamefont {Friedman},
  \citenamefont {Giacomazzo}, \citenamefont {Kyutoku}, \citenamefont
  {Markakis}, \citenamefont {Rezzolla}, \citenamefont {Shibata},\ and\
  \citenamefont {Taniguchi}}]{Read2013}%
  \BibitemOpen
  \bibfield  {author} {\bibinfo {author} {\bibfnamefont {J.~S.}\ \bibnamefont
  {Read}}, \bibinfo {author} {\bibfnamefont {L.}~\bibnamefont {Baiotti}},
  \bibinfo {author} {\bibfnamefont {J.~D.~E.}\ \bibnamefont {Creighton}},
  \bibinfo {author} {\bibfnamefont {J.~L.}\ \bibnamefont {Friedman}}, \bibinfo
  {author} {\bibfnamefont {B.}~\bibnamefont {Giacomazzo}}, \bibinfo {author}
  {\bibfnamefont {K.}~\bibnamefont {Kyutoku}}, \bibinfo {author} {\bibfnamefont
  {C.}~\bibnamefont {Markakis}}, \bibinfo {author} {\bibfnamefont
  {L.}~\bibnamefont {Rezzolla}}, \bibinfo {author} {\bibfnamefont
  {M.}~\bibnamefont {Shibata}}, \ and\ \bibinfo {author} {\bibfnamefont
  {K.}~\bibnamefont {Taniguchi}},\ }\href {\doibase 10.1103/PhysRevD.88.044042}
  {\bibfield  {journal} {\bibinfo  {journal} {Phys. Rev. D}\ }\textbf {\bibinfo
  {volume} {88}},\ \bibinfo {pages} {044042} (\bibinfo {year}
  {2013})}\BibitemShut {NoStop}%
\bibitem [{\citenamefont {Mosta}\ \emph {et~al.}(2014)\citenamefont {Mosta},
  \citenamefont {Mundim}, \citenamefont {Faber}, \citenamefont {Haas},
  \citenamefont {Noble}, \citenamefont {Bode}, \citenamefont {L{\"{o}}ffler},
  \citenamefont {Ott}, \citenamefont {Reisswig},\ and\ \citenamefont
  {Schnetter}}]{Mosta2014}%
  \BibitemOpen
  \bibfield  {author} {\bibinfo {author} {\bibfnamefont {P.}~\bibnamefont
  {Mosta}}, \bibinfo {author} {\bibfnamefont {B.~C.}\ \bibnamefont {Mundim}},
  \bibinfo {author} {\bibfnamefont {J.~A.}\ \bibnamefont {Faber}}, \bibinfo
  {author} {\bibfnamefont {R.}~\bibnamefont {Haas}}, \bibinfo {author}
  {\bibfnamefont {S.~C.}\ \bibnamefont {Noble}}, \bibinfo {author}
  {\bibfnamefont {T.}~\bibnamefont {Bode}}, \bibinfo {author} {\bibfnamefont
  {F.}~\bibnamefont {L{\"{o}}ffler}}, \bibinfo {author} {\bibfnamefont {C.~D.}\
  \bibnamefont {Ott}}, \bibinfo {author} {\bibfnamefont {C.}~\bibnamefont
  {Reisswig}}, \ and\ \bibinfo {author} {\bibfnamefont {E.}~\bibnamefont
  {Schnetter}},\ }\href {\doibase 10.1088/0264-9381/31/1/015005} {\bibfield
  {journal} {\bibinfo  {journal} {Class. Quantum Gravity}\ }\textbf {\bibinfo
  {volume} {31}},\ \bibinfo {pages} {015005} (\bibinfo {year}
  {2014})}\BibitemShut {NoStop}%
\bibitem [{\citenamefont {Dionysopoulou}\ \emph {et~al.}(2015)\citenamefont
  {Dionysopoulou}, \citenamefont {Alic},\ and\ \citenamefont
  {Rezzolla}}]{Dionysopoulou2015}%
  \BibitemOpen
  \bibfield  {author} {\bibinfo {author} {\bibfnamefont {K.}~\bibnamefont
  {Dionysopoulou}}, \bibinfo {author} {\bibfnamefont {D.}~\bibnamefont {Alic}},
  \ and\ \bibinfo {author} {\bibfnamefont {L.}~\bibnamefont {Rezzolla}},\
  }\href {\doibase 10.1103/PhysRevD.92.084064} {\bibfield  {journal} {\bibinfo
  {journal} {Phys. Rev. D}\ }\textbf {\bibinfo {volume} {92}},\ \bibinfo
  {pages} {084064} (\bibinfo {year} {2015})}\BibitemShut {NoStop}%
\bibitem [{\citenamefont {Kawamura}\ \emph {et~al.}(2016)\citenamefont
  {Kawamura}, \citenamefont {Giacomazzo}, \citenamefont {Kastaun},
  \citenamefont {Ciolfi}, \citenamefont {Endrizzi}, \citenamefont {Baiotti},\
  and\ \citenamefont {Perna}}]{Kawamura2016}%
  \BibitemOpen
  \bibfield  {author} {\bibinfo {author} {\bibfnamefont {T.}~\bibnamefont
  {Kawamura}}, \bibinfo {author} {\bibfnamefont {B.}~\bibnamefont
  {Giacomazzo}}, \bibinfo {author} {\bibfnamefont {W.}~\bibnamefont {Kastaun}},
  \bibinfo {author} {\bibfnamefont {R.}~\bibnamefont {Ciolfi}}, \bibinfo
  {author} {\bibfnamefont {A.}~\bibnamefont {Endrizzi}}, \bibinfo {author}
  {\bibfnamefont {L.}~\bibnamefont {Baiotti}}, \ and\ \bibinfo {author}
  {\bibfnamefont {R.}~\bibnamefont {Perna}},\ }\href {\doibase
  10.1103/PhysRevD.94.064012} {\bibfield  {journal} {\bibinfo  {journal} {Phys.
  Rev. D}\ }\textbf {\bibinfo {volume} {94}},\ \bibinfo {pages} {064012}
  (\bibinfo {year} {2016})}\BibitemShut {NoStop}%
\bibitem [{\citenamefont {Contopoulos}(2016)}]{Contopoulos2016}%
  \BibitemOpen
  \bibfield  {author} {\bibinfo {author} {\bibfnamefont {I.}~\bibnamefont
  {Contopoulos}},\ }\href {\doibase 10.1017/S0022377816000453} {\bibfield
  {journal} {\bibinfo  {journal} {J. Plasma Phys.}\ }\textbf {\bibinfo {volume}
  {82}},\ \bibinfo {pages} {1} (\bibinfo {year} {2016})},\ \Eprint
  {http://arxiv.org/abs/1604.03719} {arXiv:1604.03719} \BibitemShut {NoStop}%
\bibitem [{\citenamefont {Nathanail}\ \emph {et~al.}(2016)\citenamefont
  {Nathanail}, \citenamefont {Strantzalis},\ and\ \citenamefont
  {Contopoulos}}]{Nathanail2016}%
  \BibitemOpen
  \bibfield  {author} {\bibinfo {author} {\bibfnamefont {A.}~\bibnamefont
  {Nathanail}}, \bibinfo {author} {\bibfnamefont {A.}~\bibnamefont
  {Strantzalis}}, \ and\ \bibinfo {author} {\bibfnamefont {I.}~\bibnamefont
  {Contopoulos}},\ }\href {\doibase 10.1093/mnras/stv2558} {\bibfield
  {journal} {\bibinfo  {journal} {Mon. Not. R. Astron. Soc.}\ }\textbf
  {\bibinfo {volume} {455}},\ \bibinfo {pages} {4479} (\bibinfo {year}
  {2016})}\BibitemShut {NoStop}%
\bibitem [{\citenamefont {Zanotti}\ \emph {et~al.}(2010)\citenamefont
  {Zanotti}, \citenamefont {Rezzolla}, \citenamefont {{Del Zanna}},\ and\
  \citenamefont {Palenzuela}}]{Zanotti2010}%
  \BibitemOpen
  \bibfield  {author} {\bibinfo {author} {\bibfnamefont {O.}~\bibnamefont
  {Zanotti}}, \bibinfo {author} {\bibfnamefont {L.}~\bibnamefont {Rezzolla}},
  \bibinfo {author} {\bibfnamefont {L.}~\bibnamefont {{Del Zanna}}}, \ and\
  \bibinfo {author} {\bibfnamefont {C.}~\bibnamefont {Palenzuela}},\ }\href
  {\doibase 10.1051/0004-6361/201014969} {\bibfield  {journal} {\bibinfo
  {journal} {Astron. Astrophys.}\ }\textbf {\bibinfo {volume} {523}},\ \bibinfo
  {pages} {A8} (\bibinfo {year} {2010})},\ \Eprint
  {http://arxiv.org/abs/1002.4185} {arXiv:1002.4185} \BibitemShut {NoStop}%
\bibitem [{\citenamefont {Korobkin}\ \emph {et~al.}(2013)\citenamefont
  {Korobkin}, \citenamefont {Abdikamalov}, \citenamefont {Stergioulas},
  \citenamefont {Schnetter}, \citenamefont {Zink}, \citenamefont {Rosswog},\
  and\ \citenamefont {Ott}}]{Korobkin2013}%
  \BibitemOpen
  \bibfield  {author} {\bibinfo {author} {\bibfnamefont {O.}~\bibnamefont
  {Korobkin}}, \bibinfo {author} {\bibfnamefont {E.}~\bibnamefont
  {Abdikamalov}}, \bibinfo {author} {\bibfnamefont {N.}~\bibnamefont
  {Stergioulas}}, \bibinfo {author} {\bibfnamefont {E.}~\bibnamefont
  {Schnetter}}, \bibinfo {author} {\bibfnamefont {B.}~\bibnamefont {Zink}},
  \bibinfo {author} {\bibfnamefont {S.}~\bibnamefont {Rosswog}}, \ and\
  \bibinfo {author} {\bibfnamefont {C.~D.}\ \bibnamefont {Ott}},\ }\href
  {\doibase 10.1093/mnras/stt166} {\bibfield  {journal} {\bibinfo  {journal}
  {Mon. Not. R. Astron. Soc.}\ }\textbf {\bibinfo {volume} {431}},\ \bibinfo
  {pages} {349} (\bibinfo {year} {2013})}\BibitemShut {NoStop}%
\bibitem [{\citenamefont {Contopoulos}\ \emph {et~al.}(2015)\citenamefont
  {Contopoulos}, \citenamefont {Nathanail},\ and\ \citenamefont
  {Katsanikas}}]{Contopoulos2015}%
  \BibitemOpen
  \bibfield  {author} {\bibinfo {author} {\bibfnamefont {I.}~\bibnamefont
  {Contopoulos}}, \bibinfo {author} {\bibfnamefont {A.}~\bibnamefont
  {Nathanail}}, \ and\ \bibinfo {author} {\bibfnamefont {M.}~\bibnamefont
  {Katsanikas}},\ }\href {\doibase 10.1088/0004-637X/805/2/105} {\bibfield
  {journal} {\bibinfo  {journal} {Astrophys. J.}\ }\textbf {\bibinfo {volume}
  {805}},\ \bibinfo {pages} {105} (\bibinfo {year} {2015})}\BibitemShut
  {NoStop}%
\bibitem [{\citenamefont {Kouretsis}\ and\ \citenamefont
  {Tsagas}(2010)}]{Kouretsis2010}%
  \BibitemOpen
  \bibfield  {author} {\bibinfo {author} {\bibfnamefont {A.~P.}\ \bibnamefont
  {Kouretsis}}\ and\ \bibinfo {author} {\bibfnamefont {C.~G.}\ \bibnamefont
  {Tsagas}},\ }\href {\doibase 10.1103/PhysRevD.82.124053} {\bibfield
  {journal} {\bibinfo  {journal} {Phys. Rev. D}\ }\textbf {\bibinfo {volume}
  {82}},\ \bibinfo {pages} {124053} (\bibinfo {year} {2010})}\BibitemShut
  {NoStop}%
\bibitem [{\citenamefont {Kouretsis}\ and\ \citenamefont
  {Tsagas}(2013)}]{Kouretsis2013}%
  \BibitemOpen
  \bibfield  {author} {\bibinfo {author} {\bibfnamefont {A.~P.}\ \bibnamefont
  {Kouretsis}}\ and\ \bibinfo {author} {\bibfnamefont {C.~G.}\ \bibnamefont
  {Tsagas}},\ }\href {\doibase 10.1103/PhysRevD.88.044006} {\bibfield
  {journal} {\bibinfo  {journal} {Phys. Rev. D}\ }\textbf {\bibinfo {volume}
  {88}},\ \bibinfo {pages} {044006} (\bibinfo {year} {2013})}\BibitemShut
  {NoStop}%
\bibitem [{\citenamefont {Barrow}\ and\ \citenamefont
  {Tsagas}(2011)}]{Barrow2011}%
  \BibitemOpen
  \bibfield  {author} {\bibinfo {author} {\bibfnamefont {J.~D.}\ \bibnamefont
  {Barrow}}\ and\ \bibinfo {author} {\bibfnamefont {C.~G.}\ \bibnamefont
  {Tsagas}},\ }\href {\doibase 10.1111/j.1365-2966.2011.18414.x} {\bibfield
  {journal} {\bibinfo  {journal} {Mon. Not. R. Astron. Soc.}\ }\textbf
  {\bibinfo {volume} {414}},\ \bibinfo {pages} {512} (\bibinfo {year}
  {2011})}\BibitemShut {NoStop}%
\bibitem [{\citenamefont {Tsagas}(2011)}]{Tsagas2011}%
  \BibitemOpen
  \bibfield  {author} {\bibinfo {author} {\bibfnamefont {C.~G.}\ \bibnamefont
  {Tsagas}},\ }\href {\doibase 10.1103/PhysRevD.84.043524} {\bibfield
  {journal} {\bibinfo  {journal} {Phys. Rev. D}\ }\textbf {\bibinfo {volume}
  {84}},\ \bibinfo {pages} {043524} (\bibinfo {year} {2011})}\BibitemShut
  {NoStop}%
\bibitem [{\citenamefont {Dosopoulou}\ \emph {et~al.}(2012)\citenamefont
  {Dosopoulou}, \citenamefont {{Del Sordo}}, \citenamefont {Tsagas},\ and\
  \citenamefont {Brandenburg}}]{Dosopoulou2012}%
  \BibitemOpen
  \bibfield  {author} {\bibinfo {author} {\bibfnamefont {F.}~\bibnamefont
  {Dosopoulou}}, \bibinfo {author} {\bibfnamefont {F.}~\bibnamefont {{Del
  Sordo}}}, \bibinfo {author} {\bibfnamefont {C.~G.}\ \bibnamefont {Tsagas}}, \
  and\ \bibinfo {author} {\bibfnamefont {A.}~\bibnamefont {Brandenburg}},\
  }\href {\doibase 10.1103/PhysRevD.85.063514} {\bibfield  {journal} {\bibinfo
  {journal} {Phys. Rev. D}\ }\textbf {\bibinfo {volume} {85}},\ \bibinfo
  {pages} {063514} (\bibinfo {year} {2012})}\BibitemShut {NoStop}%
\bibitem [{\citenamefont {Barrow}\ \emph
  {et~al.}(2012{\natexlab{a}})\citenamefont {Barrow}, \citenamefont {Tsagas},\
  and\ \citenamefont {Yamamoto}}]{Barrow2012}%
  \BibitemOpen
  \bibfield  {author} {\bibinfo {author} {\bibfnamefont {J.~D.}\ \bibnamefont
  {Barrow}}, \bibinfo {author} {\bibfnamefont {C.~G.}\ \bibnamefont {Tsagas}},
  \ and\ \bibinfo {author} {\bibfnamefont {K.}~\bibnamefont {Yamamoto}},\
  }\href {\doibase 10.1103/PhysRevD.86.023533} {\bibfield  {journal} {\bibinfo
  {journal} {Phys. Rev. D}\ }\textbf {\bibinfo {volume} {86}},\ \bibinfo
  {pages} {023533} (\bibinfo {year} {2012}{\natexlab{a}})}\BibitemShut
  {NoStop}%
\bibitem [{\citenamefont {Barrow}\ \emph
  {et~al.}(2012{\natexlab{b}})\citenamefont {Barrow}, \citenamefont {Tsagas},\
  and\ \citenamefont {Yamamoto}}]{Barrow2012a}%
  \BibitemOpen
  \bibfield  {author} {\bibinfo {author} {\bibfnamefont {J.~D.}\ \bibnamefont
  {Barrow}}, \bibinfo {author} {\bibfnamefont {C.~G.}\ \bibnamefont {Tsagas}},
  \ and\ \bibinfo {author} {\bibfnamefont {K.}~\bibnamefont {Yamamoto}},\
  }\href {\doibase 10.1103/PhysRevD.86.107302} {\bibfield  {journal} {\bibinfo
  {journal} {Phys. Rev. D}\ }\textbf {\bibinfo {volume} {86}},\ \bibinfo
  {pages} {107302} (\bibinfo {year} {2012}{\natexlab{b}})}\BibitemShut
  {NoStop}%
\bibitem [{\citenamefont {Dosopoulou}\ and\ \citenamefont
  {Tsagas}(2014)}]{Dosopoulou2014}%
  \BibitemOpen
  \bibfield  {author} {\bibinfo {author} {\bibfnamefont {F.}~\bibnamefont
  {Dosopoulou}}\ and\ \bibinfo {author} {\bibfnamefont {C.~G.}\ \bibnamefont
  {Tsagas}},\ }\href {\doibase 10.1103/PhysRevD.89.103519} {\bibfield
  {journal} {\bibinfo  {journal} {Phys. Rev. D}\ }\textbf {\bibinfo {volume}
  {89}},\ \bibinfo {pages} {103519} (\bibinfo {year} {2014})}\BibitemShut
  {NoStop}%
\bibitem [{\citenamefont {Pen}\ and\ \citenamefont {Turok}(2016)}]{Pen2016}%
  \BibitemOpen
  \bibfield  {author} {\bibinfo {author} {\bibfnamefont {U.-L.}\ \bibnamefont
  {Pen}}\ and\ \bibinfo {author} {\bibfnamefont {N.}~\bibnamefont {Turok}},\
  }\href {\doibase 10.1103/PhysRevLett.117.131301} {\bibfield  {journal}
  {\bibinfo  {journal} {Phys. Rev. Lett.}\ }\textbf {\bibinfo {volume} {117}},\
  \bibinfo {pages} {131301} (\bibinfo {year} {2016})}\BibitemShut {NoStop}%
\bibitem [{\citenamefont {Gourgoulhon}\ \emph
  {et~al.}(2011{\natexlab{a}})\citenamefont {Gourgoulhon}, \citenamefont
  {Markakis}, \citenamefont {Uryu},\ and\ \citenamefont
  {Eriguchi}}]{GourgMUE11}%
  \BibitemOpen
  \bibfield  {author} {\bibinfo {author} {\bibfnamefont {E.}~\bibnamefont
  {Gourgoulhon}}, \bibinfo {author} {\bibfnamefont {C.}~\bibnamefont
  {Markakis}}, \bibinfo {author} {\bibfnamefont {K.}~\bibnamefont {Uryu}}, \
  and\ \bibinfo {author} {\bibfnamefont {Y.}~\bibnamefont {Eriguchi}},\
  }\href@noop {} {\bibfield  {journal} {\bibinfo  {journal} {Rev. D}\ }\textbf
  {\bibinfo {volume} {83}} (\bibinfo {year} {2011}{\natexlab{a}})}\BibitemShut
  {NoStop}%
\bibitem [{\citenamefont {Gourgoulhon}(2006)}]{Gourg06}%
  \BibitemOpen
  \bibfield  {author} {\bibinfo {author} {\bibfnamefont {E.}~\bibnamefont
  {Gourgoulhon}},\ }in\ \href {\doibase 10.1051/eas:2006106} {\emph {\bibinfo
  {booktitle} {EAS Publ. Ser.}}},\ Vol.~\bibinfo {volume} {21}\ (\bibinfo
  {publisher} {M. Rieutord and B. Dubrulle},\ \bibinfo {year} {2006})\ pp.\
  \bibinfo {pages} {43--79}\BibitemShut {NoStop}%
\bibitem [{\citenamefont {Gourgoulhon}(2013)}]{Gourgoulhon2013}%
  \BibitemOpen
  \bibfield  {author} {\bibinfo {author} {\bibfnamefont {E.}~\bibnamefont
  {Gourgoulhon}},\ }\href {\doibase 10.1007/978-3-642-00710-1_2} {\emph
  {\bibinfo {title} {{Special Relativity in General Frames: From Particles to
  Astrophysics}}}}\ (\bibinfo  {publisher} {Springer},\ \bibinfo {address}
  {Paris},\ \bibinfo {year} {2013})\BibitemShut {NoStop}%
\bibitem [{\citenamefont {Friedman}\ and\ \citenamefont
  {Stergioulas}(2013)}]{FriedmanStergioulas2013}%
  \BibitemOpen
  \bibfield  {author} {\bibinfo {author} {\bibfnamefont {J.~L.}\ \bibnamefont
  {Friedman}}\ and\ \bibinfo {author} {\bibfnamefont {N.}~\bibnamefont
  {Stergioulas}},\ }\href@noop {} {\emph {\bibinfo {title} {{Rotating
  Relativistic Stars}}}}\ (\bibinfo  {publisher} {Cambridge University Press},\
  \bibinfo {address} {New York, NY},\ \bibinfo {year} {2013})\BibitemShut
  {NoStop}%
\bibitem [{\citenamefont {Abraham}\ and\ \citenamefont
  {Marsden}(2008)}]{Abraham2008}%
  \BibitemOpen
  \bibfield  {author} {\bibinfo {author} {\bibfnamefont {R.}~\bibnamefont
  {Abraham}}\ and\ \bibinfo {author} {\bibfnamefont {J.~E.}\ \bibnamefont
  {Marsden}},\ }\href@noop {} {\emph {\bibinfo {title} {{Foundations of
  mechanics}}}}\ (\bibinfo  {publisher} {AMS Chelsea Pub.},\ \bibinfo {year}
  {2008})\ p.\ \bibinfo {pages} {826}\BibitemShut {NoStop}%
\bibitem [{\citenamefont {Poincar{\'{e}}}(1890)}]{Poincare1890}%
  \BibitemOpen
  \bibfield  {author} {\bibinfo {author} {\bibfnamefont {H.}~\bibnamefont
  {Poincar{\'{e}}}},\ }\href@noop {} {\bibfield  {journal} {\bibinfo  {journal}
  {Acta Math.}\ }\textbf {\bibinfo {volume} {13}},\ \bibinfo {pages} {1}
  (\bibinfo {year} {1890})}\BibitemShut {NoStop}%
\bibitem [{\citenamefont {Poincar{\'{e}}}(1899)}]{Poincare1899}%
  \BibitemOpen
  \bibfield  {author} {\bibinfo {author} {\bibfnamefont {H.}~\bibnamefont
  {Poincar{\'{e}}}},\ }in\ \href@noop {} {\emph {\bibinfo {booktitle}
  {Gauthier-Villars}}},\ Vol.~\bibinfo {volume} {3}\ (\bibinfo {year} {1899})\
  p.\ \bibinfo {pages} {Chapt. 26}\BibitemShut {NoStop}%
\bibitem [{\citenamefont {Cartan}(1922)}]{Cartan1922}%
  \BibitemOpen
  \bibfield  {author} {\bibinfo {author} {\bibfnamefont {{\'{E}}.}~\bibnamefont
  {Cartan}},\ }\href
  {http://neo-classical-physics.info/uploads/3/0/6/5/3065888/cartan{\_}-{\_}le%
ssons{\_}on{\_}integral{\_}invariants.pdf} {\bibfield  {journal} {\bibinfo
  {journal} {Hermann}\ } (\bibinfo {year} {1922})}\BibitemShut {NoStop}%
\bibitem [{\citenamefont {Boccaletti}\ and\ \citenamefont
  {Pucacco}(2003)}]{Boccaletti2003}%
  \BibitemOpen
  \bibfield  {author} {\bibinfo {author} {\bibfnamefont {D.}~\bibnamefont
  {Boccaletti}}\ and\ \bibinfo {author} {\bibfnamefont {G.}~\bibnamefont
  {Pucacco}},\ }\href@noop {} {\emph {\bibinfo {title} {{Theory of Orbits:
  Volume 1}}}}\ (\bibinfo  {publisher} {Springer},\ \bibinfo {year}
  {2003})\BibitemShut {NoStop}%
\bibitem [{\citenamefont {Markakis}(2014{\natexlab{a}})}]{Markakis2014a}%
  \BibitemOpen
  \bibfield  {author} {\bibinfo {author} {\bibfnamefont {C.~M.}\ \bibnamefont
  {Markakis}},\ }\href {http://arxiv.org/abs/1410.7777} {\  (\bibinfo {year}
  {2014}{\natexlab{a}})},\ \Eprint {http://arxiv.org/abs/1410.7777}
  {arXiv:1410.7777} \BibitemShut {NoStop}%
\bibitem [{\citenamefont {Woltjer}(1958)}]{Woltjer1958}%
  \BibitemOpen
  \bibfield  {author} {\bibinfo {author} {\bibfnamefont {L.}~\bibnamefont
  {Woltjer}},\ }\href@noop {} {\bibfield  {journal} {\bibinfo  {journal} {Proc.
  NUI. Acad. Sci.}\ }\textbf {\bibinfo {volume} {44}},\ \bibinfo {pages} {833}
  (\bibinfo {year} {1958})}\BibitemShut {NoStop}%
\bibitem [{\citenamefont {Moffatt}(1969)}]{MOFFAT1969}%
  \BibitemOpen
  \bibfield  {author} {\bibinfo {author} {\bibfnamefont {H.~K.}\ \bibnamefont
  {Moffatt}},\ }\href {\doibase 10.1017/S0022112069000991} {\bibfield
  {journal} {\bibinfo  {journal} {J. Fluid Mech.}\ }\textbf {\bibinfo {volume}
  {35}},\ \bibinfo {pages} {117} (\bibinfo {year} {1969})}\BibitemShut
  {NoStop}%
\bibitem [{\citenamefont {Carter}\ and\ \citenamefont
  {Khalatnikov}(1992)}]{Carter1992}%
  \BibitemOpen
  \bibfield  {author} {\bibinfo {author} {\bibfnamefont {B.}~\bibnamefont
  {Carter}}\ and\ \bibinfo {author} {\bibfnamefont {I.}~\bibnamefont
  {Khalatnikov}},\ }\href {\doibase 10.1016/0003-4916(92)90348-P} {\bibfield
  {journal} {\bibinfo  {journal} {Ann. Phys. (N. Y).}\ }\textbf {\bibinfo
  {volume} {219}},\ \bibinfo {pages} {243} (\bibinfo {year}
  {1992})}\BibitemShut {NoStop}%
\bibitem [{\citenamefont {Ioannou}\ and\ \citenamefont
  {Apostolatos}(2004)}]{Ioannou2004}%
  \BibitemOpen
  \bibfield  {author} {\bibinfo {author} {\bibfnamefont {P.~J.}\ \bibnamefont
  {Ioannou}}\ and\ \bibinfo {author} {\bibfnamefont {T.~A.}\ \bibnamefont
  {Apostolatos}},\ }\href@noop {} {\emph {\bibinfo {title} {{Elements of
  Theoretical Mechanics (in Greek)}}}},\ \bibinfo {edition} {1st}\ ed.\
  (\bibinfo  {publisher} {Leader Books},\ \bibinfo {address} {Athens},\
  \bibinfo {year} {2004})\BibitemShut {NoStop}%
\bibitem [{\citenamefont {Taub}(1959)}]{Taub59}%
  \BibitemOpen
  \bibfield  {author} {\bibinfo {author} {\bibfnamefont {A.~H.}\ \bibnamefont
  {Taub}},\ }\href {\doibase 10.1007/BF00284183} {\bibfield  {journal}
  {\bibinfo  {journal} {Arch. Ration. Mech. Anal.}\ }\textbf {\bibinfo {volume}
  {3-3}},\ \bibinfo {pages} {312} (\bibinfo {year} {1959})}\BibitemShut
  {NoStop}%
\bibitem [{\citenamefont {Christodoulou}(2007)}]{Christodoulou2007}%
  \BibitemOpen
  \bibfield  {author} {\bibinfo {author} {\bibfnamefont {D.}~\bibnamefont
  {Christodoulou}},\ }\href
  {http://www.ams.org/journals/bull/2007-44-04/S0273-0979-07-01181-0/S0273-097%
9-07-01181-0.pdf} {\bibfield  {journal} {\bibinfo  {journal} {Bull. Am. Math.
  Soc.}\ }\textbf {\bibinfo {volume} {44}},\ \bibinfo {pages} {581} (\bibinfo
  {year} {2007})}\BibitemShut {NoStop}%
\bibitem [{\citenamefont {Markakis}(2011)}]{MarkakisRBNSMF2011}%
  \BibitemOpen
  \bibfield  {author} {\bibinfo {author} {\bibfnamefont {C.}~\bibnamefont
  {Markakis}},\ }\href@noop {} {\emph {\bibinfo {title} {{Rotating and binary
  neutron stars with magnetic fields}}}}\ (\bibinfo  {publisher} {PhD
  Dissertation},\ \bibinfo {address} {University of Wisconsin-Milwaukee �},\
  \bibinfo {year} {2011})\BibitemShut {NoStop}%
\bibitem [{\citenamefont {Cauchy}(1815)}]{Cauchy1815}%
  \BibitemOpen
  \bibfield  {author} {\bibinfo {author} {\bibfnamefont {A.-L.}\ \bibnamefont
  {Cauchy}},\ }\href
  {http://gallica.bnf.fr/ark:/12148/bpt6k90181x/f14.image.r=Oeuvres completes
  d{\%}2527Augustin Cauchy.langFR} {\emph {\bibinfo {title} {{Th´eorie de la
  propagation des ondes `a la surface d'un fluide pesant d'une profondeur
  ind´efinie - Prix d'analyse math´ematique remport´e par M. Augustin-Louis
  Cauchy, ing´enieur des Ponts et Chauss´ees. (Concours de 1815)}}}},\
  \bibinfo {edition} {tome i}\ ed.\ (\bibinfo {year} {1815})\ pp.\ \bibinfo
  {pages} {5--318}\BibitemShut {NoStop}%
\bibitem [{\citenamefont {Frisch}\ and\ \citenamefont
  {Villone}(2014)}]{Frisch2014a}%
  \BibitemOpen
  \bibfield  {author} {\bibinfo {author} {\bibfnamefont {U.}~\bibnamefont
  {Frisch}}\ and\ \bibinfo {author} {\bibfnamefont {B.}~\bibnamefont
  {Villone}},\ }\href {\doibase 10.1140/epjh/e2014-50016-6} {\bibfield
  {journal} {\bibinfo  {journal} {Eur. Phys. J. H}\ }\textbf {\bibinfo {volume}
  {39}},\ \bibinfo {pages} {325} (\bibinfo {year} {2014})},\ \Eprint
  {http://arxiv.org/abs/1402.4957} {arXiv:1402.4957} \BibitemShut {NoStop}%
\bibitem [{\citenamefont {Thomson}(1869)}]{Thomson1869}%
  \BibitemOpen
  \bibfield  {author} {\bibinfo {author} {\bibfnamefont {W.}~\bibnamefont
  {Thomson}},\ }\href@noop {} {\bibfield  {journal} {\bibinfo  {journal}
  {Trans. Roy. Soc. Edinburgh.}\ }\textbf {\bibinfo {volume} {25}},\ \bibinfo
  {pages} {217} (\bibinfo {year} {1869})}\BibitemShut {NoStop}%
\bibitem [{\citenamefont {Lichnerowicz}(1955)}]{Lichnerowicz1955}%
  \BibitemOpen
  \bibfield  {author} {\bibinfo {author} {\bibfnamefont {A.}~\bibnamefont
  {Lichnerowicz}},\ }\href@noop {} {\emph {\bibinfo {title} {{Th{\'{e}}ories
  Relativistes de la Gravitation et de L'Electromagnetisme}}}}\ (\bibinfo
  {publisher} {Masson {\&} Cie},\ \bibinfo {address} {Paris},\ \bibinfo {year}
  {1955})\BibitemShut {NoStop}%
\bibitem [{\citenamefont {Prix}(2004)}]{Prix2004}%
  \BibitemOpen
  \bibfield  {author} {\bibinfo {author} {\bibfnamefont {R.}~\bibnamefont
  {Prix}},\ }\href {\doibase 10.1103/PhysRevD.69.043001} {\bibfield  {journal}
  {\bibinfo  {journal} {Phys. Rev. D}\ }\textbf {\bibinfo {volume} {69}},\
  \bibinfo {pages} {043001} (\bibinfo {year} {2004})}\BibitemShut {NoStop}%
\bibitem [{\citenamefont {Prix}(2005)}]{Prix2005}%
  \BibitemOpen
  \bibfield  {author} {\bibinfo {author} {\bibfnamefont {R.}~\bibnamefont
  {Prix}},\ }\href {\doibase 10.1103/PhysRevD.71.083006} {\bibfield  {journal}
  {\bibinfo  {journal} {Phys. Rev. D}\ }\textbf {\bibinfo {volume} {71}},\
  \bibinfo {pages} {083006} (\bibinfo {year} {2005})}\BibitemShut {NoStop}%
\bibitem [{\citenamefont {Misner}\ \emph {et~al.}(1973)\citenamefont {Misner},
  \citenamefont {Thorne},\ and\ \citenamefont {{J. A. Wheeler}}}]{MisneTW73}%
  \BibitemOpen
  \bibfield  {author} {\bibinfo {author} {\bibfnamefont {C.~W.}\ \bibnamefont
  {Misner}}, \bibinfo {author} {\bibfnamefont {K.~S.}\ \bibnamefont {Thorne}},
  \ and\ \bibinfo {author} {\bibfnamefont {.}~\bibnamefont {{J. A. Wheeler}}},\
  }\href@noop {} {\emph {\bibinfo {title} {{Gravitation}}}}\ (\bibinfo
  {publisher} {Freeman},\ \bibinfo {address} {New York},\ \bibinfo {year}
  {1973})\BibitemShut {NoStop}%
\bibitem [{\citenamefont {Wald}(1984)}]{Wald84}%
  \BibitemOpen
  \bibfield  {author} {\bibinfo {author} {\bibfnamefont {R.~M.}\ \bibnamefont
  {Wald}},\ }\href@noop {} {\emph {\bibinfo {title} {{General Relativity}}}}\
  (\bibinfo  {publisher} {University of Chicago Press},\ \bibinfo {address}
  {Chicago},\ \bibinfo {year} {1984})\BibitemShut {NoStop}%
\bibitem [{\citenamefont {Markakis}(2014{\natexlab{b}})}]{Markakis2012}%
  \BibitemOpen
  \bibfield  {author} {\bibinfo {author} {\bibfnamefont {C.}~\bibnamefont
  {Markakis}},\ }\href {\doibase 10.1093/mnras/stu715} {\bibfield  {journal}
  {\bibinfo  {journal} {Mon. Not. R. Astron. Soc.}\ }\textbf {\bibinfo {volume}
  {441}},\ \bibinfo {pages} {2974} (\bibinfo {year} {2014}{\natexlab{b}})},\
  \Eprint {http://arxiv.org/abs/1202.5228} {arXiv:1202.5228} \BibitemShut
  {NoStop}%
\bibitem [{\citenamefont {Huang}\ \emph {et~al.}(2008)\citenamefont {Huang},
  \citenamefont {Markakis}, \citenamefont {Sugiyama},\ and\ \citenamefont
  {Uryu}}]{Huang2008}%
  \BibitemOpen
  \bibfield  {author} {\bibinfo {author} {\bibfnamefont {X.}~\bibnamefont
  {Huang}}, \bibinfo {author} {\bibfnamefont {C.}~\bibnamefont {Markakis}},
  \bibinfo {author} {\bibfnamefont {N.}~\bibnamefont {Sugiyama}}, \ and\
  \bibinfo {author} {\bibfnamefont {K.}~\bibnamefont {Uryu}},\ }\href {\doibase
  10.1103/PhysRevD.78.124023} {\bibfield  {journal} {\bibinfo  {journal} {Phys.
  Rev. D}\ }\textbf {\bibinfo {volume} {78}},\ \bibinfo {pages} {124023}
  (\bibinfo {year} {2008})},\ \Eprint {http://arxiv.org/abs/0809.0673}
  {arXiv:0809.0673} \BibitemShut {NoStop}%
\bibitem [{\citenamefont {Bonazzola}\ \emph {et~al.}(1997)\citenamefont
  {Bonazzola}, \citenamefont {Gourgoulhon},\ and\ \citenamefont
  {Marck}}]{Bonazzola1997}%
  \BibitemOpen
  \bibfield  {author} {\bibinfo {author} {\bibfnamefont {S.}~\bibnamefont
  {Bonazzola}}, \bibinfo {author} {\bibfnamefont {E.}~\bibnamefont
  {Gourgoulhon}}, \ and\ \bibinfo {author} {\bibfnamefont {J.-A.}\ \bibnamefont
  {Marck}},\ }\href {\doibase 10.1103/PhysRevD.56.7740} {\bibfield  {journal}
  {\bibinfo  {journal} {Phys. Rev. D}\ }\textbf {\bibinfo {volume} {56}},\
  \bibinfo {pages} {7740} (\bibinfo {year} {1997})}\BibitemShut {NoStop}%
\bibitem [{\citenamefont {Teukolsky}(1998)}]{Teukolsky1998}%
  \BibitemOpen
  \bibfield  {author} {\bibinfo {author} {\bibfnamefont {S.~A.}\ \bibnamefont
  {Teukolsky}},\ }\href {\doibase 10.1086/306082} {\bibfield  {journal}
  {\bibinfo  {journal} {Astrophys. J.}\ }\textbf {\bibinfo {volume} {504}},\
  \bibinfo {pages} {442} (\bibinfo {year} {1998})}\BibitemShut {NoStop}%
\bibitem [{\citenamefont {Marronetti}\ \emph {et~al.}(1999)\citenamefont
  {Marronetti}, \citenamefont {Mathews},\ and\ \citenamefont
  {Wilson}}]{Marronetti1999}%
  \BibitemOpen
  \bibfield  {author} {\bibinfo {author} {\bibfnamefont {P.}~\bibnamefont
  {Marronetti}}, \bibinfo {author} {\bibfnamefont {G.~J.}\ \bibnamefont
  {Mathews}}, \ and\ \bibinfo {author} {\bibfnamefont {J.~R.}\ \bibnamefont
  {Wilson}},\ }\href {\doibase 10.1103/PhysRevD.60.087301} {\bibfield
  {journal} {\bibinfo  {journal} {Phys. Rev. D}\ }\textbf {\bibinfo {volume}
  {60}},\ \bibinfo {pages} {4} (\bibinfo {year} {1999})},\ \Eprint
  {http://arxiv.org/abs/9906088v1} {arXiv:9906088v1 [arXiv:gr-qc]} \BibitemShut
  {NoStop}%
\bibitem [{\citenamefont {Gourgoulhon}()}]{Gourgoulhon1998}%
  \BibitemOpen
  \bibfield  {author} {\bibinfo {author} {\bibfnamefont {E.}~\bibnamefont
  {Gourgoulhon}},\ }\href {http://arxiv.org/abs/gr-qc/9804054} {\ }\Eprint
  {http://arxiv.org/abs/9804054} {arXiv:9804054 [gr-qc]} \BibitemShut {NoStop}%
\bibitem [{\citenamefont {Shibata}(1998)}]{Shibata1998a}%
  \BibitemOpen
  \bibfield  {author} {\bibinfo {author} {\bibfnamefont {M.}~\bibnamefont
  {Shibata}},\ }\href {\doibase 10.1103/PhysRevD.58.024012} {\bibfield
  {journal} {\bibinfo  {journal} {Phys. Rev. D}\ }\textbf {\bibinfo {volume}
  {58}},\ \bibinfo {pages} {024012} (\bibinfo {year} {1998})}\BibitemShut
  {NoStop}%
\bibitem [{\citenamefont {Bonazzola}\ \emph {et~al.}(1999)\citenamefont
  {Bonazzola}, \citenamefont {Gourgoulhon},\ and\ \citenamefont
  {Marck}}]{Bonazzola1999}%
  \BibitemOpen
  \bibfield  {author} {\bibinfo {author} {\bibfnamefont {S.}~\bibnamefont
  {Bonazzola}}, \bibinfo {author} {\bibfnamefont {E.}~\bibnamefont
  {Gourgoulhon}}, \ and\ \bibinfo {author} {\bibfnamefont {J.-A.}\ \bibnamefont
  {Marck}},\ }\href {\doibase 10.1103/PhysRevLett.82.892} {\bibfield  {journal}
  {\bibinfo  {journal} {Phys. Rev. Lett.}\ }\textbf {\bibinfo {volume} {82}},\
  \bibinfo {pages} {892} (\bibinfo {year} {1999})}\BibitemShut {NoStop}%
\bibitem [{\citenamefont {Uryu}\ and\ \citenamefont
  {Eriguchi}(2000)}]{Uryu2000}%
  \BibitemOpen
  \bibfield  {author} {\bibinfo {author} {\bibfnamefont {K.}~\bibnamefont
  {Uryu}}\ and\ \bibinfo {author} {\bibfnamefont {Y.}~\bibnamefont
  {Eriguchi}},\ }\href {\doibase 10.1103/PhysRevD.61.124023} {\bibfield
  {journal} {\bibinfo  {journal} {Phys. Rev. D}\ }\textbf {\bibinfo {volume}
  {61}},\ \bibinfo {pages} {124023} (\bibinfo {year} {2000})}\BibitemShut
  {NoStop}%
\bibitem [{\citenamefont {Taniguchi}\ and\ \citenamefont
  {Shibata}(2010)}]{Taniguchi2010a}%
  \BibitemOpen
  \bibfield  {author} {\bibinfo {author} {\bibfnamefont {K.}~\bibnamefont
  {Taniguchi}}\ and\ \bibinfo {author} {\bibfnamefont {M.}~\bibnamefont
  {Shibata}},\ }\href {\doibase 10.1088/0067-0049/188/1/187} {\bibfield
  {journal} {\bibinfo  {journal} {Astrophys. J. Suppl. Ser.}\ }\textbf
  {\bibinfo {volume} {188}},\ \bibinfo {pages} {187} (\bibinfo {year}
  {2010})},\ \Eprint {http://arxiv.org/abs/1005.0958} {arXiv:1005.0958}
  \BibitemShut {NoStop}%
\bibitem [{\citenamefont {Price}\ \emph {et~al.}(2009)\citenamefont {Price},
  \citenamefont {Markakis},\ and\ \citenamefont
  {Friedman}}]{PriceMarkakisFriedman2009}%
  \BibitemOpen
  \bibfield  {author} {\bibinfo {author} {\bibfnamefont {R.~H.}\ \bibnamefont
  {Price}}, \bibinfo {author} {\bibfnamefont {C.}~\bibnamefont {Markakis}}, \
  and\ \bibinfo {author} {\bibfnamefont {J.~L.}\ \bibnamefont {Friedman}},\
  }\href {\doibase 10.1063/1.3166136} {\bibfield  {journal} {\bibinfo
  {journal} {J. Math. Phys.}\ }\textbf {\bibinfo {volume} {50}},\ \bibinfo
  {pages} {16} (\bibinfo {year} {2009})},\ \Eprint
  {http://arxiv.org/abs/0903.3074} {arXiv:0903.3074} \BibitemShut {NoStop}%
\bibitem [{\citenamefont {Marronetti}\ and\ \citenamefont
  {Shapiro}(2003)}]{Marronetti2003}%
  \BibitemOpen
  \bibfield  {author} {\bibinfo {author} {\bibfnamefont {P.}~\bibnamefont
  {Marronetti}}\ and\ \bibinfo {author} {\bibfnamefont {S.~L.}\ \bibnamefont
  {Shapiro}},\ }\href {\doibase 10.1103/PhysRevD.68.104024} {\bibfield
  {journal} {\bibinfo  {journal} {Phys. Rev. D}\ }\textbf {\bibinfo {volume}
  {68}},\ \bibinfo {pages} {104024} (\bibinfo {year} {2003})}\BibitemShut
  {NoStop}%
\bibitem [{\citenamefont {Tichy}(2011)}]{Tichy2011}%
  \BibitemOpen
  \bibfield  {author} {\bibinfo {author} {\bibfnamefont {W.}~\bibnamefont
  {Tichy}},\ }\href {\doibase 10.1103/PhysRevD.84.024041} {\bibfield  {journal}
  {\bibinfo  {journal} {Phys. Rev. D}\ }\textbf {\bibinfo {volume} {84}},\
  \bibinfo {pages} {024041} (\bibinfo {year} {2011})}\BibitemShut {NoStop}%
\bibitem [{\citenamefont {Tichy}(2012)}]{Tichy2012}%
  \BibitemOpen
  \bibfield  {author} {\bibinfo {author} {\bibfnamefont {W.}~\bibnamefont
  {Tichy}},\ }\href {\doibase 10.1103/PhysRevD.86.064024} {\bibfield  {journal}
  {\bibinfo  {journal} {Phys. Rev. D}\ }\textbf {\bibinfo {volume} {86}},\
  \bibinfo {pages} {064024} (\bibinfo {year} {2012})}\BibitemShut {NoStop}%
\bibitem [{\citenamefont {Tichy}(2016)}]{Tichy2016}%
  \BibitemOpen
  \bibfield  {author} {\bibinfo {author} {\bibfnamefont {W.}~\bibnamefont
  {Tichy}},\ }\href {http://arxiv.org/abs/1610.03805} {\  (\bibinfo {year}
  {2016})},\ \Eprint {http://arxiv.org/abs/1610.03805} {arXiv:1610.03805}
  \BibitemShut {NoStop}%
\bibitem [{\citenamefont {Moldenhauer}\ \emph {et~al.}(2014)\citenamefont
  {Moldenhauer}, \citenamefont {Markakis}, \citenamefont {Johnson-McDaniel},
  \citenamefont {Tichy},\ and\ \citenamefont
  {Br{\"{u}}gmann}}]{Moldenhauer2014}%
  \BibitemOpen
  \bibfield  {author} {\bibinfo {author} {\bibfnamefont {N.}~\bibnamefont
  {Moldenhauer}}, \bibinfo {author} {\bibfnamefont {C.~M.}\ \bibnamefont
  {Markakis}}, \bibinfo {author} {\bibfnamefont {N.~K.}\ \bibnamefont
  {Johnson-McDaniel}}, \bibinfo {author} {\bibfnamefont {W.}~\bibnamefont
  {Tichy}}, \ and\ \bibinfo {author} {\bibfnamefont {B.}~\bibnamefont
  {Br{\"{u}}gmann}},\ }\href {\doibase 10.1103/PhysRevD.90.084043} {\bibfield
  {journal} {\bibinfo  {journal} {Phys. Rev. D}\ }\textbf {\bibinfo {volume}
  {90}},\ \bibinfo {pages} {084043} (\bibinfo {year} {2014})}\BibitemShut
  {NoStop}%
\bibitem [{\citenamefont {Dietrich}\ \emph {et~al.}(2015)\citenamefont
  {Dietrich}, \citenamefont {Moldenhauer}, \citenamefont {Johnson-McDaniel},
  \citenamefont {Bernuzzi}, \citenamefont {Markakis}, \citenamefont
  {Br{\"{u}}gmann},\ and\ \citenamefont {Tichy}}]{Dietrich2015}%
  \BibitemOpen
  \bibfield  {author} {\bibinfo {author} {\bibfnamefont {T.}~\bibnamefont
  {Dietrich}}, \bibinfo {author} {\bibfnamefont {N.}~\bibnamefont
  {Moldenhauer}}, \bibinfo {author} {\bibfnamefont {N.~K.}\ \bibnamefont
  {Johnson-McDaniel}}, \bibinfo {author} {\bibfnamefont {S.}~\bibnamefont
  {Bernuzzi}}, \bibinfo {author} {\bibfnamefont {C.~M.}\ \bibnamefont
  {Markakis}}, \bibinfo {author} {\bibfnamefont {B.}~\bibnamefont
  {Br{\"{u}}gmann}}, \ and\ \bibinfo {author} {\bibfnamefont {W.}~\bibnamefont
  {Tichy}},\ }\href {\doibase 10.1103/PhysRevD.92.124007} {\bibfield  {journal}
  {\bibinfo  {journal} {Phys. Rev. D}\ }\textbf {\bibinfo {volume} {92}},\
  \bibinfo {pages} {124007} (\bibinfo {year} {2015})}\BibitemShut {NoStop}%
\bibitem [{\citenamefont {Padmanabhan}(2010)}]{Padmanabhan2010}%
  \BibitemOpen
  \bibfield  {author} {\bibinfo {author} {\bibfnamefont {T.}~\bibnamefont
  {Padmanabhan}},\ }\href@noop {} {\emph {\bibinfo {title} {{Gravitation:
  Foundations and Frontiers}}}}\ (\bibinfo  {publisher} {Cambridge University
  Press},\ \bibinfo {year} {2010})\ p.\ \bibinfo {pages} {728}\BibitemShut
  {NoStop}%
\bibitem [{\citenamefont {Landau}(1975)}]{Landau1975}%
  \BibitemOpen
  \bibfield  {author} {\bibinfo {author} {\bibfnamefont {L.~D.}\ \bibnamefont
  {Landau}},\ }\href@noop {} {\emph {\bibinfo {title} {{The classical theory of
  fields}}}},\ \bibinfo {edition} {4th}\ ed.\ (\bibinfo  {publisher} {Pergamon
  Press},\ \bibinfo {year} {1975})\BibitemShut {NoStop}%
\bibitem [{\citenamefont {Friedman}(1978)}]{Friedman1978a}%
  \BibitemOpen
  \bibfield  {author} {\bibinfo {author} {\bibfnamefont {J.~L.}\ \bibnamefont
  {Friedman}},\ }\href@noop {} {\bibfield  {journal} {\bibinfo  {journal}
  {Commun. Math. Phys.}\ }\textbf {\bibinfo {volume} {62}},\ \bibinfo {pages}
  {247} (\bibinfo {year} {1978})}\BibitemShut {NoStop}%
\bibitem [{\citenamefont {Katz}(1984)}]{Katz1984}%
  \BibitemOpen
  \bibfield  {author} {\bibinfo {author} {\bibfnamefont {J.}~\bibnamefont
  {Katz}},\ }\href@noop {} {\bibfield  {journal} {\bibinfo  {journal} {Proc. R.
  Soc. London A Math. Phys. Eng. Sci.}\ }\textbf {\bibinfo {volume} {391}}
  (\bibinfo {year} {1984})}\BibitemShut {NoStop}%
\bibitem [{\citenamefont {Carter}(1989)}]{Carter1989}%
  \BibitemOpen
  \bibfield  {author} {\bibinfo {author} {\bibfnamefont {B.}~\bibnamefont
  {Carter}},\ }\enquote {\bibinfo {title} {{Covariant theory of conductivity in
  ideal fluid or solid media}},}\ in\ \href {\doibase 10.1007/BFb0084028}
  {\emph {\bibinfo {booktitle} {Relativ. Fluid Dyn.}}},\ \bibinfo {editor}
  {edited by\ \bibinfo {editor} {\bibfnamefont {A.~M.}\ \bibnamefont {Anile}}\
  and\ \bibinfo {editor} {\bibfnamefont {Y.}~\bibnamefont {Choquet-Bruhat}}}\
  (\bibinfo  {publisher} {Springer},\ \bibinfo {address} {Berlin, Heidelberg},\
  \bibinfo {year} {1989})\ pp.\ \bibinfo {pages} {1--64}\BibitemShut {NoStop}%
\bibitem [{\citenamefont {Etienne}\ \emph {et~al.}(2010)\citenamefont
  {Etienne}, \citenamefont {Liu},\ and\ \citenamefont
  {Shapiro}}]{Etienneetal2010}%
  \BibitemOpen
  \bibfield  {author} {\bibinfo {author} {\bibfnamefont {Z.~B.}\ \bibnamefont
  {Etienne}}, \bibinfo {author} {\bibfnamefont {Y.~T.}\ \bibnamefont {Liu}}, \
  and\ \bibinfo {author} {\bibfnamefont {S.~L.}\ \bibnamefont {Shapiro}},\
  }\href {\doibase 10.1103/PhysRevD.82.084031} {\bibfield  {journal} {\bibinfo
  {journal} {Phys. Rev. D}\ }\textbf {\bibinfo {volume} {82}},\ \bibinfo
  {pages} {084031} (\bibinfo {year} {2010})}\BibitemShut {NoStop}%
\bibitem [{\citenamefont {Etienne}\ \emph {et~al.}(2012)\citenamefont
  {Etienne}, \citenamefont {Paschalidis}, \citenamefont {Liu},\ and\
  \citenamefont {Shapiro}}]{Etienneetal2011}%
  \BibitemOpen
  \bibfield  {author} {\bibinfo {author} {\bibfnamefont {Z.~B.}\ \bibnamefont
  {Etienne}}, \bibinfo {author} {\bibfnamefont {V.}~\bibnamefont
  {Paschalidis}}, \bibinfo {author} {\bibfnamefont {Y.~T.}\ \bibnamefont
  {Liu}}, \ and\ \bibinfo {author} {\bibfnamefont {S.~L.}\ \bibnamefont
  {Shapiro}},\ }\href {\doibase 10.1103/PhysRevD.85.024013} {\bibfield
  {journal} {\bibinfo  {journal} {Phys. Rev. D}\ }\textbf {\bibinfo {volume}
  {85}},\ \bibinfo {pages} {024013} (\bibinfo {year} {2012})}\BibitemShut
  {NoStop}%
\bibitem [{\citenamefont {Gammie}\ \emph {et~al.}(2003)\citenamefont {Gammie},
  \citenamefont {McKinney},\ and\ \citenamefont {{To ́th}}}]{Gammie2003}%
  \BibitemOpen
  \bibfield  {author} {\bibinfo {author} {\bibfnamefont {C.~F.}\ \bibnamefont
  {Gammie}}, \bibinfo {author} {\bibfnamefont {J.~C.}\ \bibnamefont
  {McKinney}}, \ and\ \bibinfo {author} {\bibfnamefont {G.~Ì.}\ \bibnamefont
  {{To ́th}}},\ }\href {\doibase 10.1086/374594} {\bibfield  {journal}
  {\bibinfo  {journal} {Astrophys. J.}\ }\textbf {\bibinfo {volume} {589}},\
  \bibinfo {pages} {444} (\bibinfo {year} {2003})},\ \Eprint
  {http://arxiv.org/abs/0301509} {arXiv:0301509 [astro-ph]} \BibitemShut
  {NoStop}%
\bibitem [{\citenamefont {Frauendiener}(2003)}]{Frauendiener2003}%
  \BibitemOpen
  \bibfield  {author} {\bibinfo {author} {\bibfnamefont {J.}~\bibnamefont
  {Frauendiener}},\ }\href {\doibase 10.1088/0264-9381/20/14/102} {\bibfield
  {journal} {\bibinfo  {journal} {Class. Quantum Gravity}\ }\textbf {\bibinfo
  {volume} {20}},\ \bibinfo {pages} {L193} (\bibinfo {year}
  {2003})}\BibitemShut {NoStop}%
\bibitem [{\citenamefont {Walton}(2005)}]{Walton2005}%
  \BibitemOpen
  \bibfield  {author} {\bibinfo {author} {\bibfnamefont {R.~A.}\ \bibnamefont
  {Walton}},\ }\href@noop {} {\bibfield  {journal} {\bibinfo  {journal} {Houst.
  J. Math.}\ }\textbf {\bibinfo {volume} {31}},\ \bibinfo {pages} {145}
  (\bibinfo {year} {2005})},\ \Eprint {http://arxiv.org/abs/0502233}
  {arXiv:0502233 [astro-ph]} \BibitemShut {NoStop}%
\bibitem [{\citenamefont {Oliynyk}\ \emph {et~al.}(2012)\citenamefont
  {Oliynyk}, \citenamefont {C},\ and\ \citenamefont {{Choquet-Bruhat
  Y}}}]{Oliynyk2012}%
  \BibitemOpen
  \bibfield  {author} {\bibinfo {author} {\bibfnamefont {T.~A.}\ \bibnamefont
  {Oliynyk}}, \bibinfo {author} {\bibfnamefont {E.~L.}\ \bibnamefont {C}}, \
  and\ \bibinfo {author} {\bibfnamefont {D.~W.-M.}\ \bibnamefont
  {{Choquet-Bruhat Y}}},\ }\href {\doibase 10.1088/0264-9381/29/15/155013}
  {\bibfield  {journal} {\bibinfo  {journal} {Class. Quantum Gravity}\ }\textbf
  {\bibinfo {volume} {29}},\ \bibinfo {pages} {155013} (\bibinfo {year}
  {2012})}\BibitemShut {NoStop}%
\bibitem [{\citenamefont {Oliynyk}(2014)}]{Oliynyk2014}%
  \BibitemOpen
  \bibfield  {author} {\bibinfo {author} {\bibfnamefont {T.~A.}\ \bibnamefont
  {Oliynyk}},\ }\href {http://arxiv.org/abs/1501.00045} {\  (\bibinfo {year}
  {2014})},\ \Eprint {http://arxiv.org/abs/1501.00045} {arXiv:1501.00045}
  \BibitemShut {NoStop}%
\bibitem [{\citenamefont {Uryu}\ \emph {et~al.}(2010)\citenamefont {Uryu},
  \citenamefont {Gourgoulhon},\ and\ \citenamefont {Markakis}}]{Uryu2010}%
  \BibitemOpen
  \bibfield  {author} {\bibinfo {author} {\bibfnamefont {K.}~\bibnamefont
  {Uryu}}, \bibinfo {author} {\bibfnamefont {E.}~\bibnamefont {Gourgoulhon}}, \
  and\ \bibinfo {author} {\bibfnamefont {C.}~\bibnamefont {Markakis}},\ }\href
  {\doibase 10.1103/PhysRevD.82.104054} {\bibfield  {journal} {\bibinfo
  {journal} {Phys. Rev. D}\ }\textbf {\bibinfo {volume} {82}} (\bibinfo {year}
  {2010}),\ 10.1103/PhysRevD.82.104054}\BibitemShut {NoStop}%
\bibitem [{\citenamefont {Gourgoulhon}\ \emph
  {et~al.}(2011{\natexlab{b}})\citenamefont {Gourgoulhon}, \citenamefont
  {Markakis}, \citenamefont {Uryu},\ and\ \citenamefont
  {Eriguchi}}]{Gourgoulhon2011}%
  \BibitemOpen
  \bibfield  {author} {\bibinfo {author} {\bibfnamefont {E.}~\bibnamefont
  {Gourgoulhon}}, \bibinfo {author} {\bibfnamefont {C.}~\bibnamefont
  {Markakis}}, \bibinfo {author} {\bibfnamefont {K.}~\bibnamefont {Uryu}}, \
  and\ \bibinfo {author} {\bibfnamefont {Y.}~\bibnamefont {Eriguchi}},\ }\href
  {\doibase 10.1103/PhysRevD.83.104007} {\bibfield  {journal} {\bibinfo
  {journal} {Phys. Rev. D}\ }\textbf {\bibinfo {volume} {83}},\ \bibinfo
  {pages} {25} (\bibinfo {year} {2011}{\natexlab{b}})},\ \Eprint
  {http://arxiv.org/abs/1101.3497} {arXiv:1101.3497} \BibitemShut {NoStop}%
\bibitem [{\citenamefont {Uryu}\ \emph {et~al.}(2014)\citenamefont {Uryu},
  \citenamefont {Gourgoulhon}, \citenamefont {Markakis}, \citenamefont
  {Fujisawa}, \citenamefont {Tsokaros},\ and\ \citenamefont
  {Eriguchi}}]{Uryu2014}%
  \BibitemOpen
  \bibfield  {author} {\bibinfo {author} {\bibfnamefont {K.}~\bibnamefont
  {Uryu}}, \bibinfo {author} {\bibfnamefont {E.}~\bibnamefont {Gourgoulhon}},
  \bibinfo {author} {\bibfnamefont {C.~M.}\ \bibnamefont {Markakis}}, \bibinfo
  {author} {\bibfnamefont {K.}~\bibnamefont {Fujisawa}}, \bibinfo {author}
  {\bibfnamefont {A.}~\bibnamefont {Tsokaros}}, \ and\ \bibinfo {author}
  {\bibfnamefont {Y.}~\bibnamefont {Eriguchi}},\ }\href {\doibase
  10.1103/PhysRevD.90.101501} {\bibfield  {journal} {\bibinfo  {journal} {Phys.
  Rev. D}\ }\textbf {\bibinfo {volume} {90}},\ \bibinfo {pages} {101501}
  (\bibinfo {year} {2014})}\BibitemShut {NoStop}%
\bibitem [{\citenamefont {Haskell}\ \emph {et~al.}(2008)\citenamefont
  {Haskell}, \citenamefont {Samuelsson}, \citenamefont {Glampedakis},\ and\
  \citenamefont {Andersson}}]{Haskelletal2008}%
  \BibitemOpen
  \bibfield  {author} {\bibinfo {author} {\bibfnamefont {B.}~\bibnamefont
  {Haskell}}, \bibinfo {author} {\bibfnamefont {L.}~\bibnamefont {Samuelsson}},
  \bibinfo {author} {\bibfnamefont {K.}~\bibnamefont {Glampedakis}}, \ and\
  \bibinfo {author} {\bibfnamefont {N.}~\bibnamefont {Andersson}},\ }\href@noop
  {} {\bibfield  {journal} {\bibinfo  {journal} {Not. R}\ }\textbf {\bibinfo
  {volume} {385}} (\bibinfo {year} {2008})}\BibitemShut {NoStop}%
\bibitem [{\citenamefont {Chern}(1996)}]{ShiingShenChern1996}%
  \BibitemOpen
  \bibfield  {author} {\bibinfo {author} {\bibfnamefont {S.-S.}\ \bibnamefont
  {Chern}},\ }\href {http://www.ams.org/notices/199609/chern.pdf} {\bibfield
  {journal} {\bibinfo  {journal} {Not. Am. Math. Soc.}\ }\textbf {\bibinfo
  {volume} {43}},\ \bibinfo {pages} {959} (\bibinfo {year} {1996})}\BibitemShut
  {NoStop}%
\bibitem [{\citenamefont {Asanov}(1985)}]{Asanov1989}%
  \BibitemOpen
  \bibfield  {author} {\bibinfo {author} {\bibfnamefont {G.~S.}\ \bibnamefont
  {Asanov}},\ }\href@noop {} {\emph {\bibinfo {title} {{Finsler Geometry,
  Relativity and Gauge Theories}}}}\ (\bibinfo  {publisher} {Springer},\
  \bibinfo {year} {1985})\BibitemShut {NoStop}%
\bibitem [{\citenamefont {Lovelock}\ and\ \citenamefont
  {Rund}(1989)}]{Lovelock1989}%
  \BibitemOpen
  \bibfield  {author} {\bibinfo {author} {\bibfnamefont {D.}~\bibnamefont
  {Lovelock}}\ and\ \bibinfo {author} {\bibfnamefont {H.}~\bibnamefont
  {Rund}},\ }\href
  {http://www.amazon.com/Tensors-Differential-Forms-Variational-Principles/dp/%
0486658406} {\emph {\bibinfo {title} {{Tensors, Differential Forms, and
  Variational Principles}}}}\ (\bibinfo  {publisher} {Dover Publications},\
  \bibinfo {year} {1989})\BibitemShut {NoStop}%
\bibitem [{\citenamefont {Randers}(1941)}]{Randers1941}%
  \BibitemOpen
  \bibfield  {author} {\bibinfo {author} {\bibfnamefont {G.}~\bibnamefont
  {Randers}},\ }\href {\doibase 10.1103/PhysRev.59.195} {\bibfield  {journal}
  {\bibinfo  {journal} {Phys. Rev.}\ }\textbf {\bibinfo {volume} {59}},\
  \bibinfo {pages} {195} (\bibinfo {year} {1941})}\BibitemShut {NoStop}%
\bibitem [{\citenamefont {Basilakos}\ \emph {et~al.}(2013)\citenamefont
  {Basilakos}, \citenamefont {Kouretsis}, \citenamefont {Saridakis},\ and\
  \citenamefont {Stavrinos}}]{Basilakos2013}%
  \BibitemOpen
  \bibfield  {author} {\bibinfo {author} {\bibfnamefont {S.}~\bibnamefont
  {Basilakos}}, \bibinfo {author} {\bibfnamefont {A.~P.}\ \bibnamefont
  {Kouretsis}}, \bibinfo {author} {\bibfnamefont {E.~N.}\ \bibnamefont
  {Saridakis}}, \ and\ \bibinfo {author} {\bibfnamefont {P.~C.}\ \bibnamefont
  {Stavrinos}},\ }\href {\doibase 10.1103/PhysRevD.88.123510} {\bibfield
  {journal} {\bibinfo  {journal} {Phys. Rev. D}\ }\textbf {\bibinfo {volume}
  {88}},\ \bibinfo {pages} {123510} (\bibinfo {year} {2013})}\BibitemShut
  {NoStop}%
\bibitem [{\citenamefont {Asanov}(1977)}]{Asanov1977}%
  \BibitemOpen
  \bibfield  {author} {\bibinfo {author} {\bibfnamefont {G.~S.}\ \bibnamefont
  {Asanov}},\ }\href@noop {} {\bibfield  {journal} {\bibinfo  {journal} {Math.
  Phys}\ }\textbf {\bibinfo {volume} {11}},\ \bibinfo {pages} {221} (\bibinfo
  {year} {1977})}\BibitemShut {NoStop}%
\bibitem [{\citenamefont {Gibbons}\ and\ \citenamefont
  {Warnick}(2011)}]{Gibbons2011a}%
  \BibitemOpen
  \bibfield  {author} {\bibinfo {author} {\bibfnamefont {G.~W.}\ \bibnamefont
  {Gibbons}}\ and\ \bibinfo {author} {\bibfnamefont {C.~M.}\ \bibnamefont
  {Warnick}},\ }\href@noop {} {\bibfield  {journal} {\bibinfo  {journal}
  {Contemp. Phys.}\ }\textbf {\bibinfo {volume} {52}} (\bibinfo {year}
  {2011})}\BibitemShut {NoStop}%
\bibitem [{\citenamefont {Gibbons}\ \emph {et~al.}(2009)\citenamefont
  {Gibbons}, \citenamefont {Herdeiro}, \citenamefont {Warnick},\ and\
  \citenamefont {Werner}}]{Gibbons2009}%
  \BibitemOpen
  \bibfield  {author} {\bibinfo {author} {\bibfnamefont {G.}~\bibnamefont
  {Gibbons}}, \bibinfo {author} {\bibfnamefont {C.}~\bibnamefont {Herdeiro}},
  \bibinfo {author} {\bibfnamefont {C.}~\bibnamefont {Warnick}}, \ and\
  \bibinfo {author} {\bibfnamefont {M.}~\bibnamefont {Werner}},\ }\href
  {\doibase 10.1103/PhysRevD.79.044022} {\bibfield  {journal} {\bibinfo
  {journal} {Phys. Rev. D}\ }\textbf {\bibinfo {volume} {79}},\ \bibinfo
  {pages} {37} (\bibinfo {year} {2009})},\ \Eprint
  {http://arxiv.org/abs/0811.2877} {arXiv:0811.2877} \BibitemShut {NoStop}%
\bibitem [{\citenamefont {Gibbons}\ and\ \citenamefont
  {Warnick}(2010)}]{Gibbons2010}%
  \BibitemOpen
  \bibfield  {author} {\bibinfo {author} {\bibfnamefont {G.}~\bibnamefont
  {Gibbons}}\ and\ \bibinfo {author} {\bibfnamefont {C.}~\bibnamefont
  {Warnick}},\ }\href {\doibase 10.1016/j.aop.2009.12.007} {\bibfield
  {journal} {\bibinfo  {journal} {Ann. Phys. (N. Y).}\ }\textbf {\bibinfo
  {volume} {325}},\ \bibinfo {pages} {909} (\bibinfo {year}
  {2010})}\BibitemShut {NoStop}%
\bibitem [{\citenamefont {Brennan}\ \emph {et~al.}(2013)\citenamefont
  {Brennan}, \citenamefont {Gralla},\ and\ \citenamefont
  {Jacobson}}]{Brennan2013}%
  \BibitemOpen
  \bibfield  {author} {\bibinfo {author} {\bibfnamefont {T.~D.}\ \bibnamefont
  {Brennan}}, \bibinfo {author} {\bibfnamefont {S.~E.}\ \bibnamefont {Gralla}},
  \ and\ \bibinfo {author} {\bibfnamefont {T.}~\bibnamefont {Jacobson}},\
  }\href {\doibase 10.1088/0264-9381/30/19/195012} {\bibfield  {journal}
  {\bibinfo  {journal} {Class. Quantum Gravity}\ }\textbf {\bibinfo {volume}
  {30}},\ \bibinfo {pages} {195012} (\bibinfo {year} {2013})}\BibitemShut
  {NoStop}%
\bibitem [{\citenamefont {Gralla}\ and\ \citenamefont
  {Jacobson}(2014)}]{Gralla2014}%
  \BibitemOpen
  \bibfield  {author} {\bibinfo {author} {\bibfnamefont {S.~E.}\ \bibnamefont
  {Gralla}}\ and\ \bibinfo {author} {\bibfnamefont {T.}~\bibnamefont
  {Jacobson}},\ }\href {\doibase 10.1093/mnras/stu1690} {\bibfield  {journal}
  {\bibinfo  {journal} {Mon. Not. R. Astron. Soc.}\ }\textbf {\bibinfo {volume}
  {445}},\ \bibinfo {pages} {2500} (\bibinfo {year} {2014})}\BibitemShut
  {NoStop}%
\bibitem [{\citenamefont {Brennan}\ and\ \citenamefont
  {Gralla}(2014)}]{Brennan2014}%
  \BibitemOpen
  \bibfield  {author} {\bibinfo {author} {\bibfnamefont {T.~D.}\ \bibnamefont
  {Brennan}}\ and\ \bibinfo {author} {\bibfnamefont {S.~E.}\ \bibnamefont
  {Gralla}},\ }\href {\doibase 10.1103/PhysRevD.89.103013} {\bibfield
  {journal} {\bibinfo  {journal} {Phys. Rev. D}\ }\textbf {\bibinfo {volume}
  {89}},\ \bibinfo {pages} {103013} (\bibinfo {year} {2014})}\BibitemShut
  {NoStop}%
\bibitem [{\citenamefont {Gralla}\ and\ \citenamefont
  {Jacobson}(2015)}]{Gralla2015}%
  \BibitemOpen
  \bibfield  {author} {\bibinfo {author} {\bibfnamefont {S.~E.}\ \bibnamefont
  {Gralla}}\ and\ \bibinfo {author} {\bibfnamefont {T.}~\bibnamefont
  {Jacobson}},\ }\href {\doibase 10.1103/PhysRevD.92.043002} {\bibfield
  {journal} {\bibinfo  {journal} {Phys. Rev. D}\ }\textbf {\bibinfo {volume}
  {92}},\ \bibinfo {pages} {043002} (\bibinfo {year} {2015})}\BibitemShut
  {NoStop}%
\bibitem [{\citenamefont {Gralla}\ \emph {et~al.}(2015)\citenamefont {Gralla},
  \citenamefont {Lupsasca},\ and\ \citenamefont {Rodriguez}}]{Gralla2015a}%
  \BibitemOpen
  \bibfield  {author} {\bibinfo {author} {\bibfnamefont {S.~E.}\ \bibnamefont
  {Gralla}}, \bibinfo {author} {\bibfnamefont {A.}~\bibnamefont {Lupsasca}}, \
  and\ \bibinfo {author} {\bibfnamefont {M.~J.}\ \bibnamefont {Rodriguez}},\
  }\href {\doibase 10.1103/PhysRevD.92.044053} {\bibfield  {journal} {\bibinfo
  {journal} {Phys. Rev. D}\ }\textbf {\bibinfo {volume} {92}},\ \bibinfo
  {pages} {044053} (\bibinfo {year} {2015})}\BibitemShut {NoStop}%
\bibitem [{\citenamefont {Gralla}\ and\ \citenamefont
  {Zimmerman}(2016)}]{Gralla2016}%
  \BibitemOpen
  \bibfield  {author} {\bibinfo {author} {\bibfnamefont {S.~E.}\ \bibnamefont
  {Gralla}}\ and\ \bibinfo {author} {\bibfnamefont {P.}~\bibnamefont
  {Zimmerman}},\ }\href {\doibase 10.1103/PhysRevD.93.123016} {\bibfield
  {journal} {\bibinfo  {journal} {Phys. Rev. D}\ }\textbf {\bibinfo {volume}
  {93}},\ \bibinfo {pages} {123016} (\bibinfo {year} {2016})}\BibitemShut
  {NoStop}%
\bibitem [{\citenamefont {Gralla}\ \emph
  {et~al.}(2016{\natexlab{a}})\citenamefont {Gralla}, \citenamefont
  {Lupsasca},\ and\ \citenamefont {Rodriguez}}]{Gralla2016a}%
  \BibitemOpen
  \bibfield  {author} {\bibinfo {author} {\bibfnamefont {S.~E.}\ \bibnamefont
  {Gralla}}, \bibinfo {author} {\bibfnamefont {A.}~\bibnamefont {Lupsasca}}, \
  and\ \bibinfo {author} {\bibfnamefont {M.~J.}\ \bibnamefont {Rodriguez}},\
  }\href {\doibase 10.1103/PhysRevD.93.044038} {\bibfield  {journal} {\bibinfo
  {journal} {Phys. Rev. D}\ }\textbf {\bibinfo {volume} {93}},\ \bibinfo
  {pages} {044038} (\bibinfo {year} {2016}{\natexlab{a}})}\BibitemShut
  {NoStop}%
\bibitem [{\citenamefont {Gralla}\ \emph
  {et~al.}(2016{\natexlab{b}})\citenamefont {Gralla}, \citenamefont
  {Lupsasca},\ and\ \citenamefont {Strominger}}]{Gralla2016b}%
  \BibitemOpen
  \bibfield  {author} {\bibinfo {author} {\bibfnamefont {S.~E.}\ \bibnamefont
  {Gralla}}, \bibinfo {author} {\bibfnamefont {A.}~\bibnamefont {Lupsasca}}, \
  and\ \bibinfo {author} {\bibfnamefont {A.}~\bibnamefont {Strominger}},\
  }\href {\doibase 10.1103/PhysRevD.93.104041} {\bibfield  {journal} {\bibinfo
  {journal} {Phys. Rev. D}\ }\textbf {\bibinfo {volume} {93}},\ \bibinfo
  {pages} {104041} (\bibinfo {year} {2016}{\natexlab{b}})}\BibitemShut
  {NoStop}%
\bibitem [{\citenamefont {Gralla}\ \emph
  {et~al.}(2016{\natexlab{c}})\citenamefont {Gralla}, \citenamefont
  {Lupsasca},\ and\ \citenamefont {Philippov}}]{Gralla2016c}%
  \BibitemOpen
  \bibfield  {author} {\bibinfo {author} {\bibfnamefont {S.~E.}\ \bibnamefont
  {Gralla}}, \bibinfo {author} {\bibfnamefont {A.}~\bibnamefont {Lupsasca}}, \
  and\ \bibinfo {author} {\bibfnamefont {A.}~\bibnamefont {Philippov}},\ }\href
  {http://arxiv.org/abs/1604.04625} {\  (\bibinfo {year}
  {2016}{\natexlab{c}})},\ \Eprint {http://arxiv.org/abs/1604.04625}
  {arXiv:1604.04625} \BibitemShut {NoStop}%
\bibitem [{\citenamefont {Comp{\`{e}}re}\ \emph {et~al.}(2016)\citenamefont
  {Comp{\`{e}}re}, \citenamefont {Gralla},\ and\ \citenamefont
  {Lupsasca}}]{Compere2016}%
  \BibitemOpen
  \bibfield  {author} {\bibinfo {author} {\bibfnamefont {G.}~\bibnamefont
  {Comp{\`{e}}re}}, \bibinfo {author} {\bibfnamefont {S.~E.}\ \bibnamefont
  {Gralla}}, \ and\ \bibinfo {author} {\bibfnamefont {A.}~\bibnamefont
  {Lupsasca}},\ }\href {http://arxiv.org/abs/1606.06727} {\  (\bibinfo {year}
  {2016})},\ \Eprint {http://arxiv.org/abs/1606.06727} {arXiv:1606.06727}
  \BibitemShut {NoStop}%
\bibitem [{\citenamefont {Gourgoulhon}(2010)}]{Gourg10}%
  \BibitemOpen
  \bibfield  {author} {\bibinfo {author} {\bibfnamefont {E.}~\bibnamefont
  {Gourgoulhon}},\ }\href {http://arxiv.org/abs/1003.5015} {\bibfield
  {journal} {\bibinfo  {journal} {An Introd. to theory rotating Relativ.
  Stars}\ } (\bibinfo {year} {2010})},\ \Eprint
  {http://arxiv.org/abs/1003.5015} {arXiv:1003.5015} \BibitemShut {NoStop}%
\bibitem [{\citenamefont {Rosswog}(2015)}]{Rosswog2015}%
  \BibitemOpen
  \bibfield  {author} {\bibinfo {author} {\bibfnamefont {S.}~\bibnamefont
  {Rosswog}},\ }\href {\doibase 10.1007/lrca-2015-1} {\bibfield  {journal}
  {\bibinfo  {journal} {Living Rev. Comput. Astrophys.}\ }\textbf {\bibinfo
  {volume} {1}},\ \bibinfo {pages} {1} (\bibinfo {year} {2015})}\BibitemShut
  {NoStop}%
\bibitem [{\citenamefont {Lasota}\ \emph {et~al.}(2014)\citenamefont {Lasota},
  \citenamefont {Gourgoulhon}, \citenamefont {Abramowicz}, \citenamefont
  {Tchekhovskoy},\ and\ \citenamefont {Narayan}}]{Lasota2014}%
  \BibitemOpen
  \bibfield  {author} {\bibinfo {author} {\bibfnamefont {J.-P.}\ \bibnamefont
  {Lasota}}, \bibinfo {author} {\bibfnamefont {E.}~\bibnamefont {Gourgoulhon}},
  \bibinfo {author} {\bibfnamefont {M.}~\bibnamefont {Abramowicz}}, \bibinfo
  {author} {\bibfnamefont {A.}~\bibnamefont {Tchekhovskoy}}, \ and\ \bibinfo
  {author} {\bibfnamefont {R.}~\bibnamefont {Narayan}},\ }\href {\doibase
  10.1103/PhysRevD.89.024041} {\bibfield  {journal} {\bibinfo  {journal} {Phys.
  Rev. D}\ }\textbf {\bibinfo {volume} {89}},\ \bibinfo {pages} {024041}
  (\bibinfo {year} {2014})}\BibitemShut {NoStop}%
\bibitem [{\citenamefont {Penrose}\ and\ \citenamefont
  {Floyd}(1971)}]{PENROSE1971}%
  \BibitemOpen
  \bibfield  {author} {\bibinfo {author} {\bibfnamefont {R.}~\bibnamefont
  {Penrose}}\ and\ \bibinfo {author} {\bibfnamefont {R.~M.}\ \bibnamefont
  {Floyd}},\ }\href {\doibase 10.1038/physci229177a0} {\bibfield  {journal}
  {\bibinfo  {journal} {Nat. Phys. Sci.}\ }\textbf {\bibinfo {volume} {229}},\
  \bibinfo {pages} {177} (\bibinfo {year} {1971})}\BibitemShut {NoStop}%
\end{thebibliography}%

\end{document}